%% file: 2012-PS.tex
\documentclass[journal=jctcce,manuscript=article]{achemso} 

\usepackage[version=3]{mhchem} 
\usepackage{comment}
\usepackage{mciteplus}
\usepackage{algorithmic} 
\usepackage{url}
\usepackage{epsfig}
\usepackage{pst-grad} 
\usepackage{pst-plot} 
\usepackage{alphalph} 
\usepackage{amsmath}
\usepackage{bm}
\usepackage{lscape}

\newcommand{\angstrom}{\textup{\AA}}

\long\def\beginpgfgraphicnamed#1#2\endpgfgraphicnamed{\includegraphics{#1}}

\renewcommand {\vec} [1] {{\bm #1}}

\author{Pablo Garc\'ia-Risue\~no}
\email{risueno@physik.hu-berlin.de}
\affiliation
[Institut f\"ur Physik, Humboldt Universit\"at zu Berlin, 12489 Berlin, Germany -
Instituto de Qu\'imica F\'isica Rocasolano (CSIC), C/ Serrano 119, 28006 Madrid, Spain]
{Humboldt Universit\"at zu Berlin and CSIC}
\author{Joseba Alberdi-Rodriguez}
\affiliation
[Dept. of Computer Architecture and Technology, Nano-Bio Spectroscopy Group and European Theoretical Spectroscopy Facility, Spanish node, University of the Basque Country UPV/EHU, M. Lardizabal, 1, 20018 Donostia/San Sebasti\'an, Spain]
{University of the Basque Country UPV/EHU}
\author{Micael J. T. Oliveira}
\affiliation
[Center for Computational Physics, University of Coimbra, Rua Larga, 3004-516 Coimbra, Portugal]
{University of Coimbra}
\author{Xavier Andrade}
\affiliation
[Dept. of Chemistry and Chemical Biology, Harvard University, 12 Oxford street, Cambridge, MA 02138, USA]
{Harvard University}
\author{Michael Pippig}
\affiliation
[Dept. of Mathematics, Chemnitz University of Technology, 09107 Chemnitz, Germany]
{Chemnitz University of Technology}
\author{Javier Muguerza}
\author{Agustin Arruabarrena}
\affiliation
[Dept. of Computer Architecture and Technology
University of the Basque Country UPV/EHU, M. Lardizabal, 1, 20018 Donostia, Spain]
{University of the Basque Country UPV/EHU}
\author{\'Angel Rubio}
\affiliation
[Nano-Bio Spectroscopy Group and European Theoretical Spectroscopy Facility, Spanish node, 
University of the Basque Country UPV/EHU, Edif. Joxe Mari Korta, Av. Tolosa 72, 
20018 Donostia/San Sebasti\'an, Spain - Centro de F\'isica de Materiales, University of the Basque Country UPV/EHU, 
20018 Donostia/San Sebasti\'an, Spain - Fritz-Haber Institut der Max-Planck Gesellschaft, 
Faradayweg 4-6, D-14195 Berlin-Dahlem, Germany]
{European Theoretical Spectroscopy Facility and Fritz-Haber Institut (MPG)}

\title{A survey of the parallel performance and the accuracy of Poisson solvers for electronic structure calculations}

\begin{document}
\begin{abstract}
  We present an analysis of different methods to calculate the
  classical electrostatic Hartree potential created by charge
  distributions.  Our goal is to provide the reader with an estimation
  on the performance ---in terms of both numerical complexity and
  accuracy--- of popular Poisson solvers, and to give an intuitive
  idea on the way these solvers operate.  Highly parallelisable
  routines have been implemented in the first-principle simulation
  code {\sc Octopus} to be used in our tests, so that reliable
  conclusions about the capability of methods to tackle large systems
  in cluster computing can be obtained from our work.
  \vspace{0.4cm}
  \vspace{0.2cm}\\
\end{abstract}

\renewcommand{\thefootnote}{\alph{footnote}} 

\tableofcontents

\section{Introduction}\label{sec:introduction}

The electrostatic interaction between charges is one of the most
important phenomena of physics and chemistry, which makes the
calculation of the energy and potential associated to a distribution
of charges one of the most common problems in scientific computing.
The calculation of potentials created by pairwise interactions is
ubiquitous in atomic and molecular simulations, and also in fields
like quantum chemistry, solid state physics, fluid dynamics, plasma
physics, and astronomy, among others.

In particular, the electrostatic interaction is a key part in the
density functional theory (DFT)~\cite{hktheorem,ksansatz} and
time-dependent density functional theory (TDDFT)~\cite{Run1984PRL}
formulations of quantum mechanics. In DFT (TDDFT) the many-body
Schr\"odinger equation is replaced by a set of single particle
equations, the Kohn--Sham (time-dependent Kohn--Sham) equations, which
include an effective potential that reproduces the interelectronic
interaction.  Such effective potential is usually divided into three
terms: Hartree potential, exchange-correlation potential and external
potential.  The Hartree term corresponds to the classical
electrostatic potential generated by electronic charge distributions
due to the (delocalised) electronic states.

The calculation of the potential associated to a charge distribution
is then an important step in most numerical implementations of DFT.
In addition, the calculation of the electrostatic potential associated
to a charge density can appear in other contexts in electronic
structure theory, like the approximation of the exchange
term~\cite{ldaperdewzunger,Umezawa2006,Andrade2011}, or the
calculation of integrals that appear in
Hartree--Fock~\cite{Hartree1957Book,Fock1964Book} or
Casida~\cite{Casida1996} theories.  It is understandable then that the
calculation of the electrostatic potential has received much interest
in the recent past within the community of electronic structure
researchers.

Since at present, the complexity of the problems we want to tackle
requires massively parallel computational
platforms~\cite{GR2012IJMPC}, every algorithm for a time-consuming
task must not only be efficient in serial, but also needs to keep its
efficiency when run in a very large number of parallel processes
(approximately 300,000 CPUs). The electrostatic interaction is
non-local, and thus the information corresponding to different points
interacting with each other can be stored in different computing units
(processor cores), with a non-negligible time for data-communication
among them. This makes critical the choice of the algorithm for the
calculation of the Hartree potential, since as well as different
accuracies, different algorithms also have different efficiencies.
Thanks to the efficiency offered by the current generation of solvers,
calculation of the Hartree potential usually contributes a minor
fraction of the computational time of a typical DFT calculation.
However, there are cases where the Poisson solver hinders the
numerical performance of electronic structure calculations. For
example, in parallel implementations of DFT it is common to distribute
the Kohn--Sham orbitals between processors~\cite{Gygi2006}; for
real-time TDDFT in particular, this is a very efficient
strategy~\cite{Alo2008PRL,And2009JCTC,And2012JPCM}. Nonetheless, since
a single Poisson equation needs to be solved independently on the
number of orbitals, the calculation of the Hartree potential becomes
an important bottleneck for an efficient
parallelisation~\cite{Alb2010Thesis, And2012JPCM} as predicted by
Amdahl's law~\cite{Amd1967}, if it is not optimally parallelised.

The objective of this article is, therefore, to analyse the relative
efficiencies and accuracies of some of the most popular methods to
calculate the Hartree potential created by charge distributions.  Our
purpose is to provide the reader with estimates on the features of
this solvers that make it possible to choose which of them is the most
appropriate for his electronic structure calculations.

We start by giving a brief theoretical introduction to the Poisson
equation problem and on different methods to solve it.  Next, we
discuss the details of our implementation and the parallel computers
we use. Following, we present the results of our numerical
experiments. We finish by stating our conclusions.  More specific
derivations and analysis are provided in the supporting info.

\section{Theoretical background}\label{sec:theory}

In the context of quantum mechanics, the electrons and their electric
charge are delocalised over space forming a continuous charge
distribution \(\rho(\vec{r})\).  Such a charge density creates an
electrostatic potential $v(\vec{r})$, which is given by
\cite{Lea2001Book}
\begin{equation}
  \label{eq:Vtotal} 
  v(\vec{r})=\int d\vec{r}\ '
  \frac{1}{4\pi\epsilon_0} \frac{\rho(\vec{r}\
    ')}{|\vec{r}-\vec{r}\ '|}  \ , 
\end{equation}
where $\epsilon_0$ is the electrical permittivity of the vacuum, i.e.\ 
\(1/4\pi\) in atomic units. When considering the electric interaction
in a medium, it is possible to approximate the polarization effects by
replacing \(\epsilon_0\) by an effective permittivity,
\(\epsilon\). In the context of electronic structure calculations this
effective permittivity is used, for example, in multiscale simulations
where part of the system is approximated by a continuous polarisable
medium~\cite{Tomasi2005}.

In 3D, it is simple to show that equation \ref{eq:Vtotal} is
equivalent to the Poisson equation\index{Poisson
  equation}~\cite{Nay1985BOOK,Cas2003JCP}
\begin{equation}
  \label{eq:ecpodif} \nabla^2
  v(\vec{r})+\frac{\rho(\vec{r})}{\epsilon_0}=0  \ .
\end{equation}
In fact, this equation provides a convenient general expression that
is valid for different dimensions and boundary conditions. For
example, to study crystalline systems, usually periodic boundary
conditions are imposed. It is also possible to simulate a molecular
system interacting with ideal metallic surfaces by choosing the
appropriate boundary conditions~\cite{Olivares2011,Watson2012}. Both
formulations of the problem, i.e.\ equations \ref{eq:Vtotal} and
\ref{eq:ecpodif}, are quite useful: while some methods to calculate
the Hartree potential are based on the former, others rely on the
latter.

In order to numerically calculate the electrostatic potential we need
to discretise the problem. To this end, we use a grid representation,
which changes the charge density and the electrostatic potential to
discrete functions, with values defined over a finite number of points
distributed over a uniform mesh. Such an approach is used in many
electronic structure codes, even when another type of discretisation
is used for the orbitals. There are several ways to calculate the
potential in a grid-based approach. Probably the simplest method is to
approximate equation~\ref{eq:Vtotal} as a sum over grid points; this
approximation, however, can produce large errors, and some correction
terms need to be considered (this is discussed in
section~\ref{sec:fmm}). Other methods to calculate the potential on a
mesh can be found using Fourier transforms or finite differences and
are based on a discretisation of the Poisson equation.

Independently of the method, the direct calculation of the potential
would take $\mathcal{O}(N^2)$ operations, with $N$ the number of
points of our grid. This is prohibitive for systems beyond some
size.  Fortunately, there exist a variety of methods that, by
exploiting the properties of the problem, reduce the cost to a linear
or quasi-linear dependency. For our survey, we have selected several
parallel implementations of some of the most popular of these methods
(parallel fast Fourier transform, interpolating scaling functions,
fast multipole method, conjugate gradients, and multigrid).  In the
next subsections we introduce them and give a brief account of their
theoretical foundations and properties.

\subsection{Parallel fast Fourier transform (PFFT)}\label{sec:pfft}
\index{Poisson solver!Fast Fourier Transform}

The Fourier transform (FT) is a powerful mathematical tool both for
analytical and numerical calculations, and certainly it can be used to
calculate the electrostatic potential created by a charge distribution
represented in an equispaced grid by operating as follows. Let
$\hat{f}(\vec{k})$ be the Fourier transform of the $f(\vec{r})$
function, and $\hat{g}^{-1}(\vec{r})$ the inverse Fourier transform of
the $g(\vec{k})$ function, with $f$ and $g$ continuous functions
defined in a tridimensional space. By construction,
$f=f(\vec{r})$. The convolution property of the Fourier transform
ensures that
$\hat{f}^{-1}\big(\hat{f}(\vec{k})\big)(\vec{r})=f(\vec{r})$.
If we apply these equations to equation \ref{eq:Vtotal}, we find that

\begin{equation}\label{eq:poisson_fourier} 
v(\vec{r}) = \hat{v}^{-1}\big(\hat{v}(\vec{k})\big)(\vec{r})=\frac{1}{4\pi \epsilon_0} \hat{v}^{-1}\big( \hat{\rho}(\vec{k}) / |\vec{k}|^2  \big) \ ,
\end{equation} 
where we have used that the Fourier transform of the function $1/|\vec{r}|$ is
\cite{Cas2003JCP}
$\hat{(1/|\vec{r}|)}(\vec{k})=1/|\vec{k}|^2=1/(k_x^2+k_y^2+k_z^2)$.

Since $\rho(\vec{r})$ is represented in discrete equispaced points
($r_{j,k,l}$) at the centre of cells whose volume equals $\Omega$, the
Fourier transform of $\rho(\vec{r})$ (i.e. $\hat{\rho}(\vec{k})$) can be calculated using its
discrete Fourier transform (i.e., the last term in the next equation):
\begin{eqnarray}\label{eq:poisson_fourier2} 
  \hat{\rho}(\vec{k}) &:=& (2\pi)^{-3/2} \int d\vec{r} \ exp(-i \vec{k} \cdot \vec{r}) \rho(\vec{r})   \\
  &\simeq& (2\pi)^{-3/2} \Omega \sum_{j,k,l} \rho(r_{j,k,l}) exp(-i (k_x j + k_y k + k_z l) )  \ .
\end{eqnarray}

The use of equation \ref{eq:poisson_fourier2} in equation
\ref{eq:poisson_fourier} results in a discretised problem, which
requires the use of a discrete Fourier transform plus an inverse
discrete Fourier transform. The expression of the potential in terms
of discrete Fourier transforms makes possible the employ of the
efficient technique FFT \cite{Coo1965mc} (see below for details), so
the problem can be sovled in $\mathcal{O}(Nlog_2 N)$ steps, $N$ being
the total number of grid points.

It is to be stressed that the use of the FT automatically imposes
periodic boundary conditions on the density, and therefore to the
potential. When finite systems are studied some scheme is required to
avoid the interaction between periodic images. The simplest strategy
is to increase the size of the real-space simulation cell and set the
charge density to zero in the new points. This moves the periodic
replicas of the density away, thus decreasing their effect on the
potential. Another strategy is to replace the $1/(\epsilon_0
\vec{k}^2)$ factor of equation \ref{eq:poisson_fourier}, known as the
kernel of the Poisson equation, by a quantity that gives the free
space potential in the simulation region.  This modified kernel has
been presented in references ~\cite{Roz2006PRB} for molecules,
one-dimensional systems, and slabs. This Coulomb cut-off technique is
very efficient and easy to implement.  In our PFFT-based solver we
combine the two approaches, doubling the size of the cell and using a
modified kernel~\cite{Cas2003JCP,Roz2006PRB}. This results in a
potential that accurately reproduces the free space results.

In one dimension, the discrete Fourier transform of a set of $n$
complex numbers $f_k$ ( $f_k$ can be, for example, the values of
$\rho$ in a set of discrete points) is given by
\begin{equation}\label{eq:dft}
  F_l = \sum_{k=0}^{n-1} f_k \exp \left( -2\pi i\tfrac{kl}{n} \right) \qquad \textrm{for } l=0,\ldots,n-1  \ ,
\end{equation}
with $i$ being the imaginary unit. The inverse discrete Fourier transform is
given by
\begin{equation}\label{dft112}
  f_k = \frac{1}{n} \sum_{l=0}^{n-1} F_l \exp \left( +2\pi i\tfrac{kl}{n} \right) \qquad \textrm{for }k=0,\ldots,n-1  \ .
\end{equation}
The definitions above enable computational savings using the fact that
both the input and output data sets ($\rho(\vec{r})$ and
$v(\vec{r})$) are real. Thus $F_{n-l}=F_l^*$, and in one direction we
just need to calculate one half of the $n$ discrete Fourier
transforms.

The Poisson problem using equations \ref{eq:dft} and \ref{dft112}
would require $\mathcal{O}(N^2)$ arithmetic operations (with e.g.\
$N=n^3$), which would not represent any improvement over the cost of
evaluating the potential directly.  However, in 1965, J.~W.~Cooley and
J.~W.~Tukey published an algorithm called fast Fourier transform
(FFT)~\cite{Coo1965mc} that exploits the special structure of equation
\ref{eq:dft} in order to reduce the arithmetic complexity.  The basic
idea of the radix-2 Cooley-Tukey FFT is to split a discrete FT of even
size $n = 2m$ into two discrete FTs of half the length; e.g., for
$l=0,\ldots,m-1$ we have
\begin{align}
  F_{2l}   &= \sum_{k=0}^{m-1} f_k \exp\left(-2\pi i \tfrac{k2l}{n}\right) + f_{k+m}\exp\left(-2\pi i \tfrac{(k+m)2l}{n}\right) \nonumber\\
  &= \sum_{k=0}^{m-1} (f_k + f_{k+m}) \exp\left(-2\pi i \tfrac{kl}{m}\right)  \ ; \\
  F_{2l+1} &= \sum_{k=0}^{m-1} f_k \exp\left(-2\pi i \tfrac{k(2l+1)}{n}\right) + f_{k+m}\exp\left(-2\pi i \tfrac{(k+m)(2l+1)}{n}\right) \nonumber\\
  &= \sum_{k=0}^{m-1} \exp\left(-2\pi i\tfrac{k}{n}\right) (f_k -
  f_{k+m}) \exp\left(-2\pi i \tfrac{kl}{m}\right)  \ .
\end{align}
If we assume $n$ to be a power of two, we can apply this splitting
recursively $\log_2 n$ times, which leads to $\mathcal{O}(n \log_2 n)$
arithmetic operations for the calculation of equation \ref{eq:dft}.
There exist analogous splitting for every divisible
sizes~\cite{VanLoan1992}, even for prime sizes~\cite{IBluestein1970}.

In three dimensions a discrete FT of size $n_1\times n_2\times n_3$
can be evaluated using 1D FFTs along each direction, yielding a fast
algorithm with arithmetic complexity \(\mathcal
O(n_1n_2n_3\log(n_1n_2n_3))\). In addition, the one-dimensional
decomposition of the 3D-FFT provides a straightforward parallelisation
strategy based on a domain decomposition strategy. We now present the
two-dimensional data decomposition that was first proposed by
H.~Q.~Ding \emph{et al.}~\cite{Ding1995} and later implemented by
M.~Eleftheriou \emph{et~al.}~\cite{ElMoFi03, ElFiRa05IBM,
  ElFiRa05Europar}.

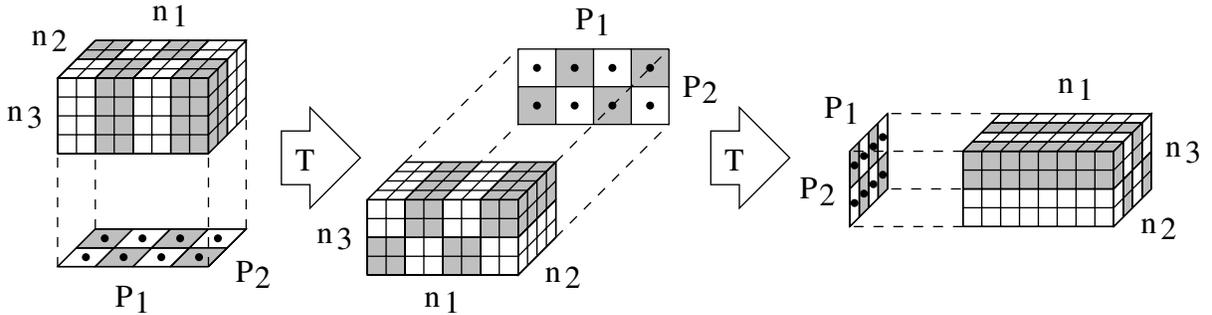
\begin{figure}[ht!]
  \centering
  \def\svgwidth{450pt} 
  \begingroup%
  \makeatletter%
  \providecommand\color[2][]{%
    \errmessage{(Inkscape) Color is used for the text in Inkscape, but the package 'color.sty' is not loaded}%
    \renewcommand\color[2][]{}%
  }%
  \providecommand\transparent[1]{%
    \errmessage{(Inkscape) Transparency is used (non-zero) for the text in Inkscape, but the package 'transparent.sty' is not loaded}%
    \renewcommand\transparent[1]{}%
  }%
  \providecommand\rotatebox[2]{#2}%
  \ifx\svgwidth\undefined%
  \setlength{\unitlength}{357.89120166bp}%
  \ifx\svgscale\undefined%
  \relax%
  \else%
  \setlength{\unitlength}{\unitlength * \real{\svgscale}}%
  \fi%
  \else%
  \setlength{\unitlength}{\svgwidth}%
  \fi%
  \global\let\svgwidth\undefined%
  \global\let\svgscale\undefined%
  \makeatother%
  \begin{picture}(1,0.26105529)%
    \put(0,0){\includegraphics[width=\unitlength]{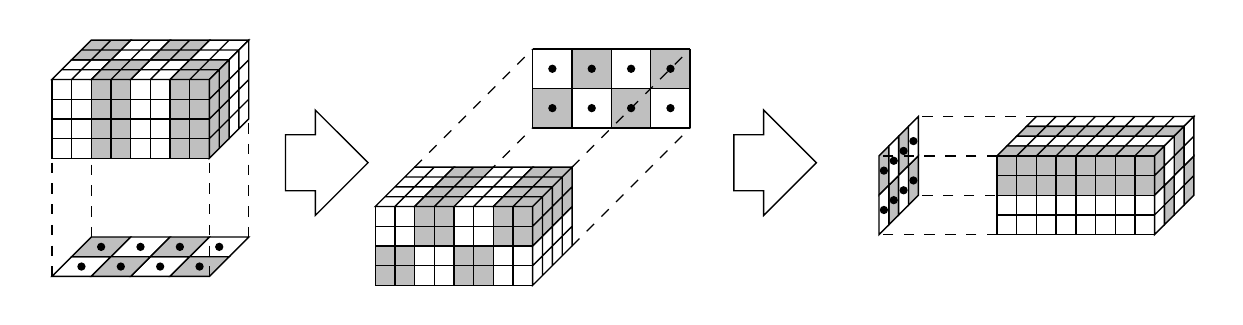}}%
    \put(0.1912808,0.0258654){\makebox(0,0)[lb]{\smash{P}}}%
    \put(0.20751479,0.02059564){\makebox(0,0)[lb]{\smash{2}}}%
    \put(0.09006289,0.00527255){\makebox(0,0)[lb]{\smash{P}}}%
    \put(0.10629688,0.00000279){\makebox(0,0)[lb]{\smash{1}}}%
    \put(0.02005838,0.22509634){\makebox(0,0)[lb]{\smash{n}}}%
    \put(0.03676178,0.21982658){\makebox(0,0)[lb]{\smash{2}}}%
    \put(0.121511,0.24568919){\makebox(0,0)[lb]{\smash{n}}}%
    \put(0.1382116,0.24041943){\makebox(0,0)[lb]{\smash{1}}}%
    \put(-0.00053447,0.16093718){\makebox(0,0)[lb]{\smash{n}}}%
    \put(0.01616893,0.15566742){\makebox(0,0)[lb]{\smash{3}}}%
    \put(0.24147483,0.11922618){\makebox(0,0)[lb]{\smash{T}}}%
    \put(0.56707937,0.18167532){\makebox(0,0)[lb]{\smash{P}}}%
    \put(0.58331335,0.17640557){\makebox(0,0)[lb]{\smash{2}}}%
    \put(0.47694863,0.23852221){\makebox(0,0)[lb]{\smash{P}}}%
    \put(0.4931854,0.23325245){\makebox(0,0)[lb]{\smash{1}}}%
    \put(0.45144367,0.02586261){\makebox(0,0)[lb]{\smash{n}}}%
    \put(0.46814427,0.02059285){\makebox(0,0)[lb]{\smash{2}}}%
    \put(0.34999105,0.00526976){\makebox(0,0)[lb]{\smash{n}}}%
    \put(0.36669445,0){\makebox(0,0)[lb]{\smash{1}}}%
    \put(0.25962839,0.05872734){\makebox(0,0)[lb]{\smash{n}}}%
    \put(0.27633179,0.05345759){\makebox(0,0)[lb]{\smash{3}}}%
    \put(0.60224089,0.11922618){\makebox(0,0)[lb]{\smash{T}}}%
    \put(0.68608481,0.16369221){\makebox(0,0)[lb]{\smash{P}}}%
    \put(0.70231879,0.15842245){\makebox(0,0)[lb]{\smash{1}}}%
    \put(0.66549196,0.09575815){\makebox(0,0)[lb]{\smash{P}}}%
    \put(0.68172594,0.09048839){\makebox(0,0)[lb]{\smash{2}}}%
    \put(0.95188067,0.06666831){\makebox(0,0)[lb]{\smash{n}}}%
    \put(0.96858407,0.06140134){\makebox(0,0)[lb]{\smash{2}}}%
    \put(0.88211086,0.18428506){\makebox(0,0)[lb]{\smash{n}}}%
    \put(0.89881426,0.1790153){\makebox(0,0)[lb]{\smash{1}}}%
    \put(0.97247352,0.13121585){\makebox(0,0)[lb]{\smash{n}}}%
    \put(0.98917692,0.1259461){\makebox(0,0)[lb]{\smash{3}}}%
  \end{picture}%
  \endgroup%

  \caption{Distribution of a three-dimensional dataset of size
    $n_1\times n_2\times n_3 = 8\times 4\times 4$ on a two-dimensional
    process grid of size $P_1\times P_2 = 4\times 2$.}
  \label{fig:fft3d_split2d}
\end{figure}

The starting point is to decompose the input data set along the first
two dimensions into equal blocks of size $\frac{n_1}{P_1}\times
\frac{n_2}{P_2}\times n_3$ and distribute these blocks on a
two-dimensional grid of $P_1 \times P_2$ processes. Therefore, each
process can perform $\frac{n_1}{P_1}\times \frac{n_2}{P_2}$
one-dimensional FFTs of size $n_3$ locally. Afterwards, a
communication step is performed that redistributes the data along
directions 1 and 3 in blocks of size $\frac{n_1}{P_1}\times n_2 \times
\frac{n_3}{P_2}$, such that the 1D-FFT along direction \(2\) can be
performed locally on each process. Then, a second communication step
is performed, that redistributes the data along direction 2 and 3 in
blocks of size $n_1 \times \frac{n_2}{P_1} \times
\frac{n_3}{P_2}$. Now, the 3D-FFT is completed by performing the
1D-FFTs along direction \(1\). This algorithm is illustrated in
\ref{fig:fft3d_split2d}. For $n_1 \ge n_2 \ge n_3$ the two-dimensional
data decomposition allows the usage of at most $n_2 \times n_3$
processes.

Since it is not trivial to implement an efficient FFT routine in
parallel, we rely on optimised implementations ~\cite{libpfft,
Pip2011,Pip2012TUC}.  Fortunately, several publicly available FFT
software libraries are available that are based on the two-dimensional
data decomposition. Among them are the FFT package from Sandia
National Laboratories \cite{libsandiafft}, the P3DFFT software
library~\cite{libp3dfft}, the 2DECOMP\&FFT
package~\cite{lib2decompfft}, and the PFFT software
library~\cite{libpfft, Pip2011,Pip2012TUC}. Other efficient
implementations exist, but unfortunately they are not distributed as
stand-alone packages~\cite{TruongDuy2012}. Our test runs are
implemented with the help of the PFFT software library
\cite{Pip2012TUC}, which utilises the FFTW \cite{FFTW05} software
package for the one-dimensional FFTs and the global communication
steps. PFFT has a similar performance to the well-known P3DFFT, as
well as some user-interface advantages \cite{Pip2011}, so we choose it
for our survey.

\subsection{Fast Fourier transform with interpolating scaling functions (ISF)}
\index{Poisson solver!Interpolating scaling functions} 

This a different solver based on the Fourier approach. It was
developed by Genovese~\emph{et al.}~\cite{Gen2006JCP} for the {\sc
  bigdft} code~\cite{Genovese2008}. Formally this solver is based on
representing the density in a basis of interpolating scaling functions
(ISF) that arise in the context of wavelet theory
\cite{Dau1998BOOK}. A representation of $\rho$ from $\rho_{j,k,l}$ can
be efficiently built by using wavelets, and efficient iterative
solvers for the Hartree potential $v$ can be taylored, e.g. from
ordinary steepest descent or conjugate gradient techniques
\cite{Goedec1998}.  A non-iterative and accurate way to calculate $v$
\cite{Gen2006JCP} is to use the fast Fourier transform (FFT) in
addition to wavelets. If we represent
\begin{equation}
  \rho(\vec{r})=\sum_{j,k,l}\rho_{j,k,l}\phi(x-j)\phi(y-k)\phi(z-l)  \ , 
\end{equation}
where $\vec{r}=x,y,z$ and $\phi$ are interpolating scaling functions
in one dimension, then equation \ref{eq:Vtotal} becomes
\begin{equation}\label{eq:VK}
  v_{m,n,o}=\sum_{j,k,l}\rho_{j,k,l} K(j,m; k,n; l,o)  \ , 
\end{equation}
where

\begin{equation}\label{eq:Kint}
  K(j,m; k,n; l,o) :=\int_V d\vec{r} \ ' \frac{\phi_{j}(x')\phi_{k}(y')\phi_{l}(z')}{|\vec{r}-\vec{r} \ '|} \ ,
\end{equation}
and $V$ indicates the total volume of the system.  Due to the
definition of the ISF, the discrete kernel $K$ satisfies $K(j,m; k,n;
l,o) = K(j - m; k-n; l-o) $, and therefore equation \ref{eq:VK} is a
convolution, which can be efficiently treated using FFTs (see the
previous section).  The evaluation of equation \ref{eq:Kint} can be
approximated by expressing the inverse of $r$ in a Gaussian basis
($1/r \simeq \sum_k \omega_k \mathrm{exp}(-p_k r^2)$), which also
enables efficient treatment. All this makes the order of this method
$N\log_2(N)$.

This method is nearly equivalent to the scheme presented in the 
previous section. ISF's main characteristic is that it uses a kernel in
equation \ref{eq:poisson_fourier} that yields an accurate free space
potential without having to enlarge the cell. For the purposes of the
discussion below we consider it a different solver since it is
distributed as a standalone package that includes its own parallel FFT
routine~\cite{Goedec1998}, so it has different scaling properties
than the solver based on the PFFT library.

\subsection{Fast multipole method (FMM)}
\index{Poisson solver!fast multipole method} \label{sec:fmm} 

The fast multipole method was first proposed in 1987 for 2D
systems~\cite{Gre1987JCOP}, and it was soon extended to 3D
problems~\cite{Gre1988BOOK}. Although only $\mathcal{O}(N)$ operations
are necessary to calculate the electrostatic potential, the first FMM
versions 
had big prefactors that 
in practice made the method competitive only for huge systems or low
accuracy calculations~\cite{Gre1999JCOP}. After thorough research, it
was possible to develop signficantly more accurate and efficient
versions of FMM~\cite{Gre1997AN,Dac2010JCP}, making it a largely
celebrated method \cite{siamtop10}.

The original FMM was devised to calculate the potential generated by a
set of discrete point-like particles, this is
\begin{equation}
  \label{eq:pointcharges}
  v(\vec{r}_i)  =  
  \frac{1}{4\pi \epsilon_0} \sum_{j=1,j\neq i}^{N}
  \frac{q_j}{\left|\vec{r}_{i}-\vec{r}_{j}\right|}  \ ,
\end{equation}
which is different from the charge-distribution problem that we are
studying in this work. While extensions of FMM to the continuous
problem exist~\cite{Whi1994CPL,Whi1996CPL,Str1996sci}, in order to
profit from the efficient parallel FMM implementations we have devised
a simple scheme to recast the continuous problem into a discrete
charge one without losing precision.

We assume that each grid point \(r_{j,k,l}\) corresponds to a charge
of magnitude \(\Omega \rho_{j,k,l}\), where $\Omega=L^3$ ( \(L\)
being the grid spacing) is the volume of the space associated to each
grid point (cell volume).  Using FMM we calculate at each point the
potential generated by this set of charges,
\(v^{\mathrm{FMM}}_{j,k,l}\). However, this is not equivalent to the
potential generated by the charge distribution, and some correction
terms need to be included (see supporting info~\cite{sm} for
the derivation of these corrections). The first correcting term
(self-interaction term) comes from the potential generated at each
point by the charge contained in the same cell
\begin{equation}
\label{vsi}
v^{\mathrm{SI}}_{j,k,l} =  2\pi \left(\frac{3}{4\pi}\right)^{2/3}L^2\rho_{j,k,l}\ .
\end{equation}
Additionally, we apply a correction to improve the accuracy of the
interaction between neighbouring points, which has the largest error
in the point-charge approximation. This correction term is derived
using a formal cubic interpolation of the density to a finer grid, obtaining
a simple finite-differences-like term
\begin{equation}\label{eq:vcorr}
\begin{aligned}
v^{\mathrm{corr.}}_{j,k,l} & =    L^2   \big(27/32 + (\alpha) 2\pi (3/4\pi)^{2/3} \big)  \rho_{j,k,l}  \\
  & + (L^2/16) \big( \rho_{j-1,k,l} + \rho_{j+1,k,l} + \rho_{j,k-1,l}  + \rho_{j,k+1,l} + \rho_{j,k,l-1} + \rho_{j,k,l+1} \big)  \\
  & - (L^2/64) \big(  \rho_{j-2,k,l} + \rho_{j+2,k,l} + \rho_{j,k-2,l} + \rho_{j,k+2,l} + \rho_{j,k,l-2} + \rho_{j,k,l+2} \big)    \ .
\end{aligned}
\end{equation}
Here $\alpha$ is a parameter to compensate the charge within the cell
$(j,k,l)$ that is counted twice. The final expression for the
potential is
\begin{equation}
  \label{vorig} 
  v_{j,k,l}= v^{\mathrm{FMM}}_{j,k,l} + v^{\mathrm{SI}}_{j,k,l} + v^{\mathrm{corr.}}_{j,k,l}\ .
\end{equation} 

Now we give a brief introduction of the FMM algorithm. More detailed
explanations on FMM can be found in
Refs.~\cite{Gre1987JCOP,Gre1988BOOK,White:1994,Gre1997AN}. The
concrete implementation we used in this work is explained in
\cite{kdfmm2010,Dac2010JCP}. To introduce the method we use spherical
coordinates in what remains of this section.

\begin{figure}[ht!]
  \centering
  \def\svgwidth{500pt} 
  \begingroup%
  \makeatletter%
  \providecommand\color[2][]{%
    \errmessage{(Inkscape) Color is used for the text in Inkscape, but the package 'color.sty' is not loaded}%
    \renewcommand\color[2][]{}%
  }%
  \providecommand\transparent[1]{%
    \errmessage{(Inkscape) Transparency is used (non-zero) for the text in Inkscape, but the package 'transparent.sty' is not loaded}%
    \renewcommand\transparent[1]{}%
  }%
  \providecommand\rotatebox[2]{#2}%
  \ifx\svgwidth\undefined%
    \setlength{\unitlength}{917.90129089bp}%
    \ifx\svgscale\undefined%
      \relax%
    \else%
      \setlength{\unitlength}{\unitlength * \real{\svgscale}}%
    \fi%
  \else%
    \setlength{\unitlength}{\svgwidth}%
  \fi%
  \global\let\svgwidth\undefined%
  \global\let\svgscale\undefined%
  \makeatother%
  \begin{picture}(1,0.39764737)%
    \put(0,0){\includegraphics[width=\unitlength]{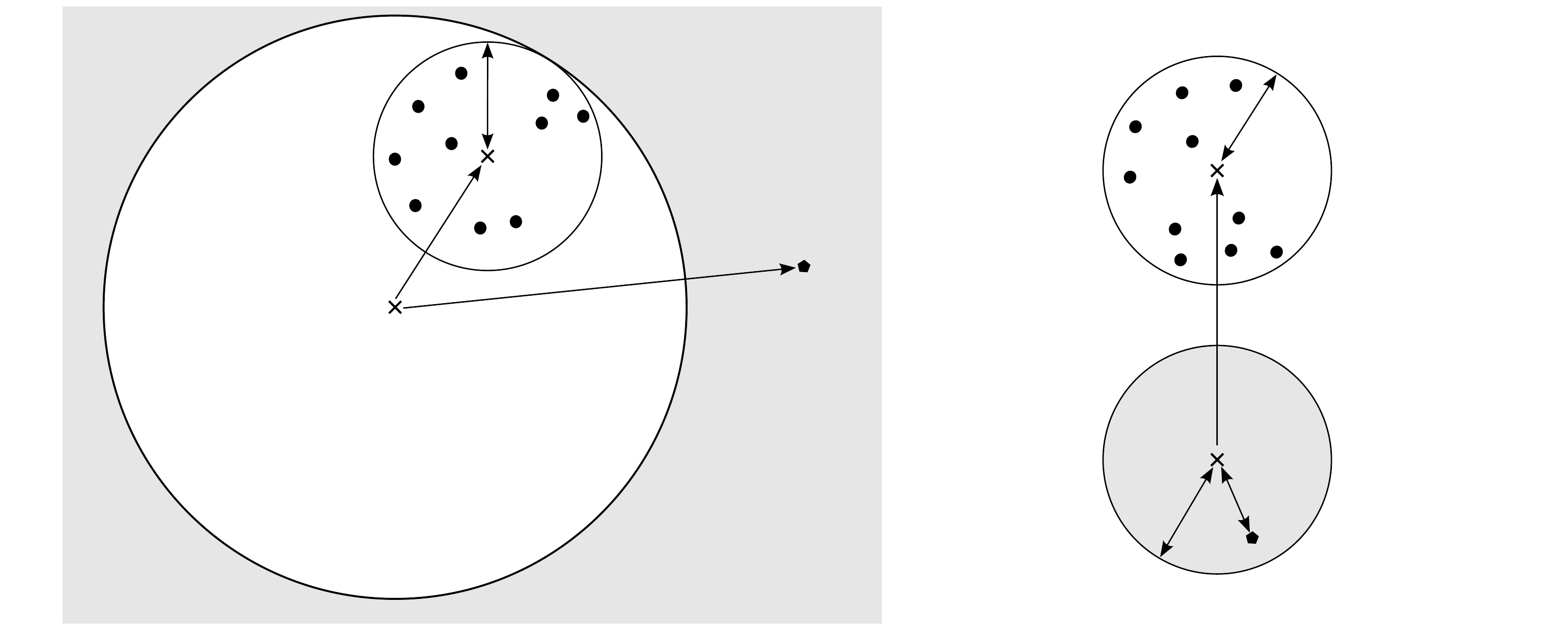}}%
    \put(-0.00013618,0.38442087){\color[rgb]{0,0,0}\makebox(0,0)[lb]{\smash{A)}}}%
    \put(0.59318665,0.38442087){\color[rgb]{0,0,0}\makebox(0,0)[lb]{\smash{B)}}}%
    \put(0.3049686,0.11536348){\color[rgb]{0,0,0}\makebox(0,0)[lb]{\smash{$D_1$}}}%
    \put(0.22816446,0.18762437){\color[rgb]{0,0,0}\makebox(0,0)[lb]{\smash{O}}}%
    \put(0.4624739,0.24018238){\color[rgb]{0,0,0}\makebox(0,0)[lb]{\smash{$P(r, \theta ,\phi )$}}}%
    \put(0.26922172,0.3396){\color[rgb]{0,0,0}\makebox(0,0)[lb]{\smash{D}}}%
    \put(0.31866498,0.28510917){\color[rgb]{0,0,0}\makebox(0,0)[lb]{\smash{$Q(\tau, \alpha, \beta )$}}}%
    \put(0.31573852,0.3319445){\color[rgb]{0,0,0}\makebox(0,0)[lb]{\smash{a}}}%
    \put(0.7055889,0.11454626){\color[rgb]{0,0,0}\makebox(0,0)[lb]{\smash{$D_0$}}}%
    \put(0.7879625,0.09811569){\color[rgb]{0,0,0}\makebox(0,0)[lb]{\smash{O}}}%
    \put(0.8045301,0.05016664){\color[rgb]{0,0,0}\makebox(0,0)[lb]{\smash{$P(r, \theta ,\phi )$}}}%
    \put(0.7902399,0.28179552){\color[rgb]{0,0,0}\makebox(0,0)[lb]{\smash{$Q(\tau, \alpha, \beta )$}}}%
    \put(0.80154222,0.31091089){\color[rgb]{0,0,0}\makebox(0,0)[lb]{\smash{a}}}%
    \put(0.72066602,0.32366827){\color[rgb]{0,0,0}\makebox(0,0)[lb]{\smash{$D_Q$}}}%
    \put(0.73692918,0.07908078){\color[rgb]{0,0,0}\makebox(0,0)[lb]{\smash{a}}}%
    \put(0.79329215,0.19218952){\color[rgb]{0,0,0}\makebox(0,0)[lb]{\smash{$\tau > 2a$}}}%
  \end{picture}%
\endgroup%

  \caption{A) A set of point charges (black circles) inside a sphere
    $D$ of radius $a$ centred at $Q=(\tau,\alpha,\beta)$ creates a
    potential outside the sphere $D_1$ of radius $(a+\tau)$ and
    centred in the origin that can be expressed with equation \ref{eq:fmmim3}.
    B) A set of point charges (black circles) inside a sphere $D_Q$ of
    radius $a$ centred at $Q=(\tau,\alpha,\beta)$ creates a potential
    inside the sphere $D_0$ of radius $a$ and centred in the origin
    that can be expressed with equation \ref{eq:fmmim5} (provided that
    $\tau>2a$).  In both A) and B), $O$ represents the origin of
    coordinates, and $P=(r,\theta,\phi)$ is the point where the
    potential is measured. In our systems, the charges lie in
    equispaced grid points.  }
  \label{fig:twospheres}
\end{figure}
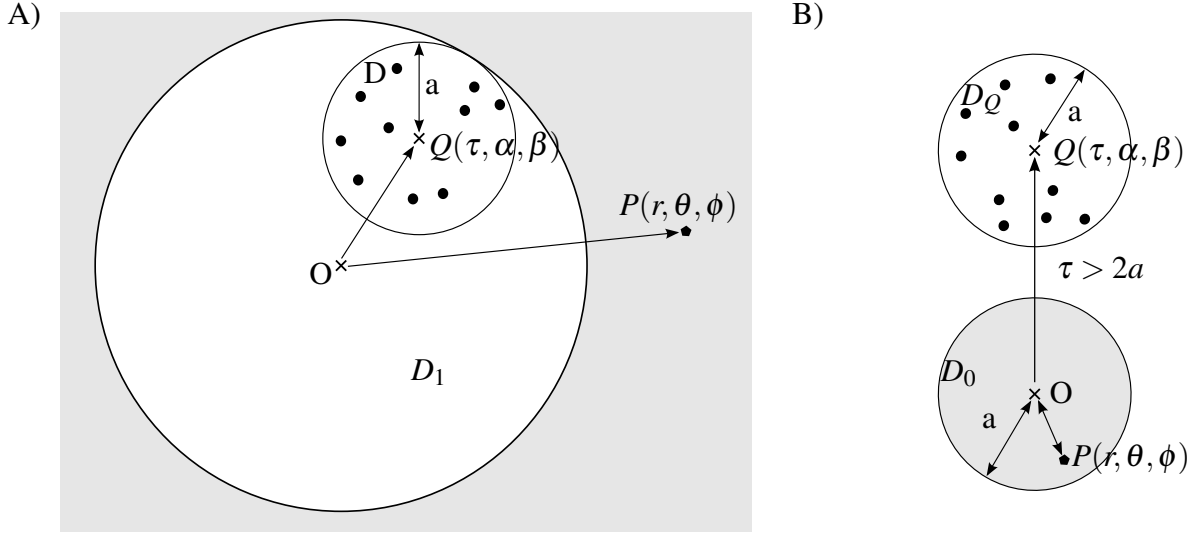

Consider a system of $l$ charges $\{ q_i, \ i=1,\ldots,l \}$ located
at points $\{ (\tau_i,\alpha_i,\beta_i), \ i=1,\ldots,l \}$ which lie
in a sphere $D$ of radius $a$ and center at $Q=(\tau,\alpha,\beta)$.
It can be proved \cite{Gre1997AN} that the electric field created by
them at a point $P=(r,\theta,\phi)$ outside $D$ is
\begin{equation}\label{eq:fmmim1}
v(P)= \sum_{n=0}^{\infty} \sum_{m=-n}^{n} \frac{O_n^m}{(r')^{n+1}} Y_n^m(\theta',\phi') \ ,
\end{equation}
where $P-Q=(r',\theta',\phi')$ and
\begin{equation}\label{eq:fmmim2}
O_n^m=\sum_{i=1}^l q_i \tau_i^n Y_n^{-m} (\alpha_i,\beta_i) \ ,
\end{equation}
with $Y_n^{m} (\alpha,\beta)$ known functions (the spherical
harmonics).  If $P$ lies outside of a sphere $D_1$ of radius $a+\tau$
(see \ref{fig:twospheres} A) the potential of \ref{eq:fmmim1} can be
re-expressed as
\begin{equation}\label{eq:fmmim3}
  v(P)= \sum_{j=0}^{\infty} \sum_{k=-j}^{j} \frac{M_j^k}{r^{j+1}} Y_j^k(\theta,\phi) \ ,
\end{equation}
where
\begin{equation}\label{eq:fmmim4}
M_j^k=\sum_{n=0}^j \sum_{m=-n}^{n} 
\frac{O_{j-n}^{k-m} \ i^{|k|-|m|-|k-m|} \sqrt{(j-n-k+m)! (j-n+k-m)!}\sqrt{(n-m)! (n+m)!} \ \tau^n \ Y_{n}^{-m} (\alpha,\beta) }{\sqrt{(j-k)! (j+k)!}} \ .
\end{equation}
Note that the ``entangled'' expression of the potential equation
\ref{eq:pointcharges}, in which the coordinates of the point where the
potential is measured and the coordinates of the charge that creates
the potential are together, has been converted to a ``factorised''
expression, in which the coordinates of the point where we measure the
potential are in terms ($Y_j^k(\theta,\phi)/r^{j+1}$) that multiply
terms ($M_j^k$) which depend on the coordinates of the charges that
create the potential. It is this factorisation which enables efficient
calculation of the potential that a set of charges creates at a given
point by using the (previously calculated) expression of the potential
created by this set of charges at other points.

If the set of $l$ charges described above is located inside a sphere
$D_Q$ of radius $a$ with center at $Q=(\tau,\alpha,\beta)$ where
$\tau>2a$ (see \ref{fig:twospheres} B) then equation \ref{eq:fmmim1}
implies that the potential created by these charges inside a sphere
$D_0$ of radius $a$ centered at the origin is given by
\begin{equation}\label{eq:fmmim5}
v(P)= \sum_{j=0}^{\infty} \sum_{k=-j}^{j} \ L_j^k \ r^{j} \ Y_j^k(\theta,\phi) \ ,
\end{equation}
where
\begin{equation}\label{eq:fmmim6}
L_j^k=\sum_{n=0}^{\infty} \sum_{m=-n}^{n} 
\frac{O_{n}^{m} \ i^{|k-m|-|k|-|m|} \sqrt{(n-m)!  (n+m)!}\sqrt{( j-k)! (j+k )!} \ Y_{j+n}^{m-k} (\alpha,\beta) }{(-1)^n \ \tau^{j+n+1} \ \sqrt{(j+n-m+k)! (j+n+m-k)!}} \ .
\end{equation}

The evaluation of the equations above requires truncation of the
infinite sums to a given order, which can be chosen to keep the error
below a given threshold.  The equations \ref{eq:fmmim3} and
\ref{eq:fmmim5} enable the efficient calculation of the potential
experienced by every charge of the system due to the influence of the
other charges.  In order to calculate it, the system is divided into a
hierarchy of boxes.  Level 0 is a single box containing the whole
system; level 1 is a set of 8 boxes containing level 0; and so on (a
box of level $\mathcal{L}$ consists of 8 boxes of level
$\mathcal{L}+1$). Different boxes at a given level do not contain
common charges. The highest level ($\mathcal{N}_l$) contains several
charges (in our case, each lying in a grid point) in every box.  The
procedure to calculate the potential in all grid points can be
summarised as follows:
\begin{itemize}
\item For every box in the highest box level $\mathcal{N}_l$ (smallest
  boxes), we calculate the potential created by the charges in that
  box using equation \ref{eq:fmmim1}.
\item We gather 8 boxes of level $\mathcal{N}_l$ to form every box of
  level $\mathcal{N}_l$-1.  We calculate the potential created by
  the charges of the $(\mathcal{N}_l-1)$-box using the potentials
  created by the eight $(\mathcal{N}_l)$-boxes that form it. To this
  end, we use equation \ref{eq:fmmim3}.
\item We repeat this procedure (we use equation \ref{eq:fmmim3} to get
  the potentials created by box $\mathcal{L}$-1 by using those of box
  $\mathcal{L}$) until all the levels are swept.
\item Then, the box hierarchy is swept in the opposite direction: from
  lower to higher levels, the equation \ref{eq:fmmim5} is used to
  calculate the potential created by the boxes (the potential given by
  equation \ref{eq:fmmim5} is valid in regions that are not equal to
  those where equation \ref{eq:fmmim3} is valid).
\item Finally, the total potential in every grid point
  ($v_{j,k,l}^{FMM}$) is calculated as an addition of three terms: the
  potentials due to nearby charges are directly calculated with the
  pairwise formula equation \ref{eq:pointcharges}, and the potentials
  due to the rest of the charges are calculated either with equation
  \ref{eq:fmmim3} or with equation \ref{eq:fmmim5}, depending on the
  relative position of the boxes which create the potential and the
  box where the potential is evaluated.
\end{itemize}
In the whole procedure above, the charge in the grid point $(j,k,l)$
is $\Omega \rho_{j,k,l}$.  This scheme corresponds to the traditional
version of FMM \cite{Gre1997AN}.  We used a slight modification of it
\cite{kdfmm2010} which not only converts multipoles between
consecutive levels, but also within every given level, which enables
further computational savings.

\subsection{Conjugate gradients}\label{sec:cg}
\index{Poisson solver!conjugate gradients}

We now present two widely used iterative methods to calculate the
electrostatic interaction: conjugate gradients and
multigrid~\cite{Saad2003}. These methods are based on finding a
solution to the Poisson equation \ref{eq:ecpodif} by starting from an
initial guess and systematically refining it so that it becomes
arbitrarily closer to the exact solution. These two methods have the
advantage that if a good initial approximation is known, only a few
iterations are required.

When using a grid representation, the Poisson equation can be
discretised using finite differences. In this scheme the Laplacian at
each grid point is approximated by a sum over the values of
neighbouring points multiplied by certain weights. High-order
expressions that include several neighbours can be used to control the
error in the approximation~\cite{Chelikowsky1994}. The
finite-difference approximation turns equation~(\ref{eq:ecpodif}) into a
linear algebraic equation
\begin{equation}
  \label{eq:linear}
  \tilde{L}\vec{x}=\vec{y}\ ,
\end{equation}
where \(\tilde{L}\) is a sparse matrix (taking advantage of the
sparsity of a system of equations can greatly reduce the numerical
complexity of its solution \cite{GR2010JCoP}), \(\vec{y}\) is known
($y=-\rho/\epsilon_0$, in this case) and \(\vec{x}\) is the quantity
we are solving for, in this case the electrostatic potential.
Equation (\ref{eq:linear}) can be efficiently solved by iterative methods
that are based on the application of the matrix-vector product without
the need to store the matrix.

Since the matrix is symmetric and positive definite, we can use the
standard conjugate gradients~\cite{Hes1952jrnbs} (CG) method. CG
builds $\vec{x}$ as a linear combination of a set of orthogonal
vectors $\vec{p_k}$. In every iteration, a new term is added
\begin{equation}
  \label{eq:sss}
  \vec{x}_{k+1}=\vec{x}_k+\alpha_{k+1}\vec{p}_{k+1}  \ , 
\end{equation}
which attempts to minimise
\begin{equation}
  f(\vec{x}):=\frac{1}{2}\vec{x}^T\tilde{L}\vec{x}-\vec{x}^T\vec{y}  \ , 
\end{equation}
whose minimum is the solution of equation \ref{eq:linear}. The term
added to the potential in iteration $k$ is built so that $\vec{x}$
moves in the direction of the gradient of $f$ but being orthogonal to
the previous terms. The gradient of $f(\vec{x})$ satisfies, \(-\nabla
f(\vec{x}) = \vec{y}-\tilde{L}\vec{x}\).  Therefore the search
direction in iteration \(k+1\) is
\begin{equation}
  \vec{p}_{k+1} = \Big( \vec{y}-\tilde{L}\vec{x}_k \Big)  - \sum_{i \leq k}   \frac{\vec{p}_{i}^T \tilde{L}( \vec{y}-\tilde{L}\vec{x}_k)}{\vec{p}_{i}^T\tilde{L}\vec{p}_{i}} \vec{p}_{i}  \ .
\end{equation}
The coefficient associated with each direction, $\alpha_{k+1}$, is
obtained from the minimization condition, yielding
\begin{equation}\label{alpha}
 \alpha_{k+1}=\left( \frac{\vec{p}_{k+1}^T( \vec{y}-\tilde{L}\vec{x}_k)}{\vec{p}_{k+1}^T\tilde{L} \vec{p}_{k+1}} \right) \ .
\end{equation}

The equations above show that the conjugate gradients method has a
linear scaling ($\mathcal{O}(N)$) with the number of points $N$ for a
given number of iterations whenever the matrix $\tilde{L}$ has a
number of non-zero entries per row that is much lower than $N$ (as is
the usual case for the Poisson equation). Bigger exponents in the
scaling can appear, however, in problems where the number of performed
iterations is chosen to keep the solution error below a given
threshold that depends on $N$ \cite{artachocg}.

The parallelisation of our CG implementation is based on the domain
decomposition approach, where the grid is divided into subregions that
are assigned to each process. Since the application of the
finite-difference Laplacian only requires near-neighbour values, only
the boundary values need to be shared between processors (see
refs.~\cite{Cas2006PSSB,And2012JPCM} for details). Since our
implementation can work with arbitrary shape grids, dividing the grids
into subdomains of equal volume while minimizing the area of the
boundaries is not a trivial problem, for this task we use the Metis
library~\cite{Metis}.

An issue that appears when using the finite-difference discretisation
to solve the Poisson equation are boundary conditions: they must be
given by setting the values of the points on the border of the
domain. For free space boundary conditions we need a secondary method
to obtain the value of the potential over these points; this
additional method can represent a significant fraction of the computational
cost and can introduce an approximation error. In our implementation
we use a multipole expansion~\cite{Flo1978PRC}.

\subsection{Multigrid}
\index{Poisson solver!multigrid} 

Multigrid~\cite{Wes1992BOOK,Zha1998JCOP,Bri1987BOOK,Bra1977MC,Trot2001Book,
  Saad2003} is a powerful method to solve elliptic partial
differential equations, such as the Poisson problem \cite{PhysRevB.81.085103}, in an iterative
fashion. Multigrid is routinely used as a solver or preconditioner for
electronic structure and other scientific
applications~\cite{Briggs1996,Bec2000RMP,Torsti2003,Mortensen2005}. In
this section we will make a brief introduction to a simple version of
the multigrid approach that is adequate for the Poisson
problem. Multigrid, however, can also be generalized to more complex
problems, like non-linear problems~\cite{Trot2001Book} and systems
where there is no direct geometric interpretation, in what is known as
algebraic multigrid~\cite{shapira2003}. It has also been extended to
solve eigenvalue problems~\cite{Mandel1989,Torsti2003}.

Multigrid is based on iterative solvers like Jacobi or
Gauss-Seidel~\cite{Saad2003}. These methods are based on a simple
iteration formula, that for equation~\ref{eq:linear} reads
\begin{equation}
  \label{eq:mgiterative}
  \vec{x}\leftarrow\vec{x} + M^{-1}(\vec{y}-\tilde{L}\vec{x})\ .
\end{equation}
The matrix \(M\) is an approximation to \(\tilde{L}\) that is simple to
invert. In the case of Jacobi, \(M\) is the diagonal of \(\tilde{L}\),
and for Gauss-Seidel, \(M\) is upper diagonal part of
\(\tilde{L}\). These methods are simple to implement, in particular in
the case of the Laplacian operator, but are quite inefficient by
themselves, so they are not practical as linear solvers for
high-performance applications. However, they have a particular
property: they are very good in removing the high-frequency error of a
solution approximation, where the frequency is defined in relation to the
grid spacing. In other words, given an approximation to the solution,
a few iterations of Jacobi or Gauss-Seidel will make the solution
smooth. In multigrid the smothing property is exploited by using a
hierarchy of grids of different spacing, and hence changing the
frequency that these smoothing operators can remove efficiently.

A fundamental concept in multigrid is the residual of a linear
equation. If we have \(\bar{x}\) as an approximate solution of
equation~\ref{eq:linear}, the associated residual, \(\vec{b}\), is defined as
\begin{equation}
  \label{eq:residue}
  \vec{b} = \vec{y} -\tilde{L}\bar{\vec{x}}\ .
\end{equation}
We can use the residual to define an alternative linear problem
\begin{equation}
  \label{eq:linear_res}
  \tilde{L}\vec{a}=\vec{b}\ .
\end{equation}
Due to the linearity of the Laplacian operator, finding \(\vec{a}\) is
equivalent to solving the original linear problem,
equation~\ref{eq:linear}, as
\begin{equation}
  \label{eq:mgcorrection}
\vec{x}=\bar{\vec{x}}+\vec{a}.
\end{equation}

If a few iterations of a smoothing operator have been applied to the
approximate solution, \(\bar{x}\), we know that the high-frequency
components of the corresponding residual \(\vec{b}\) should be
small. Then it is possible to represent \(\vec{b}\) in a grid that has,
for example, two times the spacing without too much loss of
accuracy. In this coarser grid equation~\ref{eq:linear_res} can be solved
with less computational cost. Once the solution \(\vec{a}\) is found
on the coarser grid, it can be transferred back to the original grid
and used to improve the approximation to the solution using
equation~\ref{eq:mgcorrection}.

The concept of calculating a correction term in a coarser grid is the
basis of the multigrid approach. Instead of two grids, as in our
previous example, a hierarchy of grids is used, at each level the
residual is transferred to a coarser grid where the correction term is
calculated. This is done up to the coarsest level that only contains a
few points. Then the correction is calculated and transferred back to
the finer levels.

To properly define the multigrid algorithm it is necessary to specify
the operators that transfer functions between grids of different
spacing. For transferring to a finer grid, typically a linear
interpolation is used. For transferring to a coarser grid a so-called
restriction operator is used. In a restriction operator, the value of
the coarse grid point is calculated as a weighted average of the
values of the corresponding points in the fine grid and its neighbors.

Now we introduce the multigrid algorithm in detail. Each quantity is
labeled by a super-index that identifies the associated grid level,
with \(0\) being the coarsest grid and \(L\) the finest. We denote
\(S^l\) as the smoothing operator at level \(l\), which corresponds to
a few steps (usually 2 or 3) of Gauss-Seidel or Jacobi. \(I_{l}^{m}\)
represents the transference of a function from the level \(l\) to the
level \(m\) by restriction or interpolation. Following these
conventions, we introduce the algorithm of a multigrid iteration in
fig.~\ref{fig:vcycle}. Given an initial approximation for the
solution, we perform a few steps of the smoothing operator, after
which the residual is calculated and transferred to the coarser
grid. This iteration is repeated until the coarsest level is
reached. Then we start to move towards finer grids. In each step the
approximate solution of each level is interpolated into the finer grid
and added, as a correction term, to the solution approximation of that
level, after which a few steps of smoothing are performed. Finally,
when the finest level is reached, a correction term that has
contributions from the whole grid hierarchy is added to the initial
approximation to the solution.

\begin{figure}
  \centering
  \begin{algorithmic}
  \STATE \textbf{Multigrid \textit{v}-cycle}
  \STATE Input: \(\vec{y}, \bar{\vec{x}}\)
  \STATE Output: \(\bar{\vec{x}}\)
  \STATE
  \STATE \(\vec{y}^{L}\leftarrow \vec{y}\)
  \STATE \(\vec{x}^{L}\leftarrow \bar{\vec{x}}\)
  \FOR{\(l\) from \(L\) to \(0\)}
  \IF {\(l\neq L\)} 
  \STATE \(x^l \leftarrow 0\) \COMMENT {Set initial guess to 0}
  \ENDIF
  \STATE \({\vec{x}}^l\leftarrow \mathrm{S}^l{\vec{x}^l}\) \COMMENT{Pre-smoothing}
  \IF {\(l\neq0\)}
  \STATE \(\vec{b}^l\leftarrow \vec{y}^l
  -\tilde{L}^l{\vec{x}}^l\) \COMMENT{Calculate the residual}
  \STATE \(\vec{y}^{l - 1}\leftarrow I^{l-1}_{l}\vec{b}^l\)
  \COMMENT{Transfer the residual to the coarser grid}
  \ENDIF
  \ENDFOR
  \FOR{\(l\) from \(0\) to \(L\)}
  \IF {\(l\neq 0\)}
  \STATE \(\vec{x}^{l}\leftarrow \vec{x}^{l}+I^{l}_{l-1}\vec{x}^{l-1}\)
  \COMMENT{Transfer the correction to the finer grid}
  \ENDIF
  \STATE \({\vec{x}}^l\leftarrow \mathrm{S}^l{\vec{x}^l}\)
  \COMMENT{Post-smoothing}
  \ENDFOR
  \STATE \(\bar{\vec{x}}\leftarrow \vec{x}^L\)
\end{algorithmic}
  \caption{Algorithm of a multigrid \textit{v}-cycle.}
  \label{fig:vcycle}
\end{figure}

The scheme presented in \ref{fig:vcycle} is known as a
\textit{v}-cycle, for the order in which levels are visited, some more
sophisticated strategies exist, where the coarsest level is visited
twice (a \textit{w}-cycle) or more times before coming back to the
finest level~\cite{Trot2001Book}. Usually a \textit{v}-cycle reduces
the error, measured as the norm of the residual, by around one order
of magnitude, so typically several \textit{v}-cycles are required to
find a converged solution.

When a good initial approximation is not known, an approach known as
full multigrid (FMG) can be used. In FMG the original problem is
solved first in the coarsest grid, then the solution is interpolated
to the next grid in the hierarchy, where it is used as initial
guess. The process is repeated until the finest grid is reached. It
has been shown that the cost of solving the Poisson equation by FMG
depends linearly with the number of grid points~\cite{Trot2001Book}.

Just as in the case of conjugate gradients, the parallelisation of
multigrid is based on the domain decomposition approach. However in
the case of multigrid some additional complications appear. First of
all, as coarser grids are used the domain decomposition approach
becomes less efficient as the number of points per domain is
reduced. Secondly, the Gauss-Seidel procedure used for smoothing
should be applied to each point sequentially~\cite{Trot2001Book} so
it is not suitable for domain decomposition. In our implementation we
take the simple approach of applying Gauss-Seidel in parallel over
each domain. For a large number of domains, this scheme would in fact
converge to the less-efficient Jacobi approach.

\section{Methods}\label{sec:methods}

\subsection{Implementation}\label{sec:implementation}

We chose the {\sc octopus} code
\cite{Cas2006PSSB,Mar2003CPC,And2012JPCM} (revision 8844) \footnote{
  The {\sc octopus} code is available in a public Subversion
  repository \url{http://www.tddft.org/svn/octopus/trunk}.}  for our
tests on the features of Poisson solvers in the context of quantum
mechanics.  {\sc octopus} is a program for quantum \emph{ab initio}
simulations based on density functional theory and time-dependent
density functional theory for molecules, one-dimensional systems,
surfaces, and solids under the presence of arbitrary external static
and time-dependent fields.  Its variables are represented on grids in
real space\footnote{Real space codes are at present a popular option; see
e.g. \cite{0953-8984-22-25-253202}.} (not using a basis of given functions), which feature
allows for systematic control of the discretisation error, of
particular importance for excited-state properties~\cite{Vila2010}.
Furthermore, {\sc octopus} uses a multi-level parallelisation scheme
where the data is distributed following a tree-based approach. This
scheme uses the message passing interface (MPI) library and was shown
to be quite efficient in modern parallel
computers~\cite{And2012JPCM}. We incorporated recent massively
parallelisable versions of the fast Fourier transform
\cite{Pip2012TUC} and of the fast multipole method \cite{Dac2010JCP}
into {\sc octopus}.  The goal of our tests was to measure
execution-times and accuracies of the six selected Poisson solvers, as
a function of the system size and the number of parallel
processes. Highlights in our implementation are a novel correcting
term to adapt the FMM method to charge distributions, and a smart,
fast way to manage the data flow for the PFFT.

\begin{figure}
  \centering  
  \def\svgwidth{350pt} 
  \begingroup%
  \makeatletter%
  \providecommand\color[2][]{%
    \errmessage{(Inkscape) Color is used for the text in Inkscape, but the package 'color.sty' is not loaded}%
    \renewcommand\color[2][]{}%
  }%
  \providecommand\transparent[1]{%
    \errmessage{(Inkscape) Transparency is used (non-zero) for the text in Inkscape, but the package 'transparent.sty' is not loaded}%
    \renewcommand\transparent[1]{}%
  }%
  \providecommand\rotatebox[2]{#2}%
  \ifx\svgwidth\undefined%
    \setlength{\unitlength}{678.48164159bp}%
    \ifx\svgscale\undefined%
      \relax%
    \else%
      \setlength{\unitlength}{\unitlength * \real{\svgscale}}%
    \fi%
  \else%
    \setlength{\unitlength}{\svgwidth}%
  \fi%
  \global\let\svgwidth\undefined%
  \global\let\svgscale\undefined%
  \makeatother%
  \begin{picture}(1,0.41055077)%
    \put(0,0){\includegraphics[width=\unitlength]{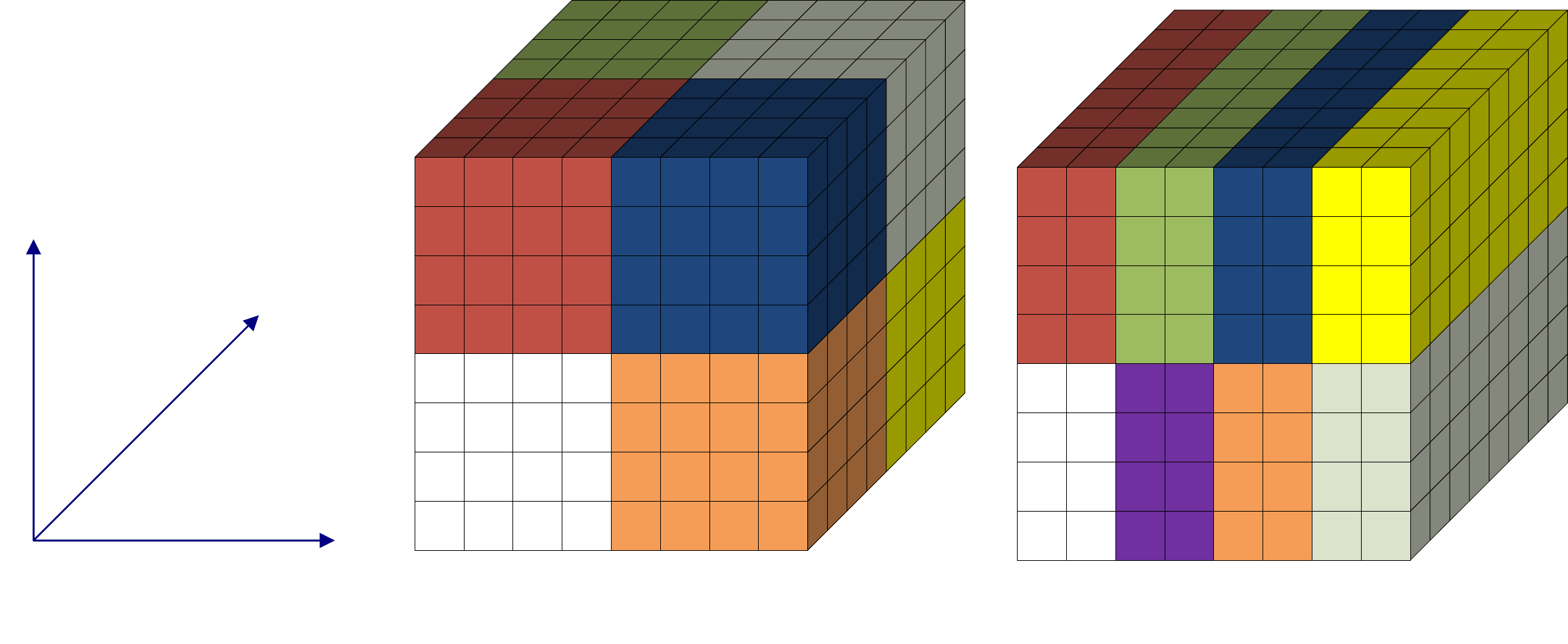}}%
    \put(0.0134379,0.12956156){\rotatebox{90}{\makebox(0,0)[lb]{\smash{Z axis}}}}%
    \put(0.08481345,0.03){\makebox(0,0)[lb]{\smash{X axis}}}%
    \put(0.06242523,0.12256063){\rotatebox{45}{\makebox(0,0)[lb]{\smash{Y axis}}}}%
    \put(0.27,0.00551841){\makebox(0,0)[lb]{\smash{A) Octopus mesh}}}%
    \put(0.67703287,0.00551841){\makebox(0,0)[lb]{\smash{B) PFFT mesh}}}%
  \end{picture}%
  \endgroup%

  \caption{Simplified domain decomposition of the simulation
    meshes. Each little cube represents a grid point ($8^3$ points in
    total) and each colour represents a partition (8 partitions). A)
    {\sc Octopus} mesh with a 3D domain decomposition; B) PFFT mesh
    with a 2D decomposition.}
  \label{fig:boxes}
\end{figure}
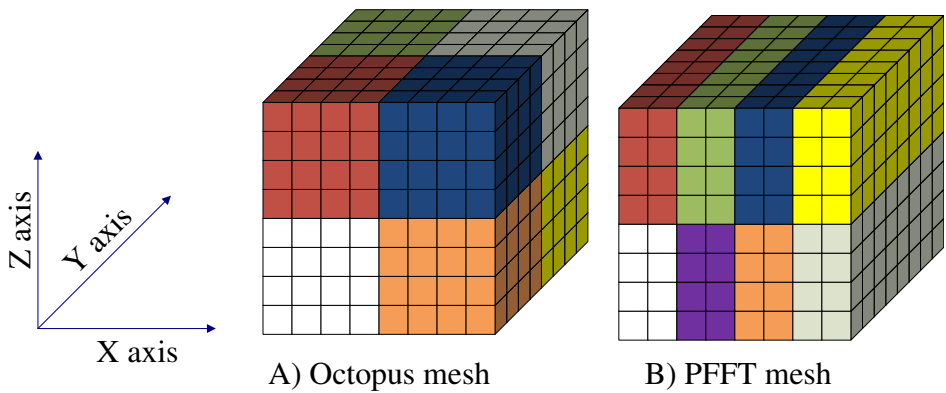

One must take into account that the execution-time of a function (like
a Poisson solver) included in a bigger program (in this case {\sc
  octopus}) is determined not only by the function itself, but also by
some features of the main program, like the way it represents the data
and the pattern for the data flow between the function and the main
program.  An example of this is the case when the FFT method (or ISF,
which is based on FFT's) are used in a code where the spatial
functions are represented in a Cartesian grid of arbitrary shape (like
{\sc octopus}).  Fast Fourier transforms require parallelepiped grids,
while the shape for the grid used in {\sc octopus} is commonly rather
amorphous to reduce the total number of points. So, whenever {\sc
  octopus} evaluates a fast Fourier transform, a parallelepiped mesh
containing the original amorphous mesh is used, and the new entries
are filled with zeroes.  The library PFFT divides such a
parallelepiped grid using a 2D grid of processes (see section
\ref{sec:pfft}) as displayed in \ref{fig:boxes} B. This means that
each process must handle the data corresponding to a column-like
dataset (i.e., a dataset that includes all the range of points in a
given direction direction; for example $\rho_{j,k,l}$ with
$j=1,\ldots,n_1/4$, $k=n_2/4+1,\ldots,2n_2/4$ and $l=1,\ldots,n_3$,
where $N=n_1 n_2 n_3$).  On the other hand, {\sc octopus} divides the
space grid into more compact domains, each handled
by a process (see \ref{fig:boxes} A), to reduce the surface
separating domains and consequently to reduce the need for information
transfer between processes. Therefore, the data that a process deals
with in the {\sc octopus} main program is not the same as the PFFT
library uses inside, and hence a data transfer is necessary.

The simplest way to carry this data-transfer out is to gather all the
data (density $\rho$ or potential $v$) in all the processes before
copying them from one grid to the other. Unfortunately, this solution
does not scale well to a large number of processors and cannot be used
in massively parallel applications. Nevertheless, we have to use this
technique with the serial FFT, because of its serial nature, and with
ISF solvers, because the kernel of the ISF solver needs to work with
entire vectors in all processes.

This problem is overcome as follows in the PFFT-based method as
implemented in {\sc octopus}.  At the initialization stage of {\sc
  octopus}, a mapping between the {\sc octopus} mesh decomposition and
the PFFT mesh decomposition is established and saved.  This mapping is
used when running the actual solver to efficiently communicate only
the strictly necessary grid data, achieving almost perfectly linear
parallel scaling. In addition, all the data-structures of the Poisson
solver (mainly density $\rho$ or potential $v$) are partitioned among
all the processes.  In this manner, memory requirement per core
decreases when more processors are available, and hence, bigger
physical systems can be simulated.

The correction for the FMM method (see section \ref{sec:fmm}) was also
implemented in a very efficient way, re-expressing its formula for
each point to avoid accessing the same memory position (which stores
the charge density of a neighbour of the point where the correction is
being calculated) more than once. Further details about the
data-transfer implementation and the correction method applied to FMM
can be found in the supporting info.

\subsection{Test on high-performance supercomputing facilities}
\label{sec:hpc}
For carrying our tests out, we have used some of the most powerful
computational resources existing at present in Europe: Curie in France
---at Commissariat \`a l'Energie Atomique (CEA)--- and two IBM Blue
Gene/P machines ---Jugene at the J\"{u}lich Supercomputing Center, and
Genius at the Rechenzentrum Garching of the Max Planck Society---.  In
addition, we have run some tests in a smaller cluster (Corvo), to
provide contrast in the lower range of the number of processors. All
four machines are considered to be representative of the current
trends of scientific High Performance Computing
\cite{Fle2012JCTC,Jia2009JCTC,Houzeaux2012CF}.

The IBM Blue Gene/P systems are based on PowerPC 450 chips at 850 MHz.
One chip consists of 4 cores, and it is soldered to a small
motherboard, together with memory (DRAM), to create a compute card
(one node). The amount of RAM per compute card is 2 GiB and they do
not have any type of hard disk. Two rows of 16 compute cards each are
plugged into one board to form a node card.  A rack holds a total of
32 node cards, and in total there are 72 racks (294,912 cores) in
J\"ulich Blue Gene/P (Jugene), and 4 racks (16,384 cores) in Garching
(Genius).  Each computing node of Blue Gene/P has 4 communication
networks: (a) a 3D torus network for point-to-point communication, 2
connections per dimension; (b) a network for collective
communications; (c) a network for barriers; and (d) a network for
control. I/O nodes, in addition, are connected to a 10 gigabit
Ethernet network.

On the other hand, the Curie supercomputer is based on Intel Xeon
X7560 processors at 2.6 GHz. Each compute node, called ``fat node'',
has 32 cores (4 chips of 8 cores each) and 128 GiB of RAM. The compute
nodes are connected with an Infiniband network. In total 11,520 cores
and almost 46 TB of RAM are available.

Finally, the configuration of the cluster Corvo, of the Spanish node
of the ETSF and Nano-Bio Spectroscopy group, is similar to the Curie
supercomputer, although in a lower scale.  Both the manufacturer (Bull
SAS) and the architecture (x86-64) are the same. More specifically,
each node of Corvo has two Intel Xeon E5645 of 6 cores and 48 GiB of
RAM. There are 960 cores in total, connected with QDR Infiniband.

\section{Results}\label{sec:results}

In this section we present the results of our tests to measure the
execution time and accuracy of the Poisson solvers discussed in
section \ref{sec:theory}. We estimate their quality in terms of
speed-up and accuracy, as well as make some remarks on the influence
of some input parameters on them. More tests and scalability
comparisons are presented in the supporting info accompanying
this paper.

\subsection{Accuracy}\label{sec:rad}

We calculate the Hartree potential created by Gaussian charge
distributions in an important part of our tests.  Such a potential can
be calculated analytically, and hence the error made by any method can
be measured by comparing the two results: numerical and analytical.
We defined two different variables to measure the accuracy of a
Poisson solver, $D$ and $F$, as follows:
\begin{subequations}
  \begin{align}
    D & := \frac{ \sum_{ijk} | v_{a}(\vec{r}_{ijk}) - v_{n}(\vec{r}_{ijk}) |}
    {\sum_{ijk} | v_{a}(\vec{r}_{ijk}) |}  \ , \label{errorDD}\\
    E_a &=\frac{1}{2}\sum_{ijk} \rho(\vec{r}_{ijk}) v_{a}(\vec{r}_{ijk}) \ , \\
    E_n &=\frac{1}{2}\sum_{ijk} \rho(\vec{r}_{ijk}) v_{n}(\vec{r}_{ijk}) \  , \\
    F & := \frac{E_{a} - E_{n} }{E_{a}} \ , \label{errorFF}
  \end{align}
\end{subequations}
where $\vec{r}_{ijk}$ are all the points of the analysed grid, $v_{a}$
is the analytically calculated potential, and $v_n$ is the potential
calculated by either the PFFT, ISF, FMM, CG, or multigrid-based
method.  $E_a$ and $E_n$ are the expressions for the Hartree energies
obtained using the potentials that were calculated analytically and
numerically, respectively. These variables provide an estimate of the
deviations of the calculated potential and energy from their exact
values. $D$ gives a comprehensive estimate of the error, for it takes
into account the calculated potential in all points. The total Hartree
energy coming from these potentials is a physically relevant variable,
so an estimate of its error is also meaningful.

In \ref{tab:errors2} we display some values of $D$ and $F$ for the
tested Poisson solvers. The input charge distributions correspond to
Gaussian distributions represented in cubic grids with variable edge
($2L_e$) and constant spacing between consecutive points ($L=0.2$
$\textrm{\AA}$). The additional {\sc Octopus} parameters we use are:
\texttt{PoissonSolverMaxMultipole = 7} (for multigrid and conjugate
gradients), \texttt{MultigridLevels = max\_levels} (i.e. use all
available multigrid levels in the multigrid calculation),
\texttt{DeltaEFMM = 0.0001} and \texttt{AlphaFMM = 0.291262136} (for
the fast multipole method). Further information on these parameters
can be found in section \ref{sec:inflinppar} and in the supplementary
material.
\begin{table}[htbp]
  \begin{center}
  
  \begin{tabular}{cccc}
    \underline{$F$ error:}&&&\\
    \begin{tabular}{ccccccc}
      $L_e$ (\AA) & PFFT & ISF & FMM & CG  & Multigrid \\ \hline
      7&    2$\cdot 10^{-3}$  & 2$\cdot 10^{-3}$   &    2$\cdot 10^{-3}$ &    2$\cdot 10^{-3}$ &         2$\cdot 10^{-3}$   \\
      10&     9$\cdot 10^{-6}$ &   9$\cdot 10^{-6}$ &   4$\cdot 10^{-6}$ &   4$\cdot 10^{-6}$ &        9$\cdot 10^{-6}$  \\
      15.8& 2$\cdot 10^{-12}$ &  3$\cdot 10^{-7}$ & 3$\cdot 10^{-6}$ & 3$\cdot 10^{-5}$  & 5$\cdot 10^{-6}$ \\
      22.1&    <2$\cdot 10^{-12}$   &  1$\cdot 10^{-11}$ &    3$\cdot 10^{-6}$  &   4$\cdot 10^{-4}$ &        -1$\cdot 10^{-6}$   \\
      25.9&     <2$\cdot 10^{-12}$ &   9$\cdot 10^{-12}$ &   -4$\cdot 10^{-6}$ &   5$\cdot 10^{-3}$  &        2$\cdot 10^{-7}$  \\
      31.7&   <4$\cdot 10^{-13}$   &  8$\cdot 10^{-12}$  &     -3$\cdot 10^{-6}$ &   1$\cdot 10^{-2}$  &         -5$\cdot 10^{-7}$   \\
    \end{tabular} \\
    \underline{$D$ error:} &&&\\
    \begin{tabular}{ccccccc}
      $L_e$ (\AA) & PFFT & ISF & FMM & CG & Multigrid \\    \hline
      7&    9$\cdot 10^{-3}$  &   4$\cdot 10^{-3}$ &    4$\cdot 10^{-3}$  &   4$\cdot 10^{-3}$    &       4$\cdot 10^{-3}$     \\
      10&     1$\cdot 10^{-4}$ & 2$\cdot 10^{-5}$   &   8$\cdot 10^{-5}$ &  3$\cdot 10^{-5}$    &    2$\cdot 10^{-5}$  \\
      15.8& 6$\cdot 10^{-5}$ & 2$\cdot 10^{-7}$ & 1$\cdot 10^{-4}$ & 5$\cdot 10^{-5}$    & 6$\cdot 10^{-6}$ \\
      22.1&     1$\cdot 10^{-5}$ & 4$\cdot 10^{-10}$   &   3$\cdot 10^{-4}$ &  3$\cdot 10^{-4}$    &    2$\cdot 10^{-6}$  \\
      25.9&      4$\cdot 10^{-7}$ &   7$\cdot 10^{-10}$ &  3$\cdot 10^{-4}$ &   5$\cdot 10^{-3}$    &       4$\cdot 10^{-7}$      \\
      31.7&     4$\cdot 10^{-8}$ &   1 $\cdot 10^{-9}$ &   2$\cdot 10^{-4}$&  1$\cdot 10^{-2}$     &     4$\cdot 10^{-7}$       \\
  \end{tabular}
  \end{tabular}
  \end{center}
  \caption{$F$ and $D$ errors of different Poisson solvers in the calculation of the Hartree 
  potential created by a Gaussian charge distribution represented on a grid of edge $L_e=15.8$ $\textrm{\AA}$ 
  and spacing
  $0.2$ \AA.}
  \label{tab:errors2}
 \end{table}
 From \ref{tab:errors2} it can be said that FFT-based methods (ISF,
 PFFT and FFT serial) give the best accuracies. Moreover, the
 multigrid and conjugate gradient solvers do not produce appropriate
 results when the set of grid points corresponds to an adapted shape
 (as is the usual case in {\sc octopus}), and must be used with
 compact shapes (such as spherical or parallelepiped).  Conjugate
 gradient methods' accuracy is acceptable for this type of
 test. Nevertheless, they show some problems if used in the
 calculation of the ground state of a set of electrons to solve the
 Kohn-Sham equation or to perform time-dependent density functional
 theory calculations (which require calculation of the Hartree
 potential).  The iterative procedure inherent to the used conjugate
 gradient method sometimes shows some convergence problems.

 Other than using $D$ and $F$ (which depend solely on the charge
 density and Hartree potential), another way to measure accuracy is to
 calculate the ground state of a given system and check the effect of
 the Poisson solver on some of its corresponding quantities. We
 calculated the ground state of a system of chlorophyll stretches
 containing 180 atoms~\cite{Liu2004} (see supporting info for
 atomic system information). To this end, we used pseudopotentials, so
 460 electrons appeared in the calculation. The grid shape was a set
 of spheres of radius 4.0 $\textrm{\AA}$ centred at the nuclei, and
 the grid spacing was 0.23 $\textrm{\AA}$. The exchange correlation
 functional was the LDA functional \cite{PhysRevB.54.1703}. In
 \ref{tab:errors3} we display the value of the Hartree energy, the
 highest eigenvalue, and the HOMO-LUMO gap. PFFT is expected to
 provide the most accurate results (by considering
 \ref{tab:errors2}). However, the differences among all five methods
 can be considered negligible (lower than the errors introduced by the
 density functional theory exchange-correlation functional). The
 maximum difference of Hartree energy divided by the number of
 electrons is less than 0.015 eV. The maximum difference in the
 HOMO-LUMO gap is less than 0.0092 eV.

\begin{table}[htbp]
  \begin{center}
    \begin{tabular}{lccc}
      &  Hartree energy (eV)       & HOMO  (eV)     &    HOMO-LUMO gap
      (eV)   \\ \hline
      PFFT  &     240820.70   &     -4.8922    &      1.4471  \\ 
      ISF   &     240821.48   &     -4.8935    &      1.4489  \\ 
      FMM   &     240817.29  &      -4.8906   &       1.4429  \\ 
      CG    &     240815.01  &      -4.9009   &       1.4521  \\
      Multigrid   &     240814.69  &      -4.8979   &       1.4506  \\
    \end{tabular}
  \end{center}
  \caption{Comparison of some quantities corresponding to the ground state of a set of chlorophyll stretches
    with 180 atoms. }
  \label{tab:errors3}
\end{table}

\subsection{Execution time}
In order to gauge the performance of the Poisson solvers, we have
measured the total execution time that the calculation of the
classical potential took for each method as a function of the number
of processes involved in the resolution (regardless of initialisation
times).  We ran one single MPI process per core on Curie and Corvo,
and only one MPI process per node on Blue Gene/P's \bibnote{Definitions
  of basic concepts on high performance computing (e.g. 'node' and
  'core') can be found in \cite{GR2012IJMPC}.}.  When possible, we ran
the same set of standard tests on each of these machines.  However, in
cases where the size of the simulated system is high, the limited
amount of memory per core becomes a serious problem. This problem is
critical in Blue Gene/P machines, which have only 2 GiB per node, i.e. 512
MiB per core, so we executed one MPI process per node, with four
OpenMP threads.  The Curie supercomputer offers 4 GiB per core, but in
some cases we could use more memory by selecting more cores and keeping
them idle.  Runs up to 4096 processes were only possible in Blue
Gene/P and Curie machines; for Corvo it was only possible to run up to 512 MPI
processes.

In our efficiency tests we calculated the potential created by
Gaussian charge distributions represented in cubic grids with edge
length $2L_e$, where $L_e$ is half the edge of the
parallelepiped mesh and the used values were 7.0, 10.0, 15.8, 22.1,
25.9 and 31.7 $\AA$ respectively (always with a spacing of 0.2
$\AA$). The smallest simulated system, with $L_e=7$, contains
$[2*L_e/spacing +1]^3=357,911$ grid points.  By reason of memory
limitations, the largest system simulated on a Blue Gene/P had
$L_e=25.9$ ($17,373,979$ points). On the Corvo and Curie machines, we were
able to run a bigger system of $L_e=31.7$ ($31,855,013$ points). For
the FFT-based methods we used an additional box, ranging from $143^3$ to
$637^3$ (up to $525^3$ on a Blue Gene/P).

\begin{landscape}
\begin{figure}
  \centering 
  \def\svgwidth{200pt} 
  \def\svgheigh{100pt}
  \begingroup%
  \makeatletter%
  \providecommand\color[2][]{%
    \errmessage{(Inkscape) Color is used for the text in Inkscape, but the package 'color.sty' is not loaded}%
    \renewcommand\color[2][]{}%
  }%
  \providecommand\transparent[1]{%
    \errmessage{(Inkscape) Transparency is used (non-zero) for the text in Inkscape, but the package 'transparent.sty' is not loaded}%
    \renewcommand\transparent[1]{}%
  }%
  \providecommand\rotatebox[2]{#2}%
  \ifx\svgwidth\undefined%
    \setlength{\unitlength}{656.03125bp}%
    \ifx\svgscale\undefined%
      \relax%
    \else%
      \setlength{\unitlength}{\unitlength * \real{\svgscale}}%
    \fi%
  \else%
    \setlength{\unitlength}{\svgwidth}%
  \fi%
  \global\let\svgwidth\undefined%
  \global\let\svgscale\undefined%
  \makeatother%
  \begin{picture}(1,1.01743438)%
    \put(0,0){\includegraphics[width=\unitlength]{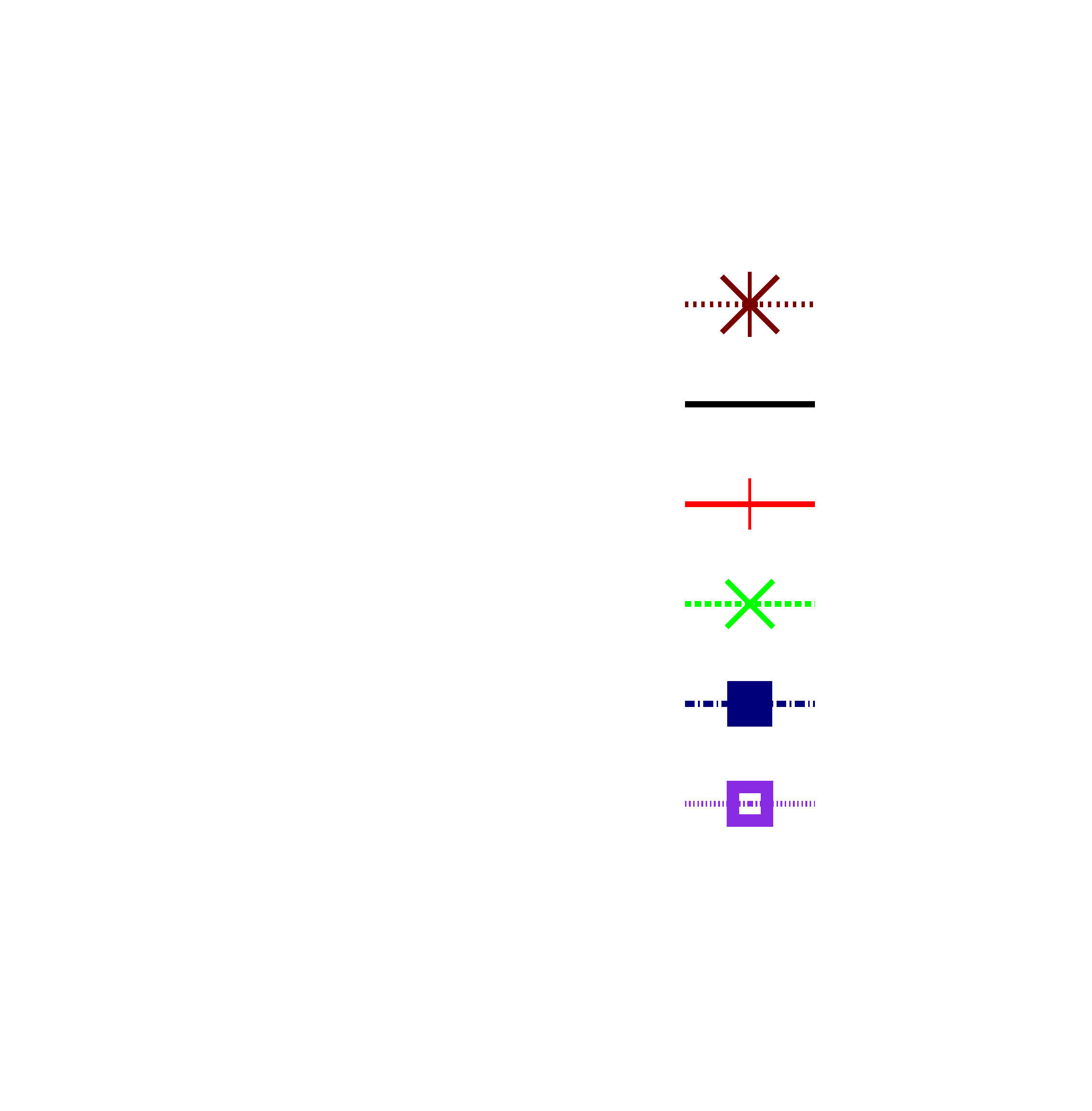}}%
    \put(0.27047111,0.63190587){\makebox(0,0)[lb]{\smash{Serial FFT}}}%
    \put(0.27047111,0.54044682){\makebox(0,0)[lb]{\smash{ISF}}}%
    \put(0.27047111,0.44898776){\makebox(0,0)[lb]{\smash{FMM}}}%
    \put(0.27047111,0.3575287){\makebox(0,0)[lb]{\smash{CG corrected}}}%
    \put(0.27047111,0.26606964){\makebox(0,0)[lb]{\smash{Multigrid}}}%
    \put(0.27047111,0.72336493){\color[rgb]{0,0,0}\makebox(0,0)[lb]{\smash{PFFT}}}%
  \end{picture}%
  \endgroup%
  \def\svgwidth{200pt} 
  \begingroup%
  \makeatletter%
  \providecommand\color[2][]{%
    \errmessage{(Inkscape) Color is used for the text in Inkscape, but the package 'color.sty' is not loaded}%
    \renewcommand\color[2][]{}%
  }%
  \providecommand\transparent[1]{%
    \errmessage{(Inkscape) Transparency is used (non-zero) for the text in Inkscape, but the package 'transparent.sty' is not loaded}%
    \renewcommand\transparent[1]{}%
  }%
  \providecommand\rotatebox[2]{#2}%
  \ifx\svgwidth\undefined%
    \setlength{\unitlength}{657.13989935bp}%
    \ifx\svgscale\undefined%
      \relax%
    \else%
      \setlength{\unitlength}{\unitlength * \real{\svgscale}}%
    \fi%
  \else%
    \setlength{\unitlength}{\svgwidth}%
  \fi%
  \global\let\svgwidth\undefined%
  \global\let\svgscale\undefined%
  \makeatother%
  \begin{picture}(1,1.01539838)%
    \put(0,0){\includegraphics[width=\unitlength]{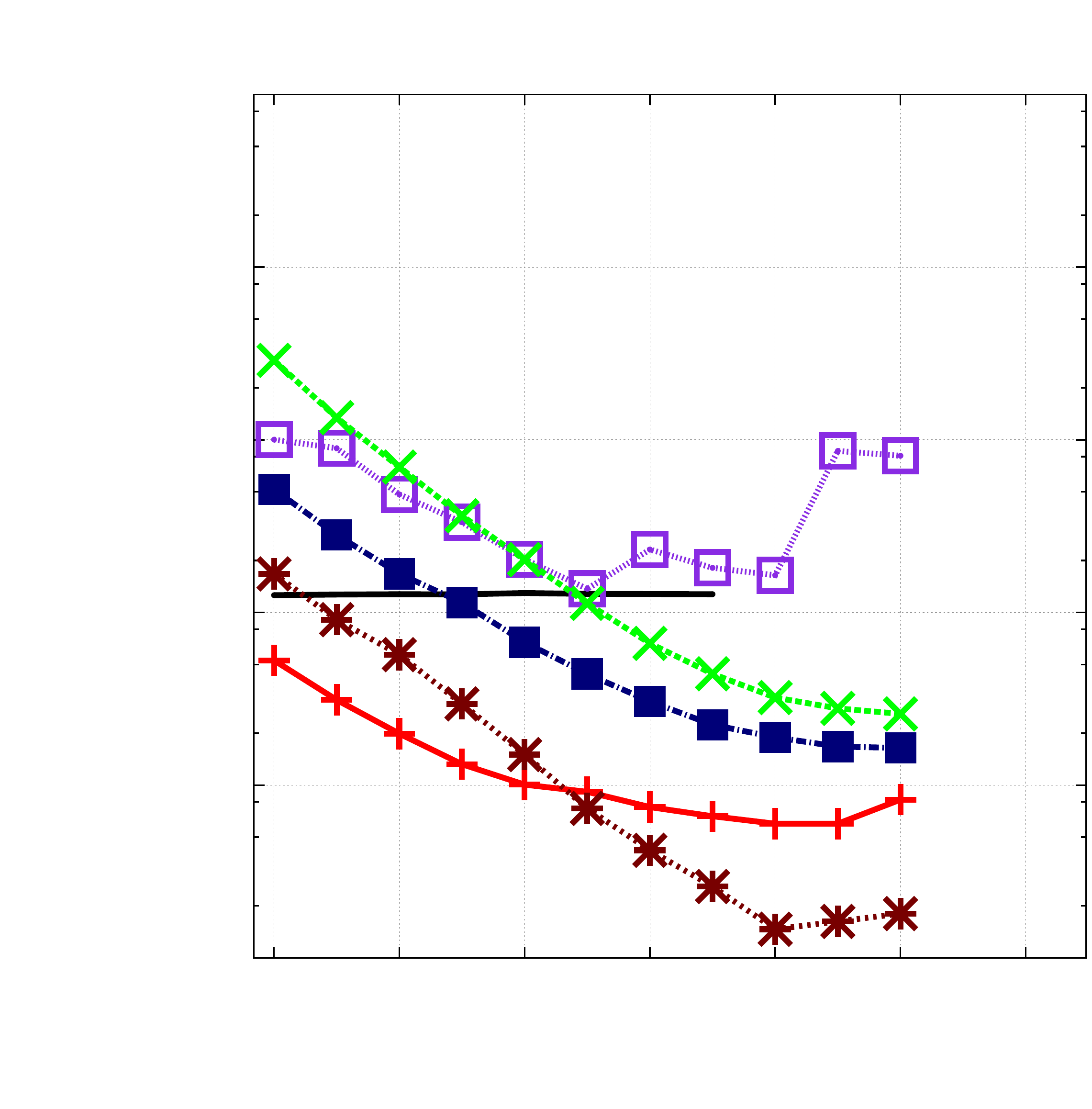}}%
    \put(0.5,1){\makebox(0,0)[lb]{\smash{$L=7.0$}}}%
    \put(0.11634984,0.12313659){\makebox(0,0)[lb]{\smash{0.01}}}%
    \put(0.14173926,0.281246){\makebox(0,0)[lb]{\smash{0.1}}}%
    \put(0.17981228,0.43935541){\makebox(0,0)[lb]{\smash{1}}}%
    \put(0.15442194,0.59746482){\makebox(0,0)[lb]{\smash{10}}}%
    \put(0.12903329,0.75557422){\makebox(0,0)[lb]{\smash{100}}}%
    \put(0.10364387,0.91368363){\makebox(0,0)[lb]{\smash{1000}}}%
    \put(0.24465387,0.07748421){\makebox(0,0)[lb]{\smash{\small 1}}}%
    \put(0.35939352,0.07748421){\makebox(0,0)[lb]{\smash{\small 4}}}%
    \put(0.46143724,0.07748421){\makebox(0,0)[lb]{\smash{\small 16}}}%
    \put(0.57617689,0.07748421){\makebox(0,0)[lb]{\smash{\small 64}}}%
    \put(0.67822213,0.07748421){\makebox(0,0)[lb]{\smash{\small 256}}}%
    \put(0.78026737,0.07748421){\makebox(0,0)[lb]{\smash{\small 1024}}}%
    \put(0.89500702,0.07748421){\makebox(0,0)[lb]{\smash{\small 4096}}}%
    \put(0.03323618,0.49425239){\rotatebox{90}{\makebox(0,0)[lb]{\smash{t (s)}}}}%
    \put(0.5159614,0.00900565){\makebox(0,0)[lb]{\smash{\small MPI proc.}}}%
  \end{picture}%
  \endgroup%
  \def\svgwidth{200pt} 
  \begingroup%
  \makeatletter%
  \providecommand\color[2][]{%
    \errmessage{(Inkscape) Color is used for the text in Inkscape, but the package 'color.sty' is not loaded}%
    \renewcommand\color[2][]{}%
  }%
  \providecommand\transparent[1]{%
    \errmessage{(Inkscape) Transparency is used (non-zero) for the text in Inkscape, but the package 'transparent.sty' is not loaded}%
    \renewcommand\transparent[1]{}%
  }%
  \providecommand\rotatebox[2]{#2}%
  \ifx\svgwidth\undefined%
    \setlength{\unitlength}{657.13989935bp}%
    \ifx\svgscale\undefined%
      \relax%
    \else%
      \setlength{\unitlength}{\unitlength * \real{\svgscale}}%
    \fi%
  \else%
    \setlength{\unitlength}{\svgwidth}%
  \fi%
  \global\let\svgwidth\undefined%
  \global\let\svgscale\undefined%
  \makeatother%
  \begin{picture}(1,1.01539838)%
    \put(0,0){\includegraphics[width=\unitlength]{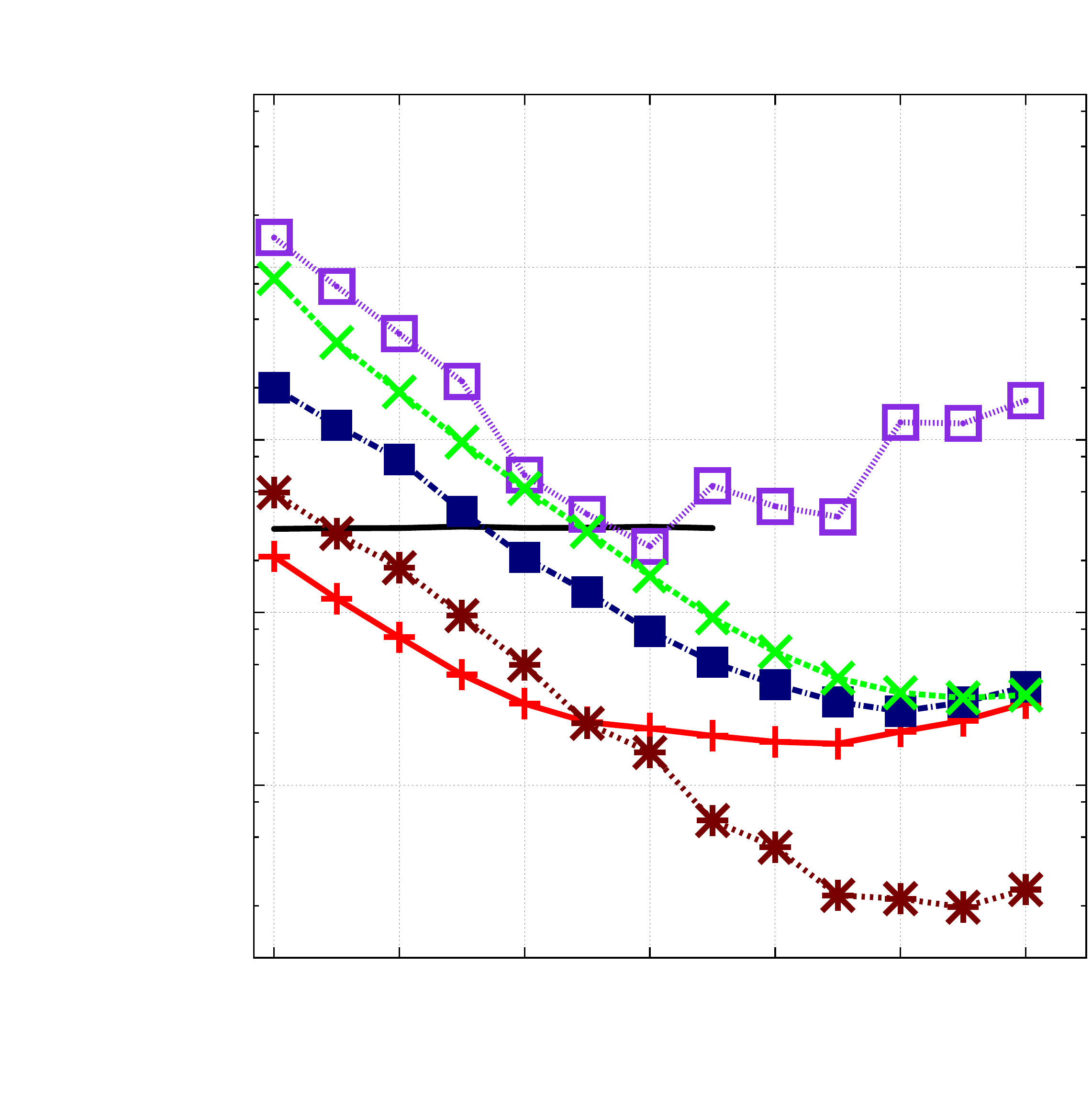}}%
    \put(0.5,1){\makebox(0,0)[lb]{\smash{$L=10.0$}}}%
    \put(0.11634984,0.12313659){\makebox(0,0)[lb]{\smash{0.01}}}%
    \put(0.14173926,0.281246){\makebox(0,0)[lb]{\smash{0.1}}}%
    \put(0.17981228,0.43935541){\makebox(0,0)[lb]{\smash{1}}}%
    \put(0.15442194,0.59746482){\makebox(0,0)[lb]{\smash{10}}}%
    \put(0.12903329,0.75557422){\makebox(0,0)[lb]{\smash{100}}}%
    \put(0.10364387,0.91368363){\makebox(0,0)[lb]{\smash{1000}}}%
    \put(0.24465387,0.07748421){\makebox(0,0)[lb]{\smash{\small 1}}}%
    \put(0.35939352,0.07748421){\makebox(0,0)[lb]{\smash{\small 4}}}%
    \put(0.46143724,0.07748421){\makebox(0,0)[lb]{\smash{\small 16}}}%
    \put(0.57617689,0.07748421){\makebox(0,0)[lb]{\smash{\small 64}}}%
    \put(0.67822213,0.07748421){\makebox(0,0)[lb]{\smash{\small 256}}}%
    \put(0.78026737,0.07748421){\makebox(0,0)[lb]{\smash{\small 1024}}}%
    \put(0.89500702,0.07748421){\makebox(0,0)[lb]{\smash{\small 4096}}}%
    \put(0.03323618,0.49425239){\rotatebox{90}{\makebox(0,0)[lb]{\smash{t (s)}}}}%
    \put(0.5159614,0.00900565){\makebox(0,0)[lb]{\smash{\small MPI proc.}}}%
  \end{picture}%
  \endgroup%
  \def\svgwidth{200pt} 
  \begingroup%
  \makeatletter%
  \providecommand\color[2][]{%
    \errmessage{(Inkscape) Color is used for the text in Inkscape, but the package 'color.sty' is not loaded}%
    \renewcommand\color[2][]{}%
  }%
  \providecommand\transparent[1]{%
    \errmessage{(Inkscape) Transparency is used (non-zero) for the text in Inkscape, but the package 'transparent.sty' is not loaded}%
    \renewcommand\transparent[1]{}%
  }%
  \providecommand\rotatebox[2]{#2}%
  \ifx\svgwidth\undefined%
    \setlength{\unitlength}{657.13989935bp}%
    \ifx\svgscale\undefined%
      \relax%
    \else%
      \setlength{\unitlength}{\unitlength * \real{\svgscale}}%
    \fi%
  \else%
    \setlength{\unitlength}{\svgwidth}%
  \fi%
  \global\let\svgwidth\undefined%
  \global\let\svgscale\undefined%
  \makeatother%
  \begin{picture}(1,1.01539838)%
    \put(0,0){\includegraphics[width=\unitlength]{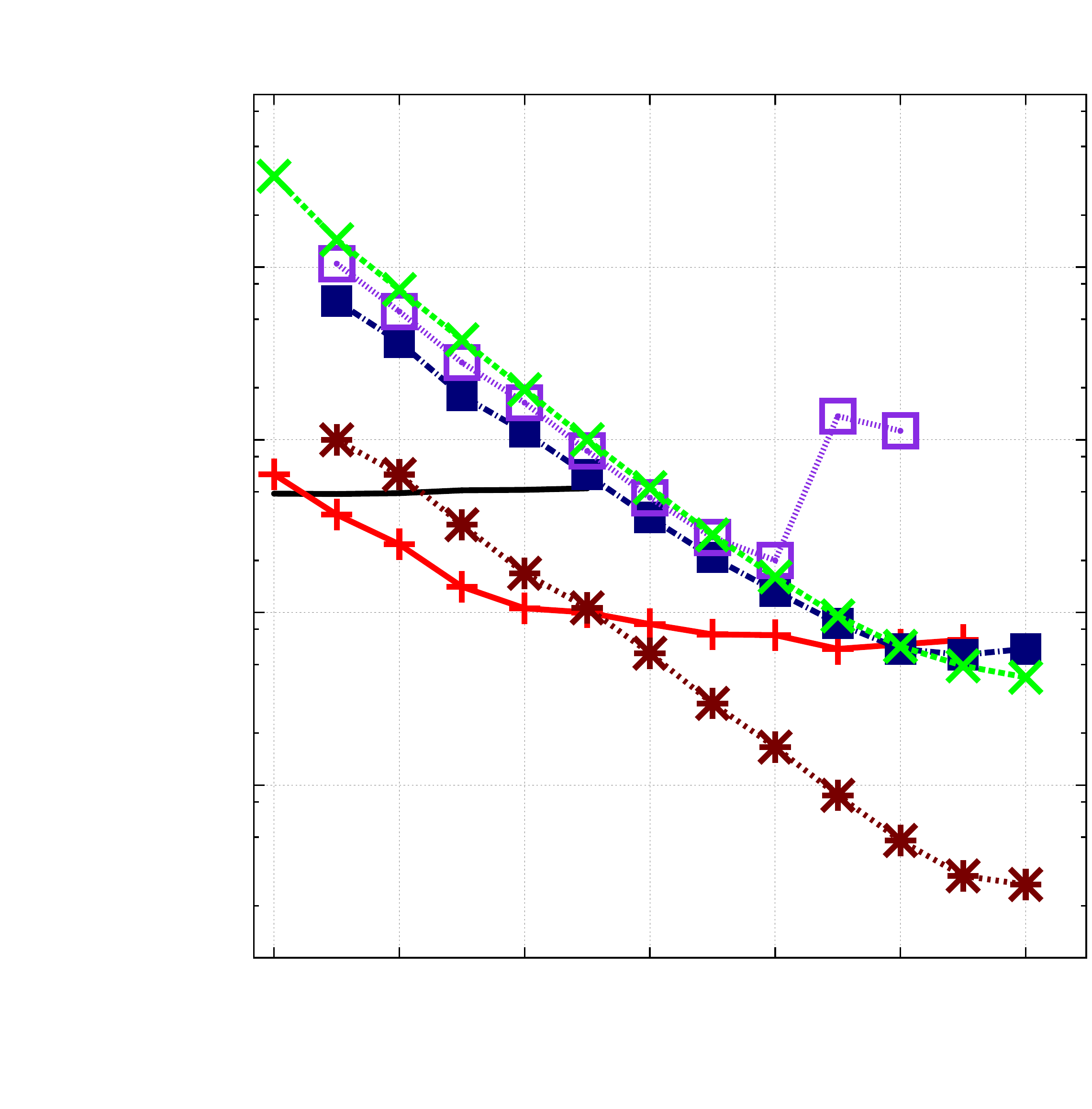}}%
    \put(0.5,0.95){\makebox(0,0)[lb]{\smash{$L=15.0$}}}%
    \put(0.11634984,0.12313659){\makebox(0,0)[lb]{\smash{0.01}}}%
    \put(0.14173926,0.281246){\makebox(0,0)[lb]{\smash{0.1}}}%
    \put(0.17981228,0.43935541){\makebox(0,0)[lb]{\smash{1}}}%
    \put(0.15442194,0.59746482){\makebox(0,0)[lb]{\smash{10}}}%
    \put(0.12903329,0.75557422){\makebox(0,0)[lb]{\smash{100}}}%
    \put(0.10364387,0.91368363){\makebox(0,0)[lb]{\smash{1000}}}%
    \put(0.24465387,0.07748421){\makebox(0,0)[lb]{\smash{\small 1}}}%
    \put(0.35939352,0.07748421){\makebox(0,0)[lb]{\smash{\small 4}}}%
    \put(0.46143724,0.07748421){\makebox(0,0)[lb]{\smash{\small 16}}}%
    \put(0.57617689,0.07748421){\makebox(0,0)[lb]{\smash{\small 64}}}%
    \put(0.67822213,0.07748421){\makebox(0,0)[lb]{\smash{\small 256}}}%
    \put(0.78026737,0.07748421){\makebox(0,0)[lb]{\smash{\small 1024}}}%
    \put(0.89500702,0.07748421){\makebox(0,0)[lb]{\smash{\small 4096}}}%
    \put(0.03323618,0.49425239){\rotatebox{90}{\makebox(0,0)[lb]{\smash{t (s)}}}}%
    \put(0.5159614,0.00900565){\makebox(0,0)[lb]{\smash{\small MPI proc.}}}%
  \end{picture}%
  \endgroup%
  \def\svgwidth{200pt} 
  \begingroup%
  \makeatletter%
  \providecommand\color[2][]{%
    \errmessage{(Inkscape) Color is used for the text in Inkscape, but the package 'color.sty' is not loaded}%
    \renewcommand\color[2][]{}%
  }%
  \providecommand\transparent[1]{%
    \errmessage{(Inkscape) Transparency is used (non-zero) for the text in Inkscape, but the package 'transparent.sty' is not loaded}%
    \renewcommand\transparent[1]{}%
  }%
  \providecommand\rotatebox[2]{#2}%
  \ifx\svgwidth\undefined%
    \setlength{\unitlength}{657.13989935bp}%
    \ifx\svgscale\undefined%
      \relax%
    \else%
      \setlength{\unitlength}{\unitlength * \real{\svgscale}}%
    \fi%
  \else%
    \setlength{\unitlength}{\svgwidth}%
  \fi%
  \global\let\svgwidth\undefined%
  \global\let\svgscale\undefined%
  \makeatother%
  \begin{picture}(1,1.01539838)%
    \put(0,0){\includegraphics[width=\unitlength]{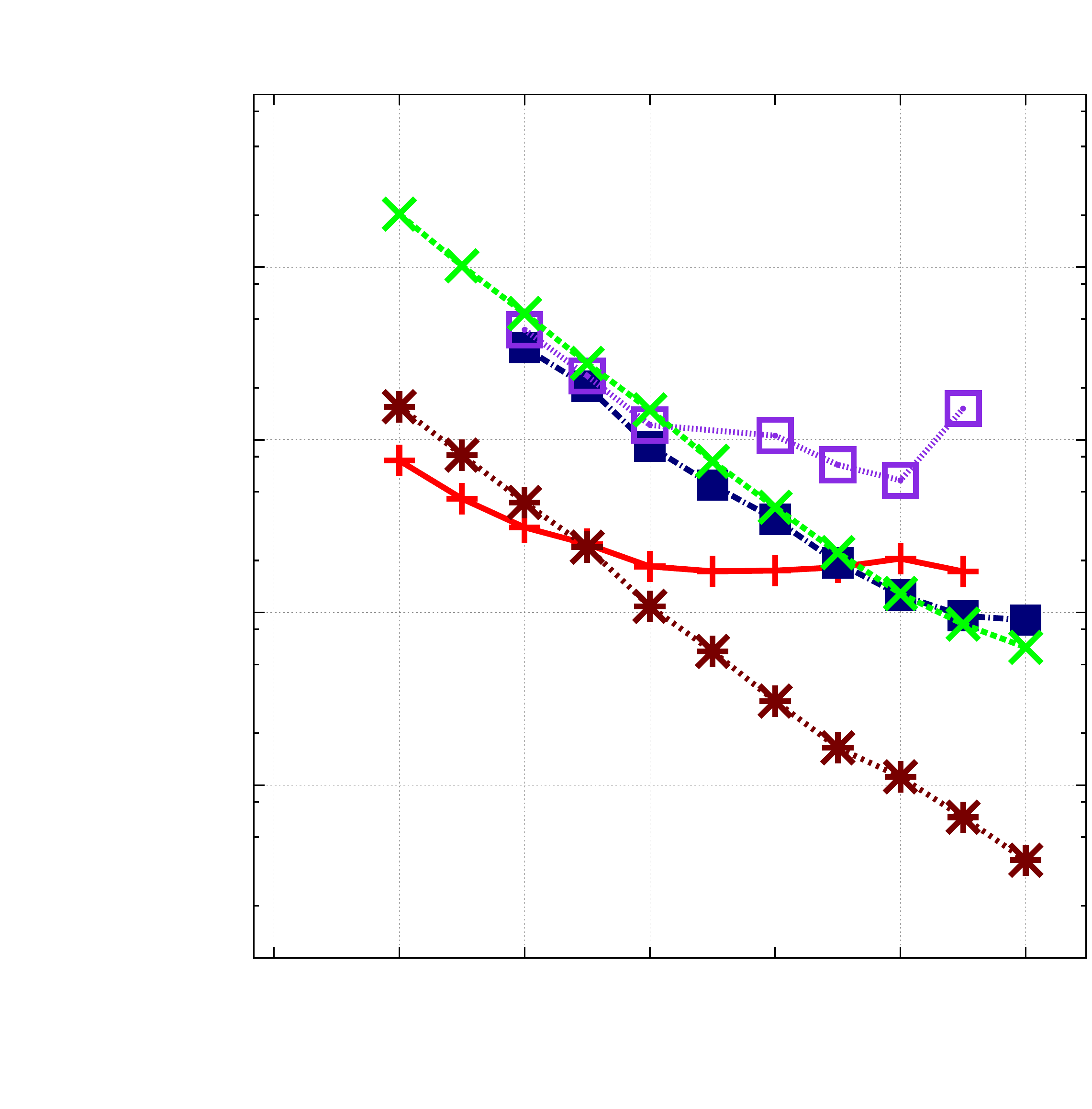}}%
    \put(0.5,0.95){\makebox(0,0)[lb]{\smash{$L=22.0$}}}%
    \put(0.11634984,0.12313659){\makebox(0,0)[lb]{\smash{0.01}}}%
    \put(0.14173926,0.281246){\makebox(0,0)[lb]{\smash{0.1}}}%
    \put(0.17981228,0.43935541){\makebox(0,0)[lb]{\smash{1}}}%
    \put(0.15442194,0.59746482){\makebox(0,0)[lb]{\smash{10}}}%
    \put(0.12903329,0.75557422){\makebox(0,0)[lb]{\smash{100}}}%
    \put(0.10364387,0.91368363){\makebox(0,0)[lb]{\smash{1000}}}%
    \put(0.24465387,0.07748421){\makebox(0,0)[lb]{\smash{\small 1}}}%
    \put(0.35939352,0.07748421){\makebox(0,0)[lb]{\smash{\small 4}}}%
    \put(0.46143724,0.07748421){\makebox(0,0)[lb]{\smash{\small 16}}}%
    \put(0.57617689,0.07748421){\makebox(0,0)[lb]{\smash{\small 64}}}%
    \put(0.67822213,0.07748421){\makebox(0,0)[lb]{\smash{\small 256}}}%
    \put(0.78026737,0.07748421){\makebox(0,0)[lb]{\smash{\small 1024}}}%
    \put(0.89500702,0.07748421){\makebox(0,0)[lb]{\smash{\small 4096}}}%
    \put(0.03323618,0.49425239){\rotatebox{90}{\makebox(0,0)[lb]{\smash{t (s)}}}}%
    \put(0.5159614,0.00900565){\makebox(0,0)[lb]{\smash{\small MPI proc.}}}%
  \end{picture}%
  \endgroup%
  \def\svgwidth{200pt} 
  \begingroup%
  \makeatletter%
  \providecommand\color[2][]{%
    \errmessage{(Inkscape) Color is used for the text in Inkscape, but the package 'color.sty' is not loaded}%
    \renewcommand\color[2][]{}%
  }%
  \providecommand\transparent[1]{%
    \errmessage{(Inkscape) Transparency is used (non-zero) for the text in Inkscape, but the package 'transparent.sty' is not loaded}%
    \renewcommand\transparent[1]{}%
  }%
  \providecommand\rotatebox[2]{#2}%
  \ifx\svgwidth\undefined%
    \setlength{\unitlength}{657.13989935bp}%
    \ifx\svgscale\undefined%
      \relax%
    \else%
      \setlength{\unitlength}{\unitlength * \real{\svgscale}}%
    \fi%
  \else%
    \setlength{\unitlength}{\svgwidth}%
  \fi%
  \global\let\svgwidth\undefined%
  \global\let\svgscale\undefined%
  \makeatother%
  \begin{picture}(1,1.01539838)%
    \put(0,0){\includegraphics[width=\unitlength]{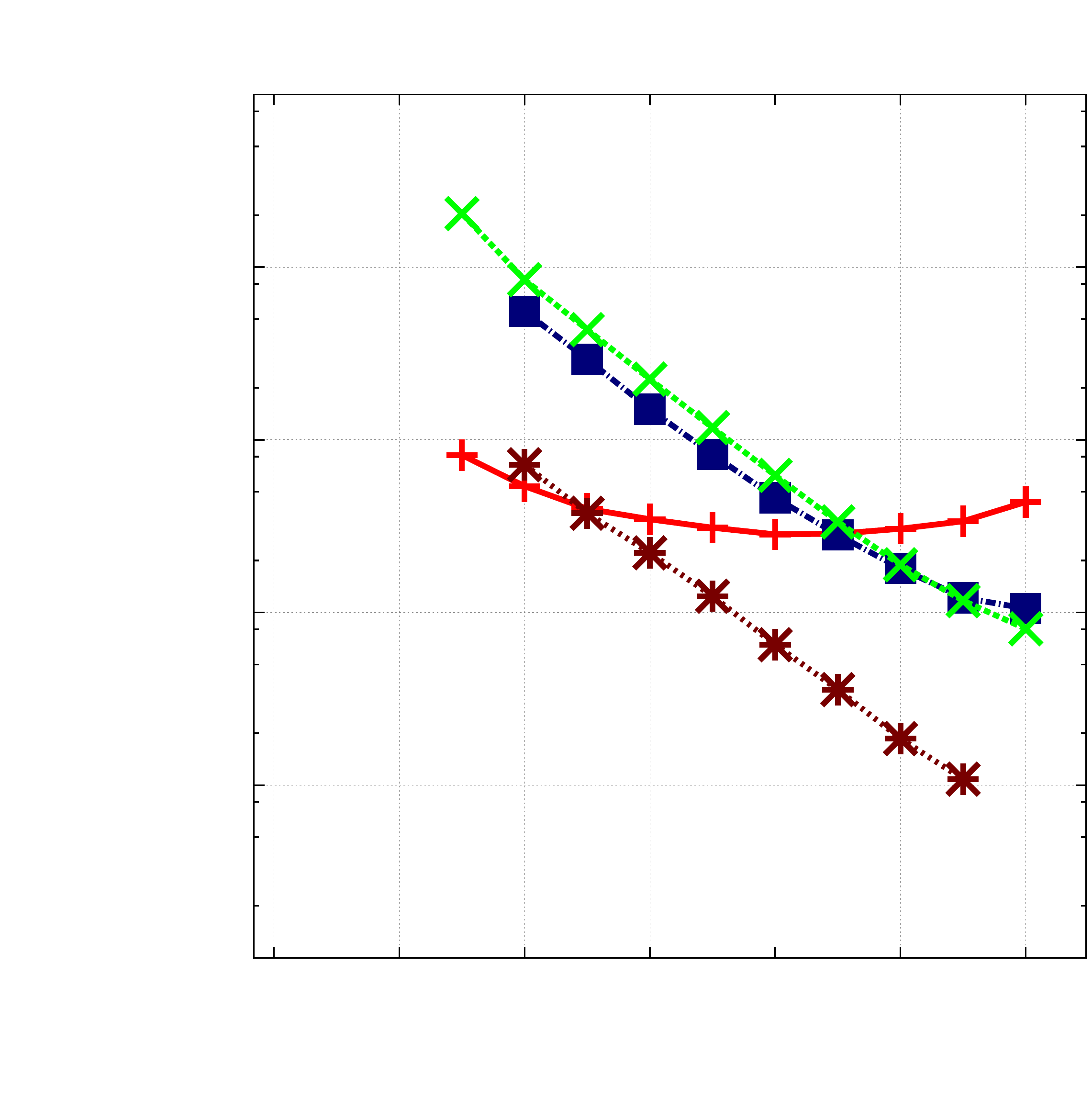}}%
    \put(0.5,0.95){\makebox(0,0)[lb]{\smash{$L=25.0$}}}%
    \put(0.11634984,0.12313659){\makebox(0,0)[lb]{\smash{0.01}}}%
    \put(0.14173926,0.281246){\makebox(0,0)[lb]{\smash{0.1}}}%
    \put(0.17981228,0.43935541){\makebox(0,0)[lb]{\smash{1}}}%
    \put(0.15442194,0.59746482){\makebox(0,0)[lb]{\smash{10}}}%
    \put(0.12903329,0.75557422){\makebox(0,0)[lb]{\smash{100}}}%
    \put(0.10364387,0.91368363){\makebox(0,0)[lb]{\smash{1000}}}%
    \put(0.24465387,0.07748421){\makebox(0,0)[lb]{\smash{\small 1}}}%
    \put(0.35939352,0.07748421){\makebox(0,0)[lb]{\smash{\small 4}}}%
    \put(0.46143724,0.07748421){\makebox(0,0)[lb]{\smash{\small 16}}}%
    \put(0.57617689,0.07748421){\makebox(0,0)[lb]{\smash{\small 64}}}%
    \put(0.67822213,0.07748421){\makebox(0,0)[lb]{\smash{\small 256}}}%
    \put(0.78026737,0.07748421){\makebox(0,0)[lb]{\smash{\small 1024}}}%
    \put(0.89500702,0.07748421){\makebox(0,0)[lb]{\smash{\small 4096}}}%
    \put(0.03323618,0.49425239){\rotatebox{90}{\makebox(0,0)[lb]{\smash{t (s)}}}}%
    \put(0.5159614,0.00900565){\makebox(0,0)[lb]{\smash{\small MPI proc.}}}%
  \end{picture}%
  \endgroup%
  
  \caption{Execution times for the calculation of the Hartree
    potential created by a Gaussian charge distribution on a Blue
    Gene/P machine as a function of six different Poisson solvers and
    of the involved number of MPI processes (each MPI process having 4
    OpenMP threads) for different simulated system sizes ($[2*L_e/0.2
    +1]^3$ points). }
  \label{fig:runs_bg}
\end{figure}
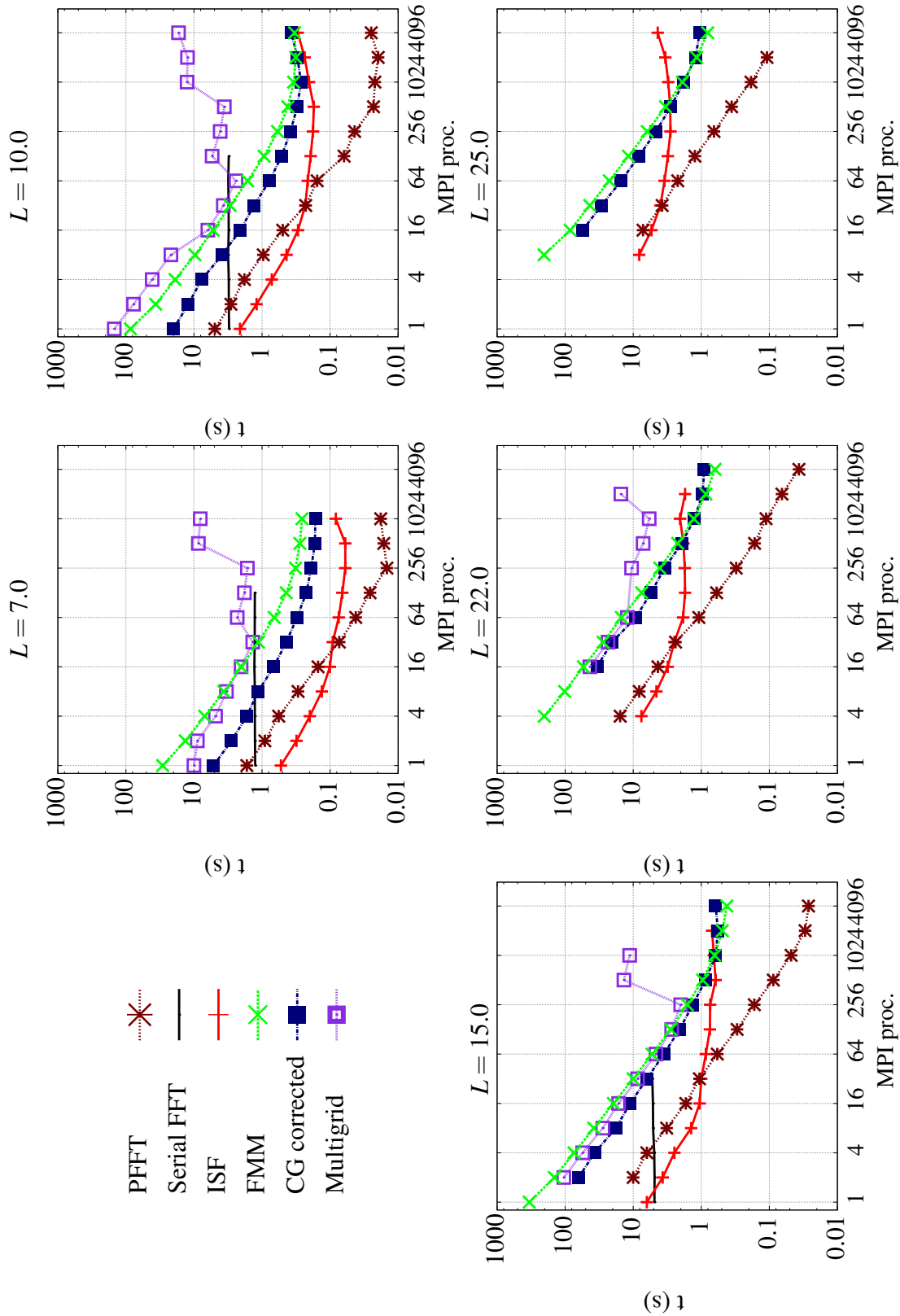
\end{landscape}

First, we present the results obtained on a Blue Gene/P machine in
\ref{fig:runs_bg}. Each graph of the figure represents a different
problem size, with a different value of $L_e$.  The X axis of
\ref{fig:runs_bg} shows the used number of MPI processes, each having
4 OpenMP threads. The Y axis is the time in seconds for the Poisson
solver.  Increasing memory requirements with higher $L_e$ made some
systems not able to fit in the available RAM memory, precluding the
opportunity to run the serial FFT of size $L_e=22.1$ and above, and
multigrid of size $L_e=25.9$ and above.  One of the most memory
consuming part of these tests was the generation of the mesh
partitions, one per problem size per processor, regardless of the
solver. Fortunately, these partitions must be done only once and they
are machine independent. Moreover, they could be read from the restart
files, which could also be transferred from one machine to
another. However, this is not true of the multigrid solver, for which
the mesh partition must be calculated for every multigrid level, which
made it, impossible, for example, to run the system of size $L_e=25.9$
due to the high memory requirement of the partitioning.

As can be observed in \ref{fig:runs_bg}, all the tests show a similar
relative behaviour with regard to the system size under simulation and
the number of MPI processes: execution times decrease with the number
of processes until saturation, and larger systems allow the efficient
use of a higher number of parallel processes, i.e., the solvers
``saturates'' at a higher number of processes.  This is especially
true for the newly implemented solvers, based on PFFT and FMM. Thus,
this behaviour leaves the way open to simulate physical systems of
more realistic size if tens of thousands of processors cores are
available.

The multigrid solver shows the poorest results, and saturates with a
relatively low number of parallel processes. The problem comes from
the fact that only a limited number of multigrid levels can be used
(obviously, every process needs to use at least one grid point to
execute correctly).  At most, we chose the closest natural number to
$\log_8(N)$ as the number of multigrid levels, where N is the total
number of grid points. In fact, if a high number of levels can be
used, then the obtained speed-ups are appreciable, but execution times
saturate, and even grow, when a few hundred of processors are used, so
the number of multigrid levels in each process is then decreased.

Similarly, the ISF solver shows a limited efficiency; its nonlinear
speed-up region appears at a relatively small number of processes.
The main problem with the ISF solver is that the data-structures of
{\sc octopus} need to be transferred just before doing actual
calculations and after finishing them. These global communication
operations are done calling \texttt{MPI\_Gatherv} and
\texttt{MPI\_Scatterv} functions, which do not scale linearly with the
number of parallel processes.

The conjugate gradient and FMM methods seem to be very efficient
approaches in terms of scalability. However, both become faster than
ISF only when thousands of processes are used to simulate large
physical systems, beyond 1024 parallel processes. The CG solver, too,
has a limited accuracy and some convergence problems when it is used
for solving the Poisson equation within TDDFT calculations in {\sc
  octopus}.

Particularly attractive is the good performance obtained with the PFFT
solver. Beginning at a low number of processes, execution times are
always lower than those of the other solvers. In fact, 32 processes
are enough to hide the overhead added by the parallelisation, and
beyond this number of processes execution times become the best. This
is because PFFT's communication patterns are so highly effective.

These tests have shown that the new implementations of the Poisson
solvers, PFFT and FMM, offer good scalability and accuracy, and could
be used efficiently when hundreds or thousands of parallel processes
are needed. Almost linear performances can be observed until a
saturation point is reached for all cases. There, the obtained
efficiency factors~\cite{GR2012IJMPC} are always above 50\% before
those congestion points. As expected, large systems, which have higher
computation needs, can make better use of a high number of processes.

\begin{figure}
  \centering
  \def\svgwidth{400pt} 
  \begingroup%
  \makeatletter%
  \providecommand\color[2][]{%
    \errmessage{(Inkscape) Color is used for the text in Inkscape, but the package 'color.sty' is not loaded}%
    \renewcommand\color[2][]{}%
  }%
  \providecommand\transparent[1]{%
    \errmessage{(Inkscape) Transparency is used (non-zero) for the text in Inkscape, but the package 'transparent.sty' is not loaded}%
    \renewcommand\transparent[1]{}%
  }%
  \providecommand\rotatebox[2]{#2}%
  \ifx\svgwidth\undefined%
    \setlength{\unitlength}{654.24082031bp}%
    \ifx\svgscale\undefined%
      \relax%
    \else%
      \setlength{\unitlength}{\unitlength * \real{\svgscale}}%
    \fi%
  \else%
    \setlength{\unitlength}{\svgwidth}%
  \fi%
  \global\let\svgwidth\undefined%
  \global\let\svgscale\undefined%
  \makeatother%
  \begin{picture}(1,0.71175441)%
    \put(0,0){\includegraphics[width=\unitlength]{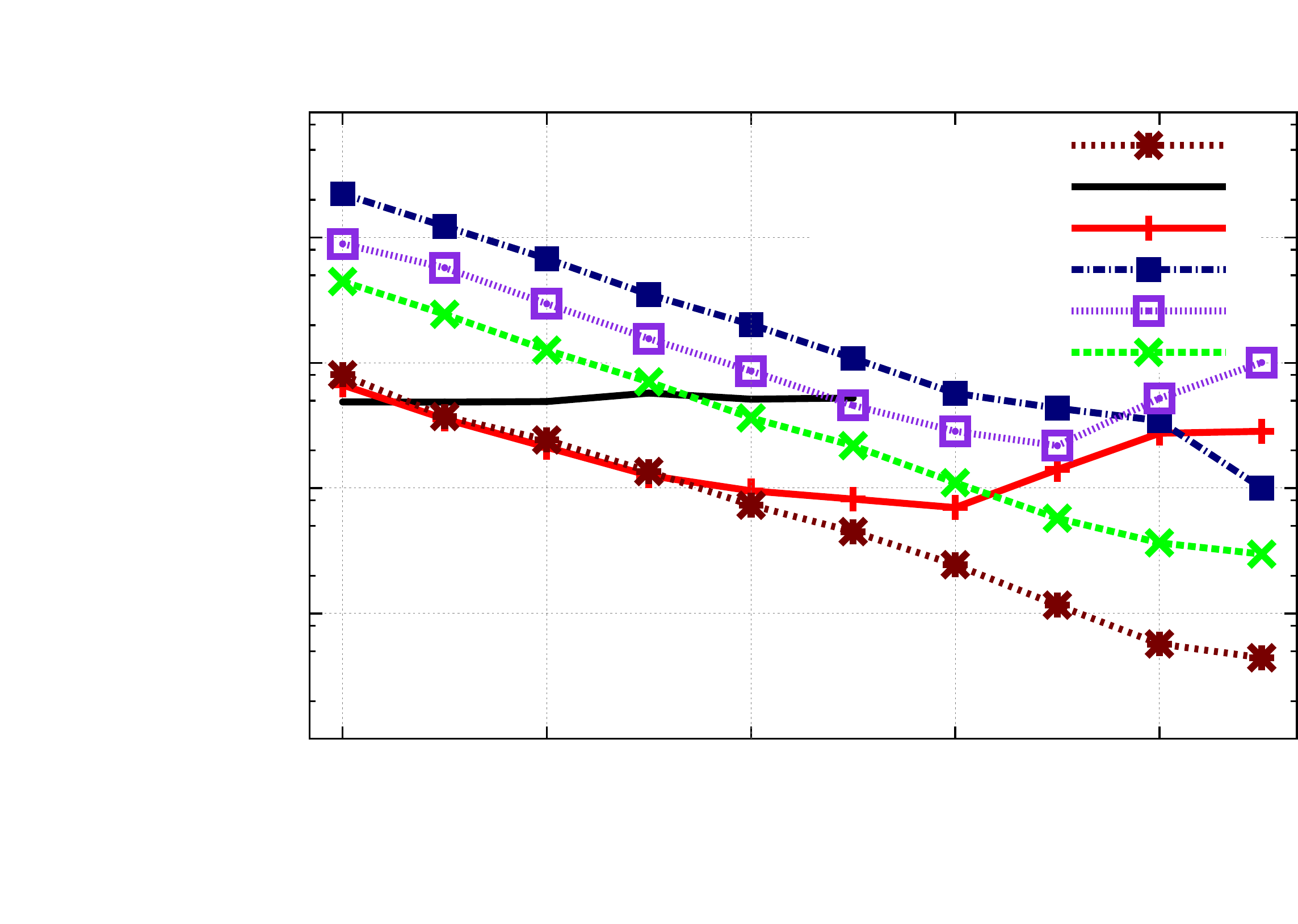}}%
    \put(0.2,0.12368224){\makebox(0,0)[rb]{\smash{0.01}}}%
    \put(0.2,0.22089415){\makebox(0,0)[rb]{\smash{0.1}}}%
    \put(0.2,0.31810606){\makebox(0,0)[rb]{\smash{1}}}%
    \put(0.2,0.41516512){\makebox(0,0)[rb]{\smash{10}}}%
    \put(0.2,0.51237703){\makebox(0,0)[rb]{\smash{100}}}%
    \put(0.2,0.60958894){\makebox(0,0)[rb]{\smash{1000}}}%
    \put(0.25276903,0.07782756){\makebox(0,0)[lb]{\smash{1}}}%
    \put(0.41112051,0.07782756){\makebox(0,0)[lb]{\smash{4}}}%
    \put(0.55687265,0.07782756){\makebox(0,0)[lb]{\smash{16}}}%
    \put(0.71522413,0.07782756){\makebox(0,0)[lb]{\smash{64}}}%
    \put(0.86082495,0.07782756){\makebox(0,0)[lb]{\smash{256}}}%
    \put(0.03338346,0.34237082){\rotatebox{90}{\makebox(0,0)[lb]{\smash{t (s)}}}}%
    \put(0.51824773,0.00904555){\makebox(0,0)[lb]{\smash{MPI proc.}}}%
    \put(0.20698161,0.67837095){\makebox(0,0)[lb]{\smash{\textbf A) Corvo (x86-64)}}}%
    \put(0.8,0.59104821){\makebox(0,0)[rb]{\smash{PFFT}}}%
    \put(0.8,0.5589501){\makebox(0,0)[rb]{\smash{Serial FFT}}}%
    \put(0.8,0.52685182){\makebox(0,0)[rb]{\smash{ISF}}}%
    \put(0.8,0.49475355){\makebox(0,0)[rb]{\smash{FMM}}}%
    \put(0.8,0.46265528){\makebox(0,0)[rb]{\smash{CG corrected}}}%
    \put(0.8,0.43055701){\makebox(0,0)[rb]{\smash{Multigrid}}}%
  \end{picture}%
  \endgroup%

  \def\svgwidth{400pt} 
  \begingroup%
  \makeatletter%
  \providecommand\color[2][]{%
    \errmessage{(Inkscape) Color is used for the text in Inkscape, but the package 'color.sty' is not loaded}%
    \renewcommand\color[2][]{}%
  }%
  \providecommand\transparent[1]{%
    \errmessage{(Inkscape) Transparency is used (non-zero) for the text in Inkscape, but the package 'transparent.sty' is not loaded}%
    \renewcommand\transparent[1]{}%
  }%
  \providecommand\rotatebox[2]{#2}%
  \ifx\svgwidth\undefined%
    \setlength{\unitlength}{654.24082031bp}%
    \ifx\svgscale\undefined%
      \relax%
    \else%
      \setlength{\unitlength}{\unitlength * \real{\svgscale}}%
    \fi%
  \else%
    \setlength{\unitlength}{\svgwidth}%
  \fi%
  \global\let\svgwidth\undefined%
  \global\let\svgscale\undefined%
  \makeatother%
  \begin{picture}(1,0.71175441)%
    \put(0,0){\includegraphics[width=\unitlength]{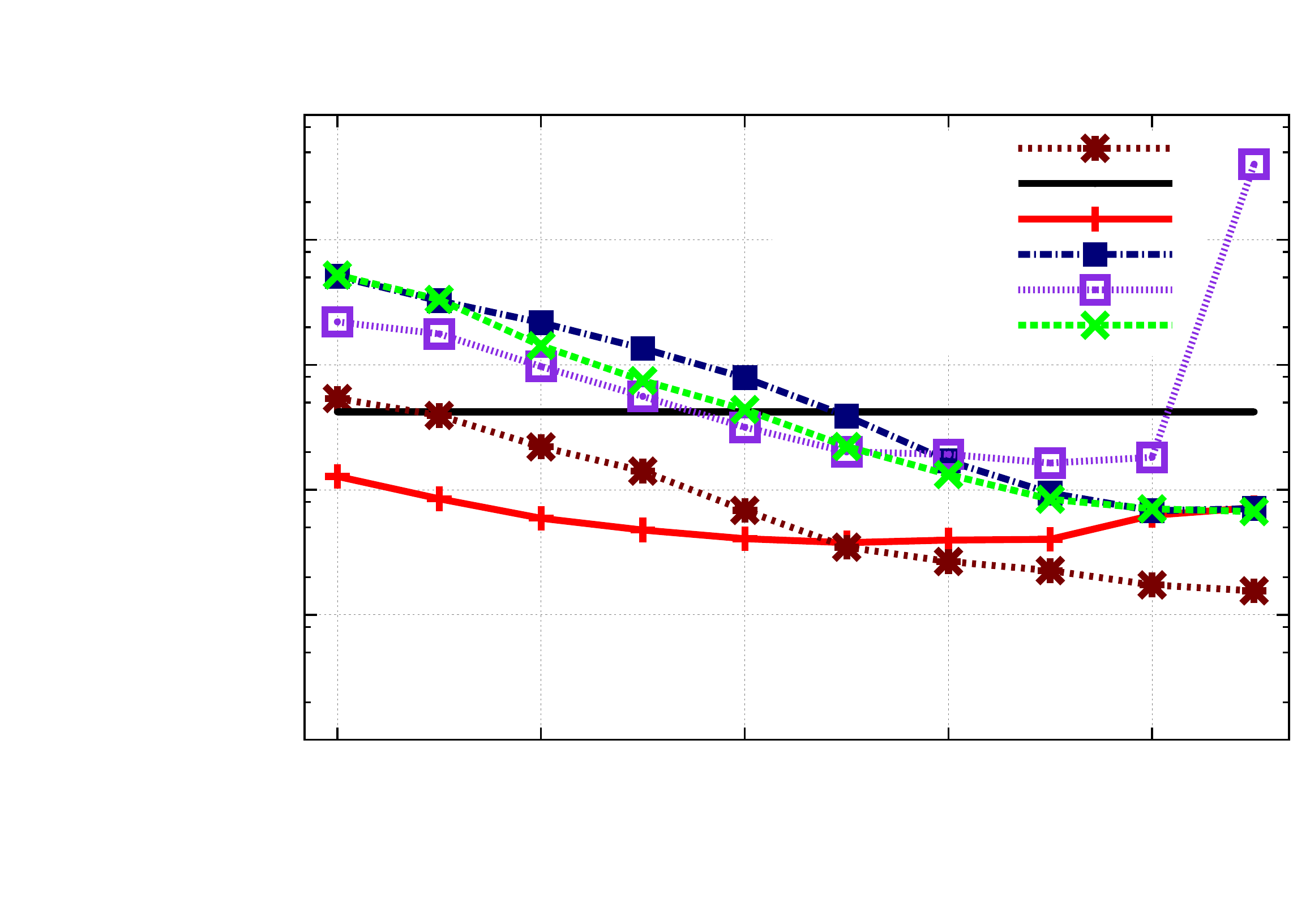}}%
    \put(0.2,0.12368224){\makebox(0,0)[rb]{\smash{0.01}}}%
    \put(0.2,0.22089415){\makebox(0,0)[rb]{\smash{0.1}}}%
    \put(0.2,0.31810606){\makebox(0,0)[rb]{\smash{1}}}%
    \put(0.2,0.41516512){\makebox(0,0)[rb]{\smash{10}}}%
    \put(0.2,0.51237703){\makebox(0,0)[rb]{\smash{100}}}%
    \put(0.2,0.60958894){\makebox(0,0)[rb]{\smash{1000}}}%
    \put(0.25276903,0.07782756){\makebox(0,0)[lb]{\smash{1}}}%
    \put(0.41112051,0.07782756){\makebox(0,0)[lb]{\smash{4}}}%
    \put(0.55687265,0.07782756){\makebox(0,0)[lb]{\smash{16}}}%
    \put(0.71522413,0.07782756){\makebox(0,0)[lb]{\smash{64}}}%
    \put(0.86082495,0.07782756){\makebox(0,0)[lb]{\smash{256}}}%
    \put(0.03338346,0.34237082){\rotatebox{90}{\makebox(0,0)[lb]{\smash{t (s)}}}}%
    \put(0.51824773,0.00904555){\makebox(0,0)[lb]{\smash{MPI proc.}}}%
    \put(0.20698161,0.67837095){\makebox(0,0)[lb]{\smash{\textbf B) Curie (x86-64)}}}%
    \put(0.77,0.59074252){\makebox(0,0)[rb]{\smash{PFFT}}}%
    \put(0.77,0.56322986){\makebox(0,0)[rb]{\smash{Serial FFT}}}%
    \put(0.77,0.53571706){\makebox(0,0)[rb]{\smash{ISF}}}%
    \put(0.77,0.50820426){\makebox(0,0)[rb]{\smash{FMM}}}%
    \put(0.77,0.48069145){\makebox(0,0)[rb]{\smash{CG corrected}}}%
    \put(0.77,0.45317865){\makebox(0,0)[rb]{\smash{Multigrid}}}%
  \end{picture}%
  \endgroup%

  \caption{Execution times of the Poisson solver of size $L_e=15.8$ in
    Corvo (A) and Curie (B) supercomputer for all the different
    solvers varying the number of MPI processes (each MPI process
    executed in a CPU processor core). }
  \label{fig:15_co_cu}
\end{figure}
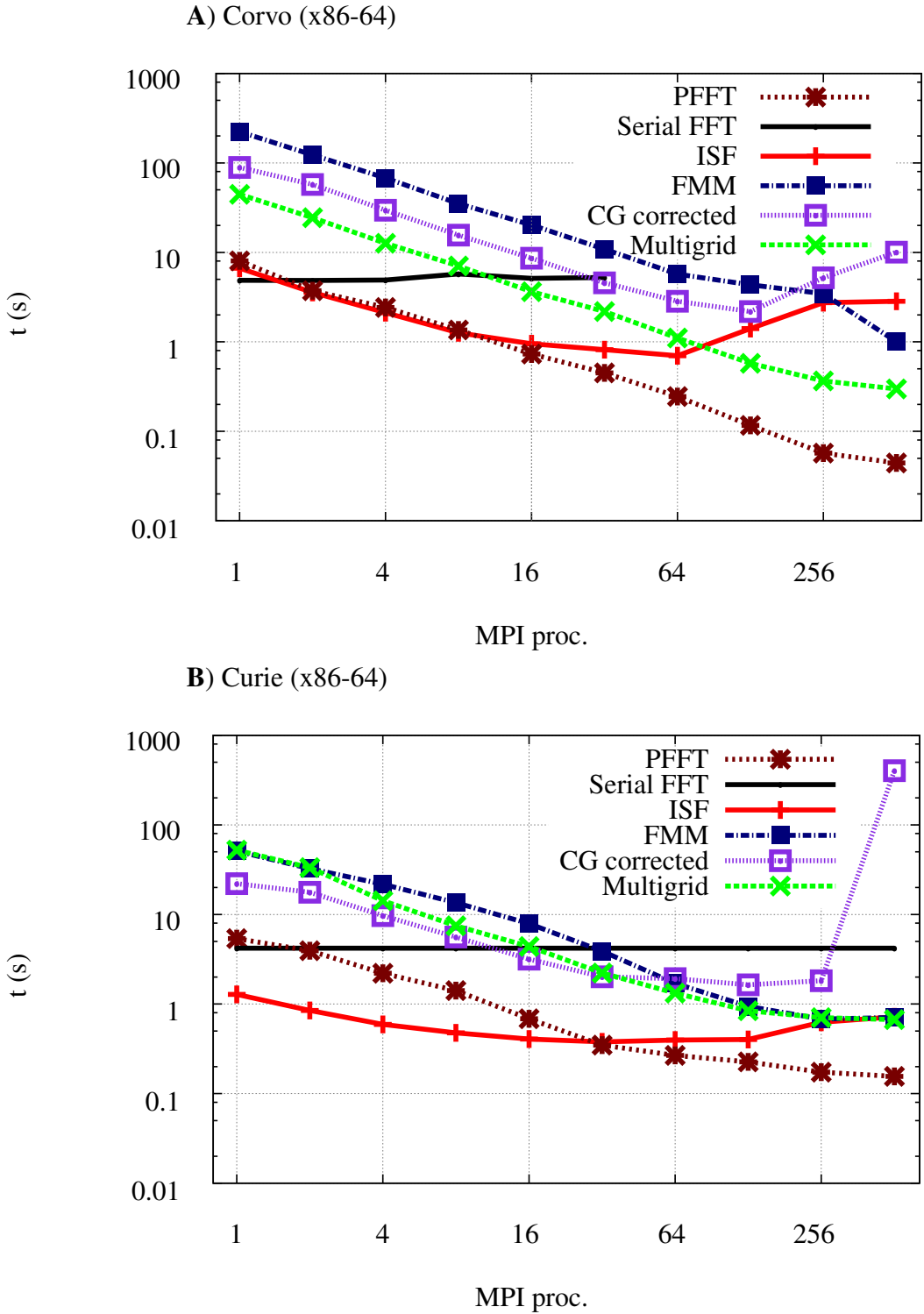

A similar overall behaviour of the different solvers, with small
differences, is also shown in the other two machines, Curie and
Corvo. \ref{fig:15_co_cu} presents the obtained results for a
particular case: $L_e=15.8$. As can be observed, the best results are
obtained again with the PFFT solver and a high number of parallel
processes. Relative performance between the solvers varies slightly
from the above analysed cases, but the conclusions are very similar.

For the serial case (only one processor), executions times are 4-5
times faster for Curie than for Corvo, probably because of the size of
the L3 cache, 24 MB in the processors of Curie (X7560 - Nehalem
processors) and only 12 MB in Corvo (E5645 - Westmere
processors). Similar performance has been reported for both processors
using standard benchmarks \footnote{ X7560:
  \url{http://browser.primatelabs.com/geekbench2/198947} \\ and E5645:
  \url{http://browser.primatelabs.com/geekbench2/389297}}, except for
the case of dot products (vector code), where Curie's processor
reaches 2.5 times more GFlop/s than Corvo's, which can also be
understood as a consequence of the L3 cache differences.

At any rate, these differences tend to disappear when using a high
number of processors, in part because the size of the problem assigned
to each processor is smaller (and so the influence of the cache size
is also lower), and in part because communication among processors, not
only data processing, must also be considered.

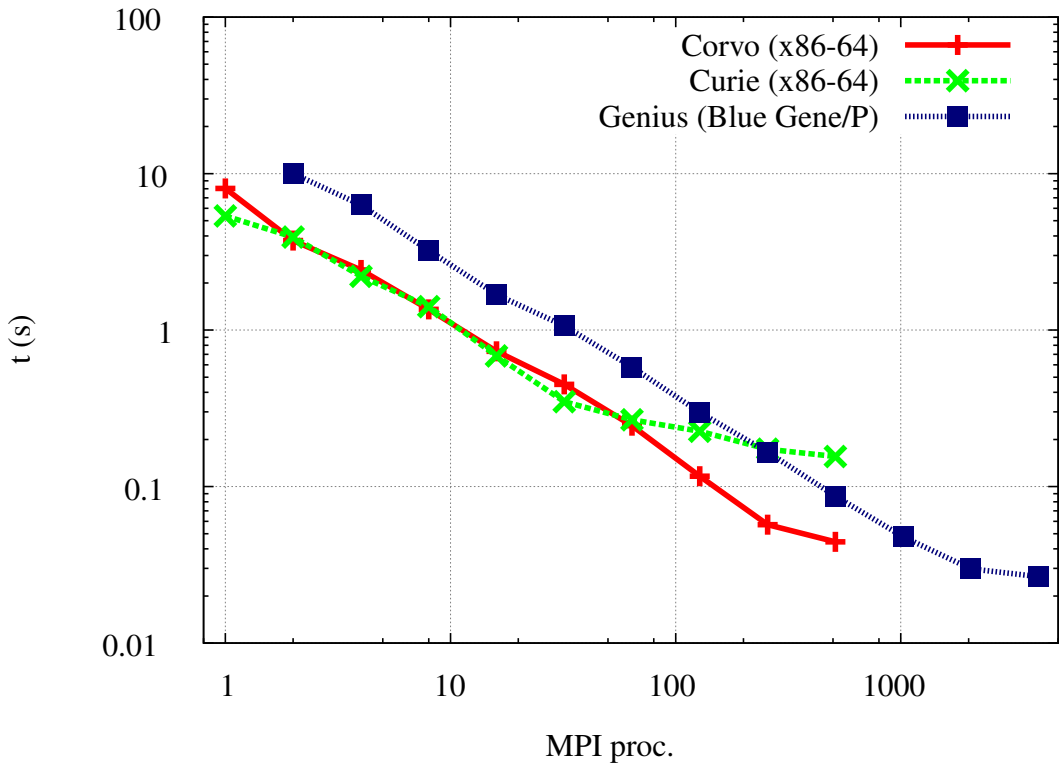
\begin{figure}
  \centering
  \def\svgwidth{400pt} 
  \begingroup%
  \makeatletter%
  \providecommand\color[2][]{%
    \errmessage{(Inkscape) Color is used for the text in Inkscape, but the package 'color.sty' is not loaded}%
    \renewcommand\color[2][]{}%
  }%
  \providecommand\transparent[1]{%
    \errmessage{(Inkscape) Transparency is used (non-zero) for the text in Inkscape, but the package 'transparent.sty' is not loaded}%
    \renewcommand\transparent[1]{}%
  }%
  \providecommand\rotatebox[2]{#2}%
  \ifx\svgwidth\undefined%
    \setlength{\unitlength}{665.27058359bp}%
    \ifx\svgscale\undefined%
      \relax%
    \else%
      \setlength{\unitlength}{\unitlength * \real{\svgscale}}%
    \fi%
  \else%
    \setlength{\unitlength}{\svgwidth}%
  \fi%
  \global\let\svgwidth\undefined%
  \global\let\svgscale\undefined%
  \makeatother%
  \begin{picture}(1,0.7208078)%
    \put(0,0){\includegraphics[width=\unitlength]{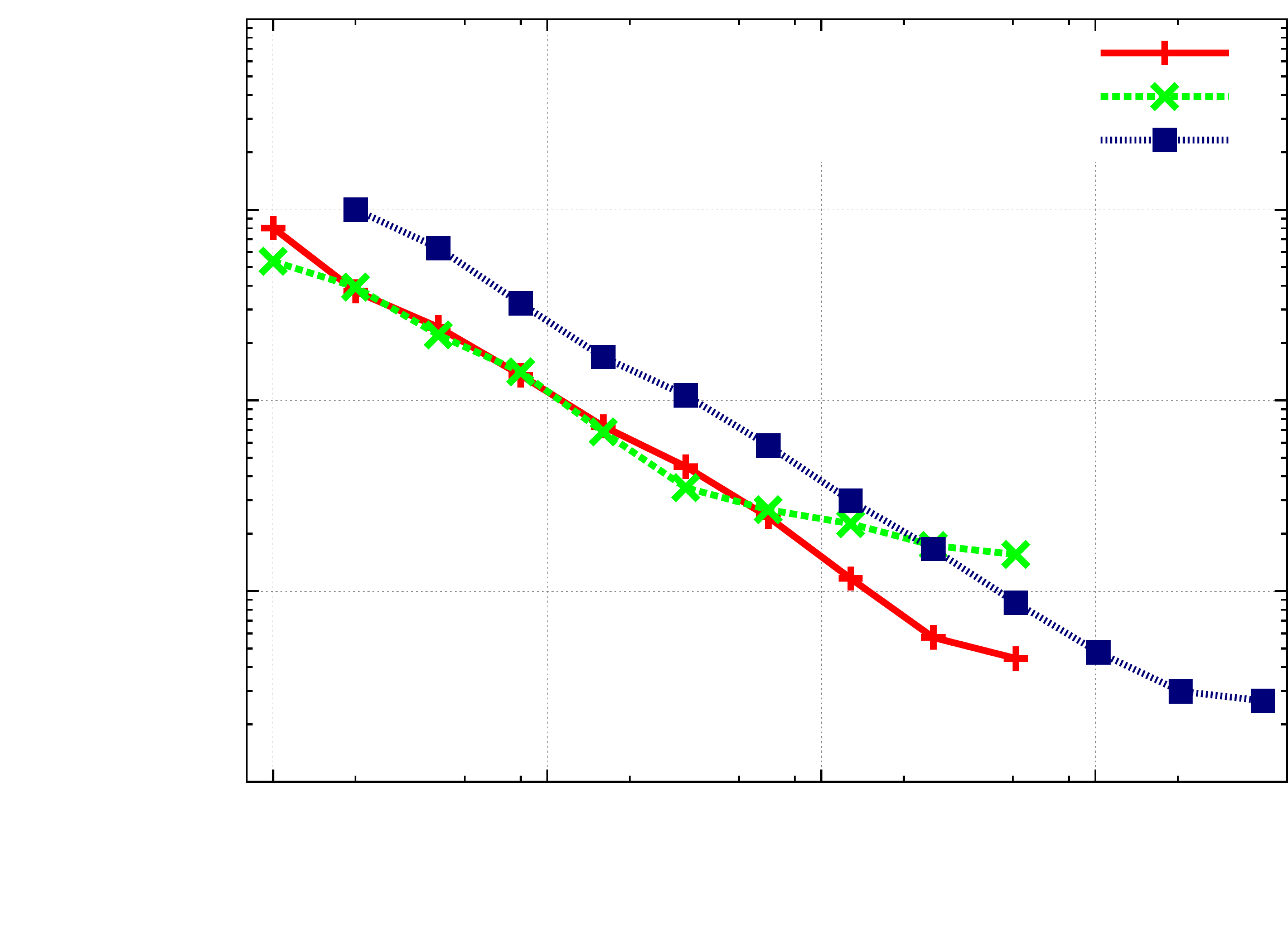}}%
    \put(0.09577845,0.10135972){\makebox(0,0)[lb]{\smash{0.01}}}%
    \put(0.11667821,0.24941978){\makebox(0,0)[lb]{\smash{0.1}}}%
    \put(0.14801734,0.39747983){\makebox(0,0)[lb]{\smash{1}}}%
    \put(0.12711758,0.54538957){\makebox(0,0)[lb]{\smash{10}}}%
    \put(0.10621931,0.69344948){\makebox(0,0)[lb]{\smash{100}}}%
    \put(0.2068265,0.06378103){\makebox(0,0)[lb]{\smash{1}}}%
    \put(0.40922384,0.06378103){\makebox(0,0)[lb]{\smash{10}}}%
    \put(0.61161968,0.06378103){\makebox(0,0)[lb]{\smash{100}}}%
    \put(0.81386521,0.06378103){\makebox(0,0)[lb]{\smash{1000}}}%
    \put(0.02735832,0.3775015){\rotatebox{90}{\makebox(0,0)[lb]{\smash{t (s)}}}}%
    \put(0.51495826,0.00741298){\makebox(0,0)[lb]{\smash{MPI proc.}}}%
    \put(0.83,0.66963977){\makebox(0,0)[rb]{\smash{Corvo (x86-64)}}}%
    \put(0.83,0.63581895){\makebox(0,0)[rb]{\smash{Curie (x86-64)}}}%
    \put(0.83,0.60199812){\makebox(0,0)[rb]{\smash{Genius (Blue Gene/P)}}}%
  \end{picture}%
  \endgroup%

  \caption{Execution times of the PFFT solver in Genius (Blue Gene/P),
    Curie and Corvo (x86-64) for a system size of $L_e=15.8$ as a
    function of the number of MPI processes.}
  \label{fig:15_pfft_ma}
\end{figure}

In order to compare more clearly the results obtained in the three
computers, \ref{fig:15_pfft_ma} shows the execution times of the PFFT
solver for the case of $L_e=15.8$ in Blue Gene, Curie and
Corvo. Execution times are higher in the Blue Gene/P computers for the
same number of processes, but scalability is much better, allowing the
efficient use of a much higher number of processes. The processors of
the Blue Gene/P machine are slower than those of Curie and Corvo, but
communication infrastructures are much better (see section
\ref{sec:hpc}). This leads us to conclude that the execution time of
the Poisson solver parallel code is closely related to the performance
of the interprocessor communication system of the parallel computer,
more so than on the performance of the processors themselves.

As stated in section \ref{sec:theory}, the numerical complexities of
PFFT and FMM scale as $\mathcal{O}(N \log N)$ and $\mathcal{O}(N)$
respectively, with $N$ the total number of points required. We have
analysed the execution order by measuring the execution time on Blue
Gene/Ps as a function of the volume of the system for PFFT and FMM in
three cases: 16, 32 and 64 parallel processes. The volume ratio among
the smallest ($L_e=7$) and the largest ($L_e=25.9$) studied systems is
about 49.  Our results agree with the $\mathcal{O}(N)$ complexity of 
the FMM method and the $\mathcal{O}(N\log N)$ complexity of the PFFT
method.  Note that computation time is appreciably higher when using
the FMM solver due to its higher prefactor (the quantity that
multiplies $N$ or $N\log N$ in the expression of the numerical
complexity). The quotient between prefactors is machine dependent; its
value is about 5.5 for Corvo, about 9 for Curie and about 12 for Blue
Gene/P. One may think that for very big values of $N$ at a constant
number of processes $P$, the growth of the term $\log N$ would
eventually make FMM more efficient than PFFT. This is strictly true,
but the huge $N$ of this eventual crossover precludes it in
practice. For example, our results using 16
cores indicate that this crossover would take place at $N>10^{53}$ in Blue Gene/P,
and at $N \simeq 4.5\cdot 10^{41}$ in Corvo.

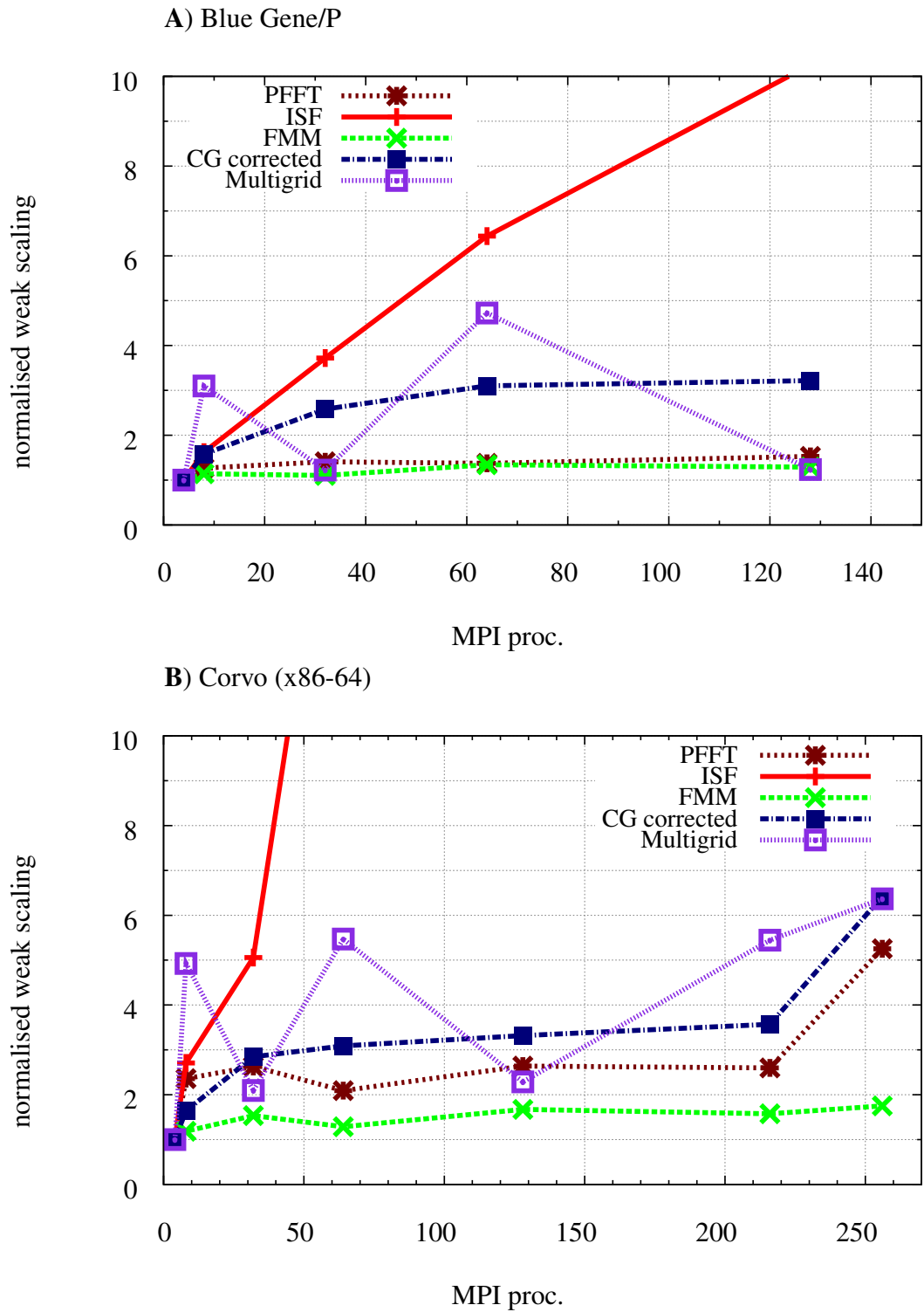
\begin{figure}
  \centering
  \def\svgwidth{400pt} 
  \begingroup%
  \makeatletter%
  \providecommand\color[2][]{%
    \errmessage{(Inkscape) Color is used for the text in Inkscape, but the package 'color.sty' is not loaded}%
    \renewcommand\color[2][]{}%
  }%
  \providecommand\transparent[1]{%
    \errmessage{(Inkscape) Transparency is used (non-zero) for the text in Inkscape, but the package 'transparent.sty' is not loaded}%
    \renewcommand\transparent[1]{}%
  }%
  \providecommand\rotatebox[2]{#2}%
  \ifx\svgwidth\undefined%
    \setlength{\unitlength}{653.87460938bp}%
    \ifx\svgscale\undefined%
      \relax%
    \else%
      \setlength{\unitlength}{\unitlength * \real{\svgscale}}%
    \fi%
  \else%
    \setlength{\unitlength}{\svgwidth}%
  \fi%
  \global\let\svgwidth\undefined%
  \global\let\svgscale\undefined%
  \makeatother%
  \begin{picture}(1,0.71215304)%
    \put(0,0){\includegraphics[width=\unitlength]{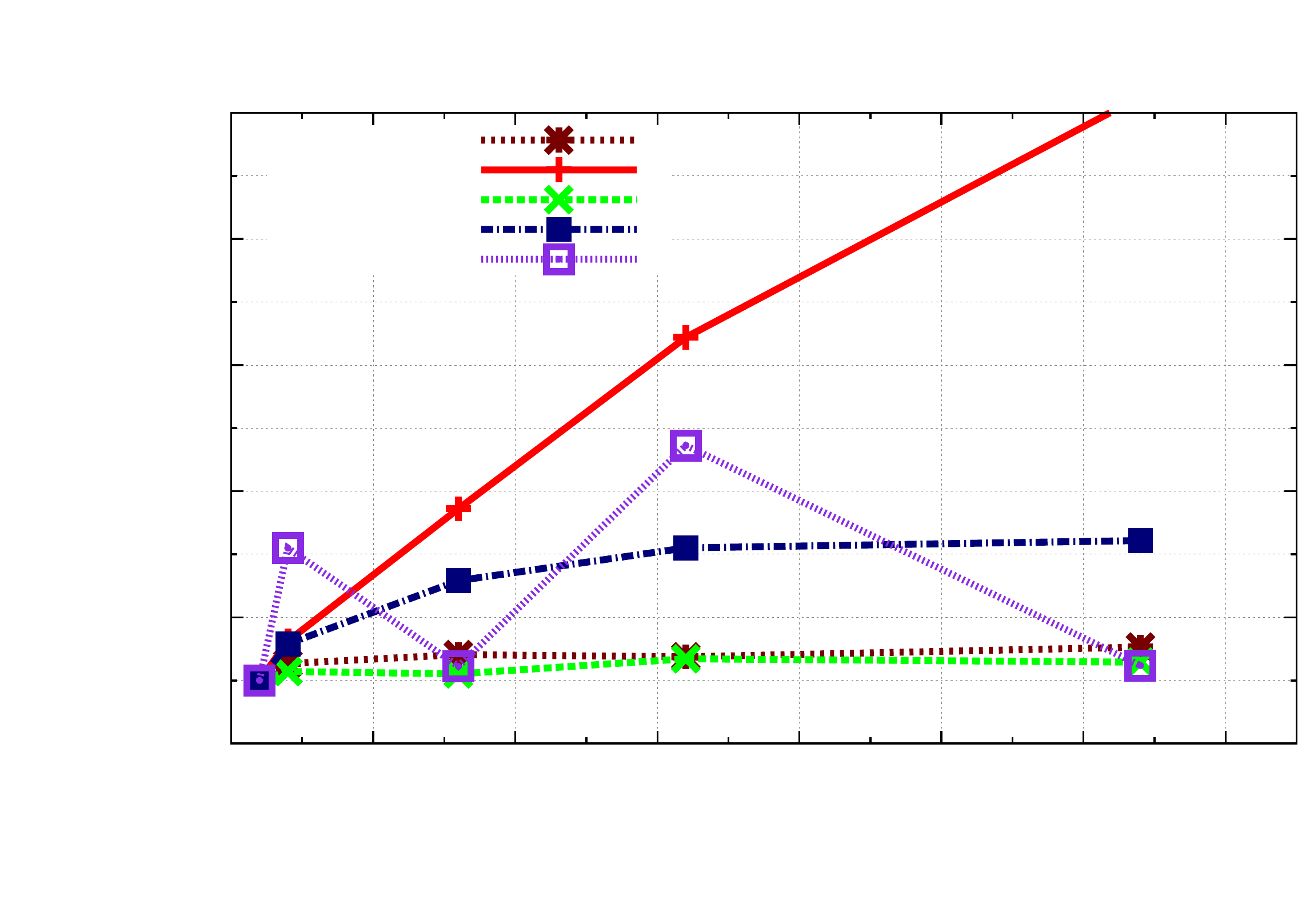}}%
    \put(0.15,0.12375151){\makebox(0,0)[rb]{\smash{0}}}%
    \put(0.15,0.22101786){\makebox(0,0)[rb]{\smash{2}}}%
    \put(0.15,.31828422){\makebox(0,0)[rb]{\smash{4}}}%
    \put(0.15,0.41539764){\makebox(0,0)[rb]{\smash{6}}}%
    \put(0.15,0.51266399){\makebox(0,0)[rb]{\smash{8}}}%
    \put(0.15,0.60993035){\makebox(0,0)[rb]{\smash{10}}}%
    \put(0.17175405,0.07787115){\makebox(0,0)[lb]{\smash{0}}}%
    \put(0.26849584,0.07787115){\makebox(0,0)[lb]{\smash{20}}}%
    \put(0.37799695,0.07787115){\makebox(0,0)[lb]{\smash{40}}}%
    \put(0.48749807,0.07787115){\makebox(0,0)[lb]{\smash{60}}}%
    \put(0.59684625,0.07787115){\makebox(0,0)[lb]{\smash{80}}}%
    \put(0.69358957,0.07787115){\makebox(0,0)[lb]{\smash{100}}}%
    \put(0.80309069,0.07787115){\makebox(0,0)[lb]{\smash{120}}}%
    \put(0.9125918,0.07787115){\makebox(0,0)[lb]{\smash{140}}}%
    \put(0.03284209,0.2){\rotatebox{90}{\makebox(0,0)[lb]{\smash{normalised weak scaling}}}}%
    \put(0.49044971,0.00905062){\makebox(0,0)[lb]{\smash{MPI proc.}}}%
    \put(0.17900926,0.67875088){\makebox(0,0)[lb]{\smash{\textbf A) Blue Gene/P}}}%
    \put(0.35,0.59596727){\makebox(0,0)[rb]{\smash{\small{PFFT}}}}%
    \put(0.35,0.57302725){\makebox(0,0)[rb]{\smash{\small ISF}}}%
    \put(0.35,0.55008707){\makebox(0,0)[rb]{\smash{\small FMM}}}%
    \put(0.35,0.52714689){\makebox(0,0)[rb]{\smash{\small{CG corrected}}}}%
    \put(0.35,0.50420671){\makebox(0,0)[rb]{\smash{\small Multigrid}}}%
  \end{picture}%
  \endgroup%
  \def\svgwidth{400pt} 
  \begingroup%
  \makeatletter%
  \providecommand\color[2][]{%
    \errmessage{(Inkscape) Color is used for the text in Inkscape, but the package 'color.sty' is not loaded}%
    \renewcommand\color[2][]{}%
  }%
  \providecommand\transparent[1]{%
    \errmessage{(Inkscape) Transparency is used (non-zero) for the text in Inkscape, but the package 'transparent.sty' is not loaded}%
    \renewcommand\transparent[1]{}%
  }%
  \providecommand\rotatebox[2]{#2}%
  \ifx\svgwidth\undefined%
    \setlength{\unitlength}{653.87460938bp}%
    \ifx\svgscale\undefined%
      \relax%
    \else%
      \setlength{\unitlength}{\unitlength * \real{\svgscale}}%
    \fi%
  \else%
    \setlength{\unitlength}{\svgwidth}%
  \fi%
  \global\let\svgwidth\undefined%
  \global\let\svgscale\undefined%
  \makeatother%
  \begin{picture}(1,0.71215304)%
    \put(0,0){\includegraphics[width=\unitlength]{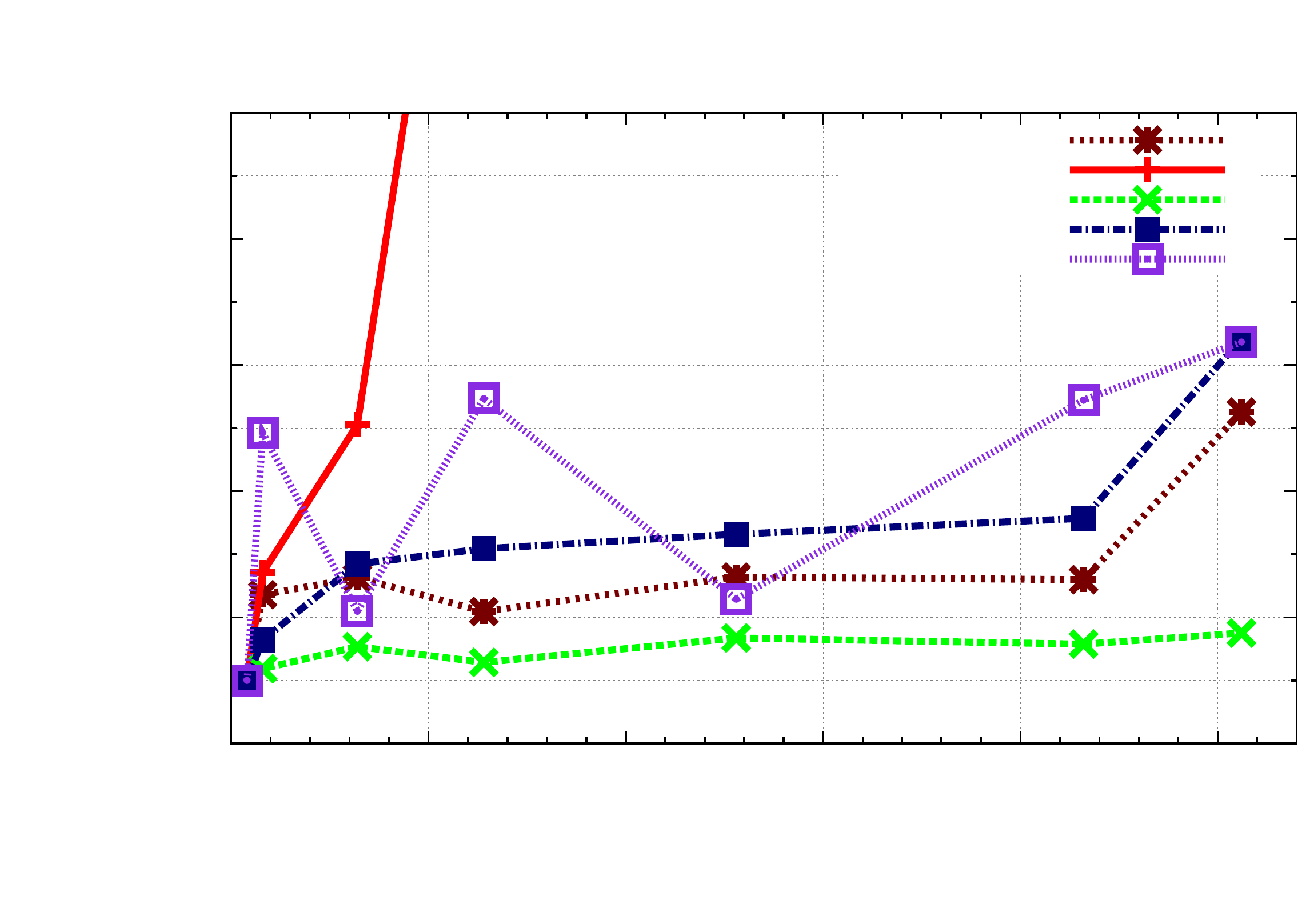}}%
    \put(0.15,0.12375151){\makebox(0,0)[rb]{\smash{0}}}%
    \put(0.15,0.22101786){\makebox(0,0)[rb]{\smash{2}}}%
    \put(0.15,.31828422){\makebox(0,0)[rb]{\smash{4}}}%
    \put(0.15,0.41539764){\makebox(0,0)[rb]{\smash{6}}}%
    \put(0.15,0.51266399){\makebox(0,0)[rb]{\smash{8}}}%
    \put(0.15,0.60993035){\makebox(0,0)[rb]{\smash{10}}}%
    \put(0.17175405,0.07787115){\makebox(0,0)[lb]{\smash{0}}}%
    \put(0.31101163,0.07787115){\makebox(0,0)[lb]{\smash{50}}}%
    \put(0.45042368,0.07787115){\makebox(0,0)[lb]{\smash{100}}}%
    \put(0.6024406,0.07787115){\makebox(0,0)[lb]{\smash{150}}}%
    \put(0.75445751,0.07787115){\makebox(0,0)[lb]{\smash{200}}}%
    \put(0.90647442,0.07787115){\makebox(0,0)[lb]{\smash{250}}}%
    \put(0.03284209,0.2){\rotatebox{90}{\makebox(0,0)[lb]{\smash{normalised weak scaling}}}}%
    \put(0.49044971,0.00905062){\makebox(0,0)[lb]{\smash{MPI proc.}}}%
    \put(0.17900926,0.67875088){\makebox(0,0)[lb]{\smash{\textbf B) Corvo (x86-64)}}}%
    \put(0.8,0.59596727){\makebox(0,0)[rb]{\smash{\small PFFT}}}%
    \put(0.8,0.57302725){\makebox(0,0)[rb]{\smash{\small ISF}}}%
    \put(0.8,0.55008707){\makebox(0,0)[rb]{\smash{\small FMM}}}%
    \put(0.8,0.52714689){\makebox(0,0)[rb]{\smash{\small CG corrected}}}%
    \put(0.8,0.50420671){\makebox(0,0)[rb]{\smash{\small Multigrid}}}%
  \end{picture}%
  \endgroup%

  \caption{Normalised weak-scaling of the measured Poisson solvers in
    Blue Gene/P (A) and Corvo (B). Each MPI process has roughly
    125,000 points and selected system sizes are given by $L_e$ equal to
    7.9, 10, 15.8, 20.1, 25.1 and 31.7 (the latter only in Corvo)}.
  \label{fig:co_wk}
\end{figure}

Apart from the speed-up and execution time of a concrete parallel
execution, another very useful quality measure is the weak-scaling.
Weak-scaling measures the ability of an algorithm to scale with the
number of parallel processes at a fixed computing load, comparing the
execution time of a problem of size $N$ using $P$ processes with the
execution time of the same problem but of size $kN$ using $kP$
processes. In this way, each process will use the same amount of
computational resources, regardless of the number of processors
used. If the algorithm is mainly constricted by computational needs,
then execution times should be very similar in both cases. On the
other hand, if communication needs dominates the parallel execution,
then execution times will grow with $P$, usually faster than
linearly. For these tests we have adapted the system sizes of the
parallelepiped meshes to 7.9, 10, 15.8, 20.1 and 25.1 (also 31.7 in
Corvo) for the Poisson solver on Blue Gene/P and Corvo. Since the
number of grid points increases like the cube of $L_e$, we did
parallel runs using 4, 8, 32, 64 and 128 MPI processes (256 processes
in Corvo). So, the number of grid points processed in each parallel
MPI process is roughly 125,000. \ref{fig:co_wk} shows the obtained
results, with data taken from the profiling output.

On one hand, the weak-scaling factor of the ISF solver increases
sharply with the number of parallel processes, certainly because of
the communication needs of the method, and confirming the conclusions
we have obtained previously. The conjugate gradient and multigrid
methods show an important increase in the weak-scaling factor with the
number of processes, and this can be used to predict that parallel
execution will ``saturate'' at a moderate number of processes (as we
had observed in \ref{fig:runs_bg}). In contrast, FMM shows the best
results in the analysed range: weak-scaling increases very moderately,
so we can conclude that communication overheads are quite acceptable
and that the method will offer good speed-up results when using
thousands of processes.

PFFT also gives very good results in the Blue Gene/P system. They are
similar to those of FMM, so similar conclusions can be
stated. Nonetheless, the results obtained in Corvo up to 256 processes
seem to indicate that communication will play an important role if a
high number of processes are to be used. In these cases, the
efficiency of the communication network and protocols of the parallel
computer will play an important role in the obtained parallel
performance. This can already be observed in \ref{fig:co_wk}: the
weak-scaling factor is markedly better and bounded in Blue Gene/P, a
system with a very high performance communication network. Also
limited to the Corvo system, a sharp increment in the weak-scaling
factor is observed when passing from 216 (within only one Infiniband
switch, $12 \times 18 = 216$ cores) to 256 cores, that needs
communication with processes in a second switch.

From all our tests, we conclude that the communication needs of the
Poisson solvers fit better in the network of a Blue Gene/P machine
than in that of a machine with x86-64 processors and Infiniband
network (e.g. Corvo).

\subsection{Influence of the input parameters}\label{sec:inflinppar}
In this section we discuss several input parameters that we used in
our tests of Poisson solvers. These parameters are
the \texttt{BoxShape} (i.e., the spatial structure of the points where
the Hartree potential was calculated), the \texttt{DeltaEFMM}
(accuracy tolerance of the energy calculated with
FMM), \texttt{MultigridLevels} (number of stages in the grid hierarchy
of the multigrid solver) and \texttt{PoissonSolverMaxMultipole} (the
order of the multipole expansion of the charges whose potential is
analytically calculated when using multigrid or conjugate gradient
solvers).

{\sc octopus} is able to handle different grid shapes, like spherical,
parallelepiped or \emph{minimum} shapes (the \emph{minimum} mesh shape
is the addition of spheres centred in each nucleus of the test
system).  Apart from the mesh shape and size, the grid is defined by
its spacing parameter, i.e., the distance between consecutive grid
points.  The
\emph{minimum} mesh shape option in {\sc octopus} produces
fair results with FMM. Serial FFT, PFFT, and ISF (which is based on
FFT) solvers create a parallelepiped mesh ---which contains the
original minimum one--- to calculate the Hartree potential. If
multigrid or conjugate gradients are used with a minimum mesh, it is
frequent to find a serious loss of accuracy. This is because minimal
boxes are usually irregular, and the multipole expansion
(whose terms are based on spherical harmonics, which are rather
smooth) cannot adapt well to arbitrary charge values in irregular
meshes\footnote{See the explanation of PoissonSolverMaxMultipole below
for more information.}.  These effects of box shape made us
choose cubic meshes for our tests. In any case, one should take into
account that if non-parallelepiped meshes are used, FMM suffers
less overhead than the solvers based on reciprocal space (FFT, PFFT and
ISF), since FMM can work directly with that data-representation and
does not need to transfer data to a parallelepiped
representation. Examples of this can be viewed in
\ref{tab:boxshapes}. There, it can be seen that the execution
time is essentially the same regardless of the box shape for PFFT,
while spherical and minimal box shapes are much more efficient than
parallelepiped shape for FMM (with a ratio of about 1.67).

Spherical meshes are still valid for multigrid and conjugate gradient
methods.  We have compared the relative accuracy of cubic and
spherical meshes for FMM, multigrid, and CG methods containing the
same number of points.  When FMM is used, both errors
decrease while the mesh size is increased, being relatively equal for
both representations, until an accuracy plateau is reached for big
meshes. The same behaviour is shown for the multigrid solver and by
the CG solver up to a given volume for a given system.

\begin{table}[htbp]
  \begin{center}
    \begin{tabular}{cccccccc}
       & \multicolumn{3}{c}{PFFT}  &   \multicolumn{3}{c}{FMM}  \\
       \cline{2-4}\cline{6-8}
$R || L_e (\angstrom)$ & Minimal	    & Sphere & 	Parallelepiped	& &  Minimal	 &  Sphere	 & Parallelepiped	\\  
 \hline
15.8 & 1.001  & 	1.032 & 	1.030 & &	3.371 & 	3.5413 & 	5.945	\\  
22.1 & 	2.535 & 	2.6641 & 	2.667 & &	10.500 & 	10.447 & 	16.922    \\  
25.9 & 	4.321 & 	4.265 & 	4.393 & &	16.321 & 	16.212 & 	27.861	\\  
31.7 & 	8.239 & 	8.034 & 	 8.271 && 	27.821 & 	28.707 & 	48.206	\\  
    \end{tabular}
  \end{center}
  \caption{Comparison of total times required for the calculation of the Hartree potential as a function of the box shape and radius ($R$) or
semi-edge length ($L_e$) for PFFT and FMM solvers. }
  \label{tab:boxshapes}
 \end{table}

 The implementations of multigrid and conjugate gradient solvers we
 used in our tests are based on a multipole expansion. Each value of
 the input charge density represented on a grid ($\rho_{ijk}$) is
 expressed with an analytic term via this multipole expansion, plus a
 numeric term. The Hartree potential created by the analytic part is
 calculated analytically, while the Hartree potential created by the
 numerical term is calculated numerically with either multigrid or
 conjugate gradient methods.  Since the use of the analytic term makes
 the numerical term smaller, numerical errors are expected to be
 reduced. Therefore, the smaller the difference between the charge
 density and its multipole expansion, the smaller the potential
 numerically calculated (with multigrid or conjugate gradient
 solvers), and the higher the accuracy of the total Hartree potential
 calculation.  An order for the multipole expansion (the input
 parameter \texttt{PoissonSolverMaxMultipole}, PSMM) must be chosen.
 It is to be stressed that arbitrarily higher values of PSMM do not
 lead to more accurate results; rather, there exists a PSMM that
 minimises the error for each problem.  We ran some tests to obtain
 this optimal value.  In them, we calculated the ground-state of a
 180-atom chlorophyll system using {\sc octopus}, varying the value of
 PSMM. This molecule's size is expected to be large enough to be
 representative of a wide range of molecular sizes. The ground state
 calculations permit one to evaluate the final Hartree energy, as well
 as the HOMO-LUMO gap, which is defined as the difference between the
 highest occupied eigenvalue and the lowest unoccupied eigenvalue in a
 density functional theory calculation, and its value can be used for
 estimations of simulation accuracy \cite{Wan2011GFP}. \ref{tab:psmm}
 shows the obtained results. For $PSMM= 8, 9,$ and 10, the
 accumulation of errors was big enough for calculations not to
 converge (this is because beyond a given order, the spherical
 harmonics are too steep to describe the $\rho$, which is rather
 smooth).  From the data of \ref{tab:psmm}, we chose $PSSM=7$ for our
 simulations (with $PSSM=7$ the HOMO-LUMO gap is equivalent to the
 reference value given by ISF, and the Hartree energy is the closest
 one).

\begin{table}[htbp]
  \begin{center}
    \begin{tabular}{cccccccc}
       & \multicolumn{3}{c}{Hartree energy (eV)}  & &  \multicolumn{3}{c}{HOMO-LUMO gap}  \\
       \cline{2-4}\cline{6-8}
      PSMM & ISF       & CG        & Multigrid & & ISF     & CG     & Multigrid \\ 
      \hline
      4    & 240821.47 & 240813.51 & 240813.41 & & 1.4489  & 1.2179 & 1.2178    \\ 
      6    & 240821.47 & 240813.93 & 240813.95 & & 1.4489  & 1.4340 & 1.4353    \\
      7    & 240821.47 & 240815.01 & 240814.69 & & 1.4489  & 1.4520 & 1.4506    \\ 
    \end{tabular}
  \end{center}
  \caption{Ground state values of the Hartree energy and the HOMO-LUMO
    gap as a function of the PSMM (input parameter of Multigrid and
    Conjugate gradients solvers). Reference values are given by the ISF
    solver.}
  \label{tab:psmm}
 \end{table}

The implementation we used for FMM \cite{Dac2010JCP} allows one to tune
the relative error of the calculations.  Its expression is the
quotient $(E_{ref}-E_n)/E_n$, i.e. the variable \texttt{DeltaEFMM},
where $E_n$ is the Hartree energy calculated with the FMM method and
$E_{ref}$ is an estimation of what its actual value is.  We chose for
our calculations a relative error of $10^{-4}$. Note that this error
corresponds only to the pairwise term of the Hartree potential, before
the correction for charge distribution (see section \ref{sec:fmm} and
supporting info) is applied.

\section{Conclusions}\label{sec:conclusions}

In this paper we analysed the relative performance of several popular
methods (parallel fast Fourier transform, ISF, fast multipole method,
conjugate gradients and multigrid) as implemented in the {\sc Octopus}
code for the calculation of the classical Hartree potential created by
a charge distribution.  The first part of the paper presents the
fundamentals of these calculation methods. In the second part, we
summarise the computational aspects of the tests we carried out to
measure the methods' relative performances.  These tests were run on
three kinds of supercomputers, which we chose as representative of
present-day high performance architectures. In the tests, we focused
on measuring accuracies, execution times, speed-ups and weak-scalings.
Test runs involved up to 4,096 parallel processes (each containing 4
OpenMP threads), and solved system sizes from about 350,000 grid
points to about 32,000,000 grid points.

Our results (section \ref{sec:results}) show that PFFT is the most
efficient option when a high number of parallel processes are to be
used.  The current implementation of PFFT is very efficient, and
should be the default option to study large physical systems using a
number of cores beyond a certain threshold (when PFFT becomes faster
than ISF). The fact that the charges are located at equispaced points
makes FFT-based methods suitable for our problem.  In special cases
when charges are not lying in equispaced points (e.g.\ when
curvilinear coordinates are used), FMM should be chosen instead, since
it works and is accurate regardless of the charge density's spatial
location.  The FMM solver also shows good performance and scaling, yet
its execution times are greater and its accuracy lower than those of
the PFFT solver on the analysed machines.  In contrast to ISF, FMM
scales almost linearly up to high values of the number of processes,
but since its numerical complexity has a larger prefactor, it would be
competitive with ISF when the number of parallel processes increases
significantly, a scenario in which the performance of ISF
``collapses''.  The performance of the conjugate gradients solver has
a trend similar to that of FMM, as does multigrid for low values of
the number of processes.  Weak-scaling tests show that communication
overheads are the smallest for the FMM solver, while they are
acceptable for PFFT, CG, and multigrid solvers, and they only
significantly increases in the case of ISF, as we might expect from
the data of \ref{fig:runs_bg}.

FFT-based methods (PFFT, serial FFT, and ISF) are more accurate than
FMM, conjugate gradients, and multigrid. Hence, they should be chosen
if accurate calculations of electrostatics are required.  However,
according to our tests, the accuracy of all the analysed solvers is
expected to be appropriate for density functional theory and
time-dependent density functional theory calculations (where the
calculation of the Hartree potential is an essential step) because
these have other sources of error that will typically have a much
stronger impact (section \ref{sec:rad}).  Nevertheless, neither
multigrid nor conjugate gradients can reach acceptable accuracy if the
data set is not represented on a compact (spherical or parallelepiped)
grid.

The competitive features of PFFT and FMM solvers make them suitable to
perform calculations involving hundreds or thousands of processors to
obtain electronic properties of large physical systems.  ISF should be
chosen only whenever a low number of parallel processes is to be used.

In the future, we plan to improve the accuracy and efficiency of the
Poisson solvers implemented in {\sc Octopus} (keeping their
suitability for both periodic boundaries and open systems) by taking
advantage in the sparsity of the input data
\cite{Hassanieh:2012:NOS:2213977.2214029} and by including screened
interactions (in the spirit of \cite{screenedpoisson}).

\acknowledgement We would like to express our gratitude to
Jos\'e Luis Alonso, Ivo Kabadshow and David Cardamone for
support and illuminating advice. We acknowledge the PRACE Research
Infrastructure resource Jugene based on Germany at Forschungszentrum
J\"ulich and Curie based in France at CEA, funded by GENCI and the
Rechenzentrum Garching (RZG) of the Max Planck Society for the usage
of IBM Blue Gene/P. We specially acknowledge Laurent Nguyen for the
support given for the Curie Supercomputer, Heiko Appel for the help
given with the Genius supercomputer in Garching.  We also acknowledge
financial support from the European Research Council Advanced Grant
DYNamo (ERC-2010-AdG-Proposal No. 267374), Spanish Grants (FIS2011-
65702-C02-01 and PIB2010US-00652), ACI-Promociona (ACI2009-1036),
general funding for research groups UPV/EHU (ALDAPA, GIU10/02), Grupos
Consolidados UPV/EHU del Gobierno Vasco (IT-319-07) and European
Commission project CRONOS (280879-2 CRONOS
CP-FP7). P. Garc\'ia-Risue\~no is funded by the Humboldt Universit\"at
zu Berlin.  J. Alberdi-Rodriguez acknowledges the scholarship of the
University of the Basque Country UPV/EHU. This work was partly
supported by the BMBF under grant 01IH08001B.



\bibliography{biblio}

\end{document}


\begin{abstract}
  In this document, we include some information to complement our
  paper.  First we provide some remarks on the way the data transfer
  for PFFT \cite{libpfft} and FMM \cite{Dac2010JCP} is carried out in
  our implementations. Then we add some remarks on the efficiency of
  the algorithms.  Finally, we discuss the correction term we devised
  to adapt the FMM method (originally devised to deal with point
  charges) to charge distributions.
\end{abstract}

\begin{figure*}[!ht]
\centering
\centering
\def\svgwidth{400pt} 
\input{chlorophyll_molecule.tex}
\caption{Different chunks of the chlorophyll molecule of the spinach.}
\label{fig:chlorophyll}
\end{figure*}
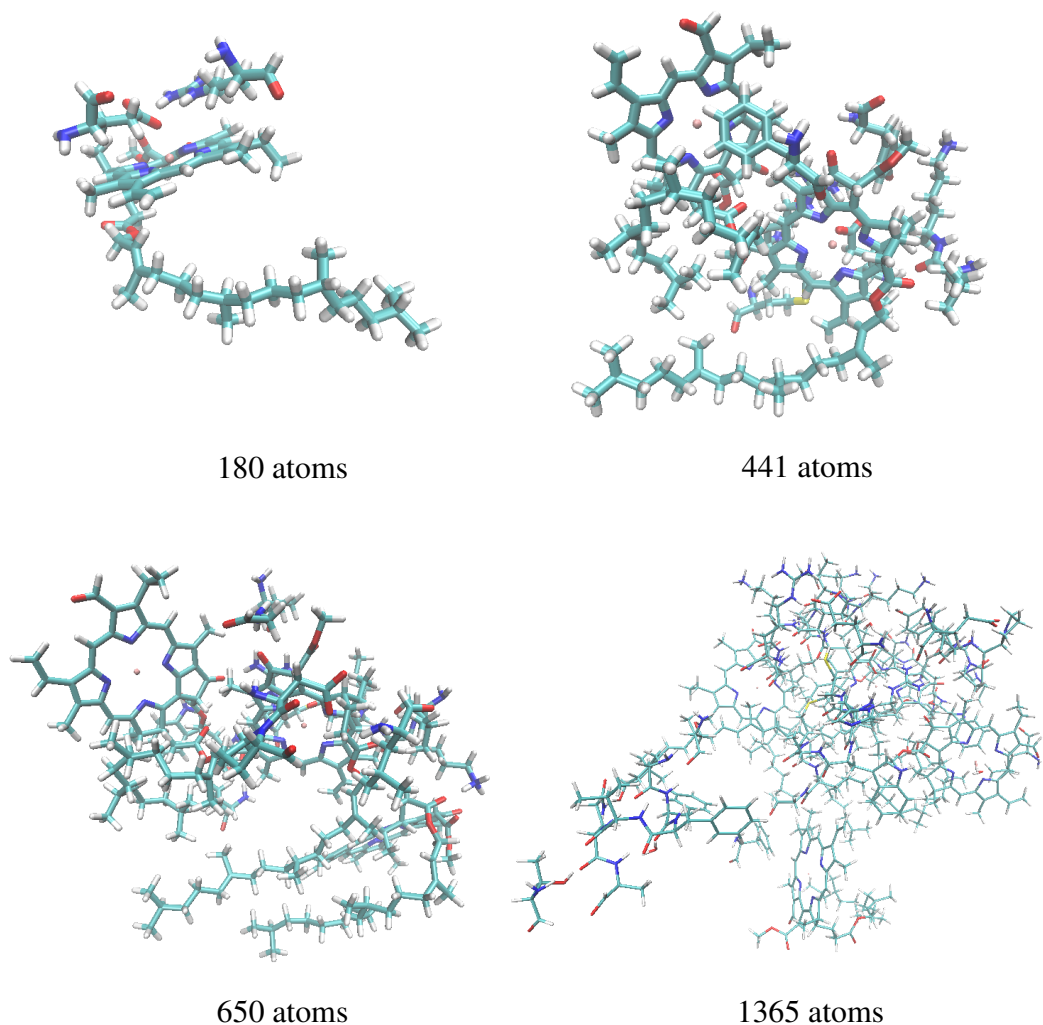

For testing purposes we have used different portions of the
chlorophyll molecule of the spinach photosynthetic
unit~\cite{Liu2004}, that it is a quite remarkable and complex
system. Our test systems consisted of 180, 441, 650 or 1365 atoms, and
contained several chlorophyll units (see figure
\ref{fig:chlorophyll}). The space is represented with cubic grids with
edge length $2L_e$ containing these molecules, where $L_e$ is the half
of the edge of the parallelepiped mesh and the used values are 15.8,
22.1, 25.9 and 31.7 respectively.

\section{Communication patterns}\label{sec:comm_pat}
As stated in our paper, the way the data of variables is transferred
from the main program ( {\sc octopus} in this case) and the Poisson
solver can be critical for its efficiency.  The data-transfer between
the box used for PFFT and that one used by {\sc octopus} (which
contain different sets of points, see section 3
methods) has been encapsulated in an specific Fortran module. Each of
those box representations corresponds to an MPI group. At a given
moment, data points have to be send from one group (\emph{sender I})
to the other (\emph{receiver O}). Since both groups stores the same
global grid, although in a different way, each point stored in a given
process of one group is also stored in one process of the other group.
For example, if point $n$ is stored in process $x_i$ of group I and in
process $y_o$ of group O, it should be sent from process $x_i$ to
process $y_o$. Unfortunately, MPI does not allow to send information
between different groups unless they are disjoint, which is not the
current case. This means communication will have to be done using only
one of the groups (senders or receivers group). This is not a problem,
because we can determine the rank of the receiver process in the I
group through the rank in the \texttt{MPI\_COMM\_WORLD} global
communicator.  Then, point $n$ is sent from process $x_i$ to process
$y_i$.

We have implemented a routine that determines to which process each
point should be sent. This information is then used to put the data in
the ``correct order''. Then, a simple call to the
\texttt{MPI\_Alltoall} function is enough for the communication step.
It is important to note that, using this technique, each process only
transfers each point once. Therefore, the total amount of information
that must be sent between all the processes is equal to the number of
points in the grid, and it is independent of the number of processes.

The PFFT library requires two communication steps in addition to the
box transformation. Required communication needs are two
\texttt{MPI\_Alltoall} calls for every calculated FFT. In total, six
\texttt{MPI\_Alltoall} calls are needed in every Poisson solver.

Regarding to the FMM library, three MPI global communication functions
have to be executed: \texttt{MPI\_Allgather}, \texttt{MPI\_Allreduce}
and \texttt{MPI\_Alltoall}. Additionally, synchronization between
different FMM levels has to be done using \texttt{MPI\_Barrier}
\cite{kabadshow2011}.

\section{More comments on execution-time}

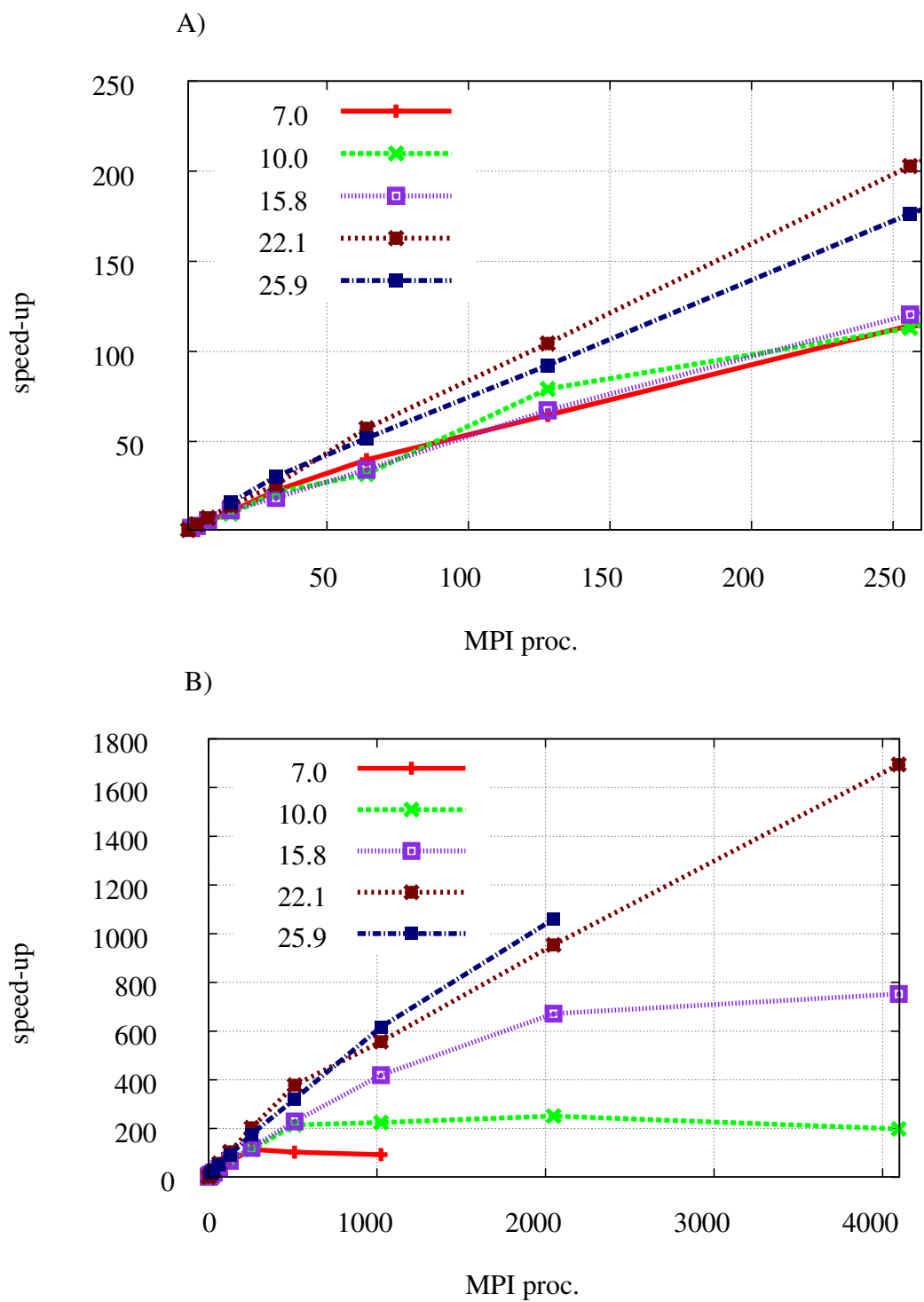
\begin{figure}
  \def\svgwidth{400pt} 
  \input{graph_bg_pfft_sp_1.tex}
  \def\svgwidth{400pt} 
  \input{graph_bg_pfft_sp.tex}
  \caption{ Speed-up of the PFFT Poisson solver in a Blue Gene/P
    computer for different system sizes (given by $L_e$, semi-legth
    of the parallelepiped edge). Largest systems saturate
    with more processes than the smallers ones. A) linear speed-up
    is shown up to 256 processes, independently of the simulated
    system. B) saturation point is higher when the simulated
    system is bigger.}
  \label{fig:pfft_sp_bg}
\end{figure}

Our tests showed that the novel implementations of PFFT
\cite{Pip2012TUC} and FMM \cite{Dac2010JCP} offer a good scalability
and accuracy, and are competitive if thousands of parallel processes
are available.  \ref{fig:pfft_sp_bg} shows the speed-ups obtained
using PFFT for the different system sizes in a BlueGene/P
supercomputer. Almost linear performances can be observed until
saturation for all the cases, and the obtained efficiencies have been
always above 50\% for just nearly all these points. As expected, large
systems, which have higher computation needs, can make a better use of
a high number of processes. Tests run in Corvo and Curie machines show
similar trends (although with efficiency problems of PFFT for some
values of MPI proc.).

\section{Correction terms for FMM}
\subsection{General remarks}
\begin{figure*}[!ht]
\centering
\centering
\def\svgwidth{200pt} 
\input{scheme.tex}
\caption{Scheme of how the inclusion of semi-neighbours of point P
  (pink points) helps to improve the accuracy of the integration to
  calculate the Hartree potential. A) Scheme of the function whose
  integration will be approximated by a summation, without considering
  semi-neighbours. B) id., considering semi-neighbours of point P; The
  error made by the approximation is proportional to the yellow
  surface in A) and B). C) 2D scheme of the grid: green points are
  grid points, while pink points are semi-neighbours of P.}
\label{fig:scheme}
\end{figure*}
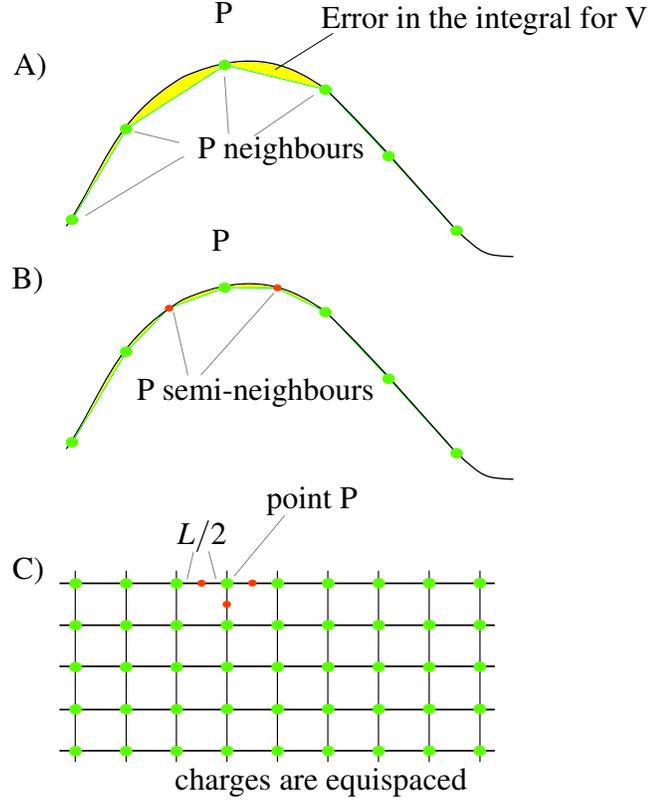

The fast multipole method
(FMM)\cite{Gre1987JCOP,Gre1988BOOK,Gre1997AN,Gre1999JCOP,Dac2010JCP}
was devised to efficiently calculate pairwise potentials created by
pointlike charges, like pairwise Coulomb potential. In the literature
it is possible to find some modifications of the traditional FMM which
deal with charges which are modelled as Gaussian
functions\cite{Str1996sci}. Such modifications of FMM can be used into
LCAO codes as Gaussian\cite{Frisch2009} or FHI-Aims. However, they are
not useful when the charge distribution is represented through a set
of discrete charge density values.  The Fast Multipole Method
presented in \cite{kdfmm2010,Dac2010JCP} belongs to the family of FMM
methods which calculate the electrostatic potential created by a set
of pointlike charges. This method is very accurate and efficient, but
it needs some modifications to work in programs like {\sc octopus}
\cite{Cas2006PSSB,Mar2003CPC}, where the 3D grid points actually
represent charge densities.  As stated in the section 'Theoretical
background' of the paper, the electronic density is a $\mathbb{R}^3
\rightarrow \mathbb{R}$ field, where values of the $\mathbb{R}^3$ set
correspond to a equispaced grid (see \ref{fig:scheme} C) ). The
variable $\rho_{j,k,l}$ is the charge density at the portion of volume
(cell) centred in the point $(j,k,l)$. Each cell is limited by the
planes bisecating the lines that join two consecutive grid points, and
its volume is $\Omega=L^3$, being $L$ the spacing between consecutive
grid points.  The density $\rho_{j,k,l}$ is always negative and it is
expected to vary slowly among nearby points.

The term $ v^{\mathrm{SI}}$ in equation 16 of the paper can be
calculated analytically as follows assuming that the cell is a sphere
of volume $\Omega$:
\begin{eqnarray}\label{vsisupp}
v^{\mathrm{SI}}(\vec{r_0}) &=&
\int_{\Omega}d\vec{r}\frac{\rho(\vec{r})}{|\vec{r}-\vec{r_0}|} =
\rho(\vec{r_0})\int_{\Omega}d\vec{r}\frac{1}{|\vec{r}-\vec{r_0}|}
\nonumber \\ &\simeq& \rho(\vec{r_0}) \int_0^{R}dr \int_0^{2\pi}d\phi
\int_0^{\pi}d\theta \frac{r^2sen(\theta)}{r} \nonumber \\ &=&
\rho(\vec{r_0}) 2\pi L^2 \left(\frac{3}{4\pi}\right)^{2/3}\ ,
\end{eqnarray}
where we have used the approximation of constant charge density within
the cell.  One may expect this approximate way to proceed to be less
accurate than the numerical integration of $1/r$ in a cubic cell (what
is also efficient, since the integration through an arbitrary size
cube is proportional to the integral through a cube of unit
volume). However, it happens the converse: the difference between both
methods is small (about 1\% of difference between integrals) but, due
to error cancellations, the analytical method is slightly more
accurate when calculating potentials.

The term $v^{\mathrm{corr.}}_{j,k,l}$ in (equation 16) is included to
calculate more accurately the potential created by the charge in cells
nearby to $(j,k,l)$. To devise a expression for it, we consider that
charge distribution is similar to sets of Gaussians centred in the
atoms of the system. For Gaussian distributions, the greatest
concavity near the centre of the Gaussian makes the influence of
neighbouring points to be major for the potential. As we can see in
the scheme of \ref{fig:scheme} A)-B), considering semi-neighbours of
point $P$ (in $\vec{r}_0:=(j,k,l)$), i.e., points whose distance to
$P$ is not $L$, but $L/2$, the integral of equation 16 of the
paper can be calculated in a much more accurate way.

\subsection{Method 1: 6-neighbours correction}
\begin{figure}[!ht]
  \centering
  \def\svgwidth{450pt}
  \input{cells_img.tex}
  \caption{2D example of the position of cells containing semi
    neighbours. Assume the centre of the plots is $\vec{r_0}$, the
    point where we want to calculate the correcting term for the
    potential. The volume of semi-neighbour cells is $L^2/4$ in 2D,
    and $\Omega/8$ in 3D. One half of the semi-neighbour cell occupies
    the volume of a neighbour cell (the cell whose centre is $L$ away
    from $\vec{r_0}$). The other half of the semi-neighbour cell
    occupies the space of the $\vec{r_0}$-centred cell itself.}
  \label{fig:cells}
\end{figure}
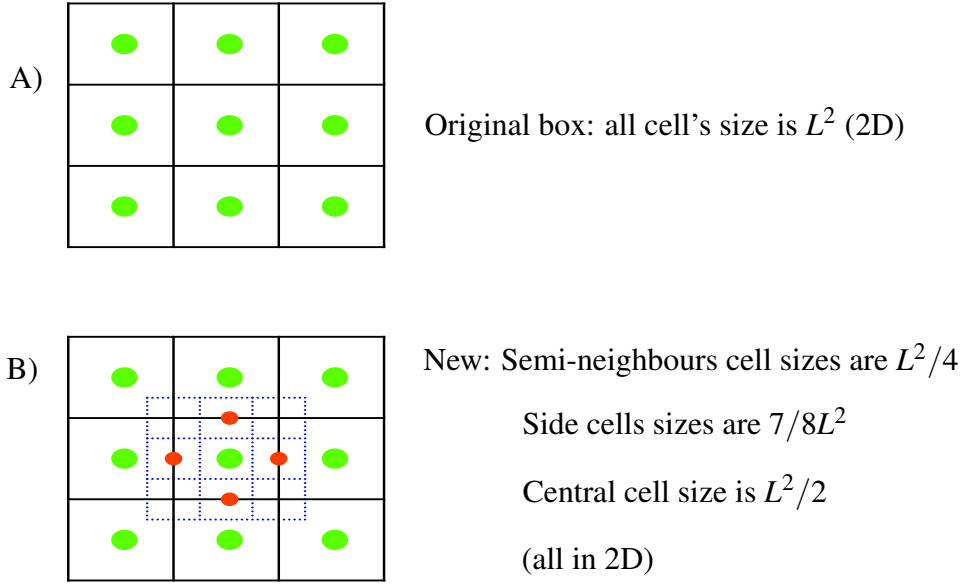

We build a corrective term by calculating the charge in the 6
semi-neighbours of every point of the mesh $\vec{r_0}$ (see
\ref{fig:cells} for a intuitive scheme). The total correction term is the
potential created by the semi-neighbours ($v^{\mathrm{corr.+}}$) minus
the potential created by the charge lying in the volume of the
semi-neighbour cells that was already counted in $v^{FMM}$ or in
$v^{\mathrm{SI}}$ ($v^{\mathrm{corr.-}}$):
\begin{equation}\label{masmenos}
  v^{\mathrm{corr.}}(\vec{r_0})=v^{\mathrm{corr.+}}(\vec{r_0})-v^{\mathrm{corr.-}}(\vec{r_0}) \ .
\end{equation}
In order to calculate $v^{\mathrm{corr.+}}$, we use the formula por
the 3rd degree interpolation polynomial:
\begin{subequations}
\begin{align}
f(0) &= \frac{(-1)}{16} f\left(\frac{-3}{2}L\right)+\frac{9}{16} f\left(\frac{-L}{2}\right)+\frac{9}{16}f\left(\frac{L}{2}\right)-\frac{(-1)}{16} f\left(\frac{3}{2}L\right) \ , \\
f\left(\frac{-L}{2}\right) &=\frac{(-1)}{16} f\left(-2L\right)+ \frac{9}{16}f\left(-L\right)+\frac{9}{16}f\left(0\right)-\frac{(-1)}{16} f\left(L\right) \ ,\\
f\left(\frac{L}{2}\right) &=\frac{(-1)}{16}f\left(-L\right)+\frac{9}{16}f\left(0\right)+\frac{9}{16} f\left(L\right)-\frac{(-1)}{16} f\left(2L\right) \ . 
\end{align}
\end{subequations}
So, the semi-neighbours of $\vec{r_0}=(x_0,y_0,z_0)$ are
\begin{subequations}
\begin{align}
\rho(x_0-L/2,y_0,z_0) =&\frac{-1}{16} \rho \left(x_0-2L,y_0,z_0\right)+\frac{9}{16} \rho\left(x_0-L,y_0,z_0\right) \nonumber \\ 
& +\frac{9}{16} \rho\left(x_0,y_0,z_0\right)-\frac{1}{16} \rho\left(x_0+L,y_0,z_0\right) \ ; \\
\rho(x_0+L/2,y_0,z_0) =&\frac{-1}{16} \rho\left(x_0-L,y_0,z_0\right)+\frac{9}{16} \rho\left(x_0,y_0,z_0\right)  \nonumber \\ 
& +\frac{9}{16} \rho\left(x_0+L,y_0,z_0\right)-\frac{1}{16} \rho\left(x_0+2L,y_0,z_0\right) \ ; \\
\rho(x_0,y_0-L/2,z_0) =&\frac{-1}{16} \rho\left(x_0,y_0-2L,z_0\right)+\frac{9}{16} \rho\left(x_0,y_0-L,z_0\right)  \nonumber \\ 
& +\frac{9}{16} \rho\left(x_0,y_0,z_0\right)-\frac{1}{16} \rho\left(x_0,y_0+L,z_0\right) \ ; \\
\rho(x_0,y_0+L/2,z_0) =&\frac{-1}{16} \rho\left(x_0,y_0-L,z_0\right)+\frac{9}{16} \rho\left(x_0,y_0,z_0\right)  \nonumber \\  
& +\frac{9}{16} \rho\left(x_0,y_0+L,z_0 \right)-\frac{1}{16} \rho\left(x_0,y_0+2L,z_0 \right) \ ; \\
\rho(x_0,y_0,z_0-L/2) =&\frac{-1}{16} \rho\left(x_0,y_0,z_0-2L\right)+\frac{9}{16} \rho\left(x_0,y_0,z_0-L \right)  \nonumber \\ 
& +\frac{9}{16} \rho\left(x_0,y_0,z_0 \right)-\frac{1}{16} \rho\left(x_0,y_0,z_0+L \right) \ ; \\
\rho(x_0,y_0,z_0+L/2) =&\frac{-1}{16} \rho\left(x_0,y_0,z_0-L\right)+\frac{9}{16} \rho\left(x_0,y_0,z_0 \right) \nonumber \\ 
& +\frac{9}{16} \rho\left(x_0,y_0,z_0+L\right)-\frac{1}{16} \rho\left(x_0,y_0,z_0+2L \right) \ .
\end{align}
\end{subequations}
We consider all this six charges to be homogeneously distributed in
cells whose volume is $\Omega/8$. The distance between the centre of
these small cells and the centre of the cell whose $v$ we are
calculating (i.e., the cell centred in $\vec{r_0}$) is $L/2$. So the
first part of $v^{\mathrm{corr.}}(\vec{r_0})$ is
\begin{eqnarray}\label{corr1}
v^{\mathrm{corr.+}}(\vec{r}_0)&=& \Big(\rho(x_0-L/2,y_0,z_0)+\rho(x_0+L/2,y_0,z_0)+\ldots+ \nonumber \\
&+& \rho(x_0,y_0,z_0+L/2)  \Big) \left(\frac{\Omega}{8} \right) \left(\frac{1}{L/2}\right)  \ .
\end{eqnarray}
Since we have created these new 6 cells, we must subtract the
potential created by their corresponding volume from that created by
the cells whose volume is partly occupied by these new cells. This
potential is:

\begin{eqnarray}\label{corr2}
v^{\mathrm{corr.-}}(\vec{r}_0)&=&  \Big(\rho(x_0-L,y_0,z_0)+\rho(x_0+L,y_0,z_0)+\rho(x_0,y_0-L,z_0)+\rho(x_0,y_0+L,z_0) \nonumber \\
   &+&\rho(x_0,y_0,z_0-L)+\rho(x_0,y_0,z_0+L)  \Big) \left(\frac{\Omega}{16}\right)\left(\frac{1}{L}\right) + \alpha \ v^{\mathrm{SI}}(\vec{r_0}) \ .
\end{eqnarray}
The aim of the term $\alpha \ v^{\mathrm{SI}}$ (i.e., the variable 
\texttt{AlphaFMM} in {\sc octopus}) is to compensate the
errors arising from the assumption that the charge is concentrated at
the centre of the cells and reduced cells.  The value of $\alpha$ is
tuned to minimize the errors in the potentials.

It is worth to re-express as follows the correction terms of
eqs. \ref{masmenos}, \ref{corr1} and \ref{corr2} avoiding to call to
every variable more than once for the sake of getting higher
computational efficiency:

\begin{eqnarray}
  v^{\mathrm{SI}}(\vec{r_0}) +  v^{\mathrm{corr.}}(\vec{r_0}) =  L^2  \Big[ & \rho(x_0,y_0,z_0) \big(27/32 + (1-\alpha) 2\pi (3/4\pi)^{2/3} \big) \qquad \qquad \qquad \\
  &+(1/16) \Big( \rho(x_0-L,y_0,z_0) + \rho(x_0+L,y_0,z_0) + \rho(x_0,y_0-L,z_0) \qquad \quad  \nonumber \\
  & + \rho(x_0,y_0+L,z_0) + \rho(x_0,y_0,z_0-L) + \rho(x_0,y_0,z_0+L)  \nonumber \\
  &   - (1/4) \big(  \rho(x_0-2L,y_0,z_0) + \rho(x_0+2L,y_0,z_0) + \rho(x_0,y_0-2L,z_0)  \nonumber \\
  & + \rho(x_0,y_0+2L,z_0) + \rho(x_0,y_0,z_0-2L) + \rho(x_0,y_0,z_0+2L) \big) \Big) \Big] \nonumber \ .
\end{eqnarray}

We ran tests using the error formula $E:=\sqrt{\sum_{i} (v^{\mathrm{Exact}}(\vec{r}_i) -
  v^{\mathrm{FMM}}(\vec{r}_i))^2}$, with the index $i$ running for all points of the
system.  The inclusion of the correcting term introduced in this
section typically reduced $E$ in a factor about 50.

\subsection{Method 2: 124-neighbours correction}
This method is similar to the one explained in the previous section,
but with two differences
\begin{itemize}
\item It uses 3D interpolation polynomials, instead of 1D
  polynomials. Then, it considers $5^3-1=124$ neighbours in a cube of
  edge 5$L$ centred in $\vec{r_0}$ to calculate the corrective term
  for $V(\vec{r_0})$
\item The interpolation polynomials representing $\rho(x,y,z)$ are
  numerically integrated (after their division by $r$). This is, we
  calculate
\begin{eqnarray}\label{vsiii}
v^{\mathrm{corr.+}}(\vec{r_0}) &=& \int_{125\Omega}d\vec{r}\frac{\rho(\vec{r})}{|\vec{r}-\vec{r_0}|} \simeq
\int_{125\Omega}d\vec{r}\frac{Pol(\vec{r})}{|\vec{r}-\vec{r_0}|} \ .
\end{eqnarray}
\end{itemize}
The integration is to be performed between -5/2$L$ and 5/2$L$ for $x$, $y$ and $z$. The interpolation 
polynomial (with 125 support points) $Pol(\vec{r})$ is
\begin{equation}\label{pol}
Pol(\vec{r}) = \sum_{i=1}^5  \sum_{j=1}^5  \sum_{k=1}^5 \rho(x_0+(i-3)L,y_0+(j-3)L,z_0+(k-3)L) \alpha_i(x) \alpha_j(y) \alpha_k(z) \ , \\
\end{equation}
being
\begin{subequations}
\begin{align}
& \alpha_1(\xi) := \frac{\xi^4}{24} - \frac{\xi^3}{12} - \frac{\xi^2}{24} + \frac{\xi}{12}  \ , \\
& \alpha_2(\xi) :=  -\frac{\xi^4}{6} +\frac{\xi^3}{6}+\frac{2\xi^2}{3}-\frac{2\xi}{3}  \ , \\ 
& \alpha_3(\xi) :=   \frac{\xi^4}{4} - \frac{5\xi^2}{4} +1  \ ,\\
& \alpha_4(\xi) :=  -\frac{\xi^4}{6} -\frac{\xi^3}{6}+\frac{2\xi^2}{3}+\frac{2\xi}{3}  \ , \\
& \alpha_5(\xi) :=  \frac{\xi^4}{24} + \frac{\xi^3}{12} - \frac{\xi^2}{24} - \frac{\xi}{12}  \ . 
\end{align}
\end{subequations}
The quotient of the polynomials $\alpha_i(x) \alpha_j(y) \alpha_k(z)$
divided by $|\vec{r}-\vec{r_0}|$ can be numerically integrated through
the cubic cell of edge 5$L$ and centred in $x_0$. Such integrals can
indeed be tabulated, because equation (\ref{vsiii}) implies
$v^{\mathrm{corr.+}}(\vec{r_0})=v^{\mathrm{corr.+}}(\vec{r_0})|_{L=1}\cdot
L^2$. Terms of $\alpha_i(x) \alpha_j(y) \alpha_k(z) /
|\vec{r}-\vec{r_0}| $ are often odd functions whose integral is
null. The non-zero integrals taking part in (\ref{vsisupp}) (with $L=1$)
can be easily calculated numerically.

Therefore
\begin{eqnarray}\label{vsii}
v^{\mathrm{corr.+}}(\vec{r_0}) &=& \sum_{i=1}^5 \sum_{j=1}^5 \sum_{k=1}^5 \rho(x_0+(i-3)L,y_0+(j-3)L,z_0+(k-3)L) \cdot \nonumber \\
&& \int_{125\Omega} d\vec{r} \frac{ \alpha_i(x) \alpha_j(y) \alpha_k(z)} {|\vec{r}-\vec{r_0}|} \nonumber \\
&=& \sum_{i=1}^5 \sum_{j=1}^5 \sum_{k=1}^5 \rho(x_0+(i-3)L,y_0+(j-3)L,z_0+(k-3)L) \cdot \nonumber \\
&&\quad \sum_{l=1}^5 \sum_{m=1}^5 \sum_{n=1}^5 \alpha_{i,l}\alpha_{j,m}\alpha_{k,n} \int_{125\Omega} d\vec{r} \frac{ x^{l-1} y^{m-1} z^{n-1}} {|\vec{r}-\vec{r_0}|} \nonumber \\
&=& \sum_{i=1}^5 \sum_{j=1}^5 \sum_{k=1}^5 \rho(x_0+(i-3)L,y_0+(j-3)L,z_0+(k-3)L) \cdot \nonumber \\
&& \quad \sum_{l=1}^5 \sum_{m=1}^5 \sum_{n=1}^5 \alpha_{i,l}\alpha_{j,m}\alpha_{k,n} \beta(l-1,m-1,n-1) L^2 \ ,
\end{eqnarray}
where $\alpha_{i,l}$ is the coefficient of $\xi^{l-1}$ if $\alpha_{i}(\xi)$ and 
\begin{equation}
\beta(l,m,n):=\int_{-1/2}^{1/2}dx \int_{-1/2}^{1/2}dy \int_{-1/2}^{1/2}dz \frac{x^l y^m z^n}{\sqrt{x^2+y^2+z^2} }\ .
\end{equation}
In this case, $v^{\mathrm{corr.-}}$ is equal to all the contributions
to $V(\vec{r_0})$ due to charges whose position $(x,y,z)$ satisfies
\begin{equation}
|x-x_0| <= 2L; \ |y-y_0| <= 2L; \ |z-z_0| <= 2L \ ,
\end{equation}
including self-interaction integral.

This way to calculate $v^{\mathrm{corr.+}}$ is not inefficient,
because only 27 integrals are not null, and both $\alpha$ and $\beta$
are known. In order to calculate $v^{\mathrm{corr.+}}(\vec{r_0})$ we
need 125 products and additions, what is essentially the same number
of operations which is required in order to calculate the potential
created in $\vec{r_0}$ by the neighbouring points (whose calculation
can be removed and then saved). Nevertheless, results using this
correction method were worse than that obtained using the first
method, so only that one was implemented into the standard version of
{\sc octopus}.

\bibliography{biblio}

%% file: chlorophyll_molecule.tex
\begingroup%
  \makeatletter%
  \providecommand\color[2][]{%
    \errmessage{(Inkscape) Color is used for the text in Inkscape, but the package 'color.sty' is not loaded}%
    \renewcommand\color[2][]{}%
  }%
  \providecommand\transparent[1]{%
    \errmessage{(Inkscape) Transparency is used (non-zero) for the text in Inkscape, but the package 'transparent.sty' is not loaded}%
    \renewcommand\transparent[1]{}%
  }%
  \providecommand\rotatebox[2]{#2}%
  \ifx\svgwidth\undefined%
    \setlength{\unitlength}{1193.60000648bp}%
    \ifx\svgscale\undefined%
      \relax%
    \else%
      \setlength{\unitlength}{\unitlength * \real{\svgscale}}%
    \fi%
  \else%
    \setlength{\unitlength}{\svgwidth}%
  \fi%
  \global\let\svgwidth\undefined%
  \global\let\svgscale\undefined%
  \makeatother%
  \begin{picture}(1,0.99543723)%
    \put(0,0){\includegraphics[width=\unitlength]{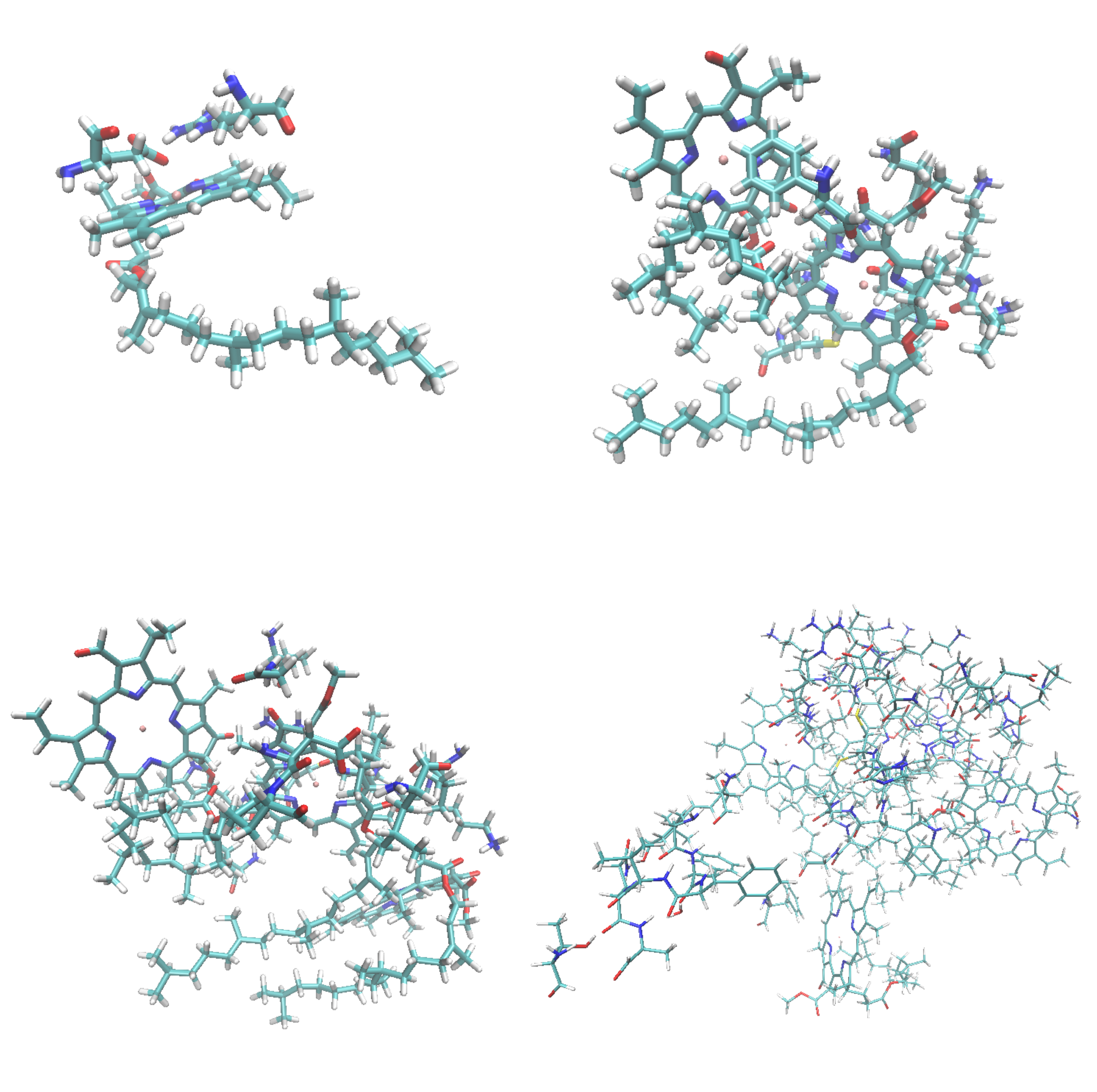}}%
    \put(0.20574837,0.51516151){\color[rgb]{0,0,0}\makebox(0,0)[lb]{\smash{180 atoms}}}%
    \put(0.70213596,0.51524987){\color[rgb]{0,0,0}\makebox(0,0)[lb]{\smash{441 atoms}}}%
    \put(0.20601673,0.00018981){\color[rgb]{0,0,0}\makebox(0,0)[lb]{\smash{650 atoms}}}%
    \put(0.69745933,0.00018981){\color[rgb]{0,0,0}\makebox(0,0)[lb]{\smash{1365 atoms}}}%
  \end{picture}%
\endgroup%

%% file: graph_bg_pfft_sp_1.tex
\begingroup%
  \makeatletter%
  \providecommand\color[2][]{%
    \errmessage{(Inkscape) Color is used for the text in Inkscape, but the package 'color.sty' is not loaded}%
    \renewcommand\color[2][]{}%
  }%
  \providecommand\transparent[1]{%
    \errmessage{(Inkscape) Transparency is used (non-zero) for the text in Inkscape, but the package 'transparent.sty' is not loaded}%
    \renewcommand\transparent[1]{}%
  }%
  \providecommand\rotatebox[2]{#2}%
  \ifx\svgwidth\undefined%
    \setlength{\unitlength}{664.65990686bp}%
    \ifx\svgscale\undefined%
      \relax%
    \else%
      \setlength{\unitlength}{\unitlength * \real{\svgscale}}%
    \fi%
  \else%
    \setlength{\unitlength}{\svgwidth}%
  \fi%
  \global\let\svgwidth\undefined%
  \global\let\svgscale\undefined%
  \makeatother%
  \begin{picture}(1,0.70059708)%
    \put(0,0){\includegraphics[width=\unitlength]{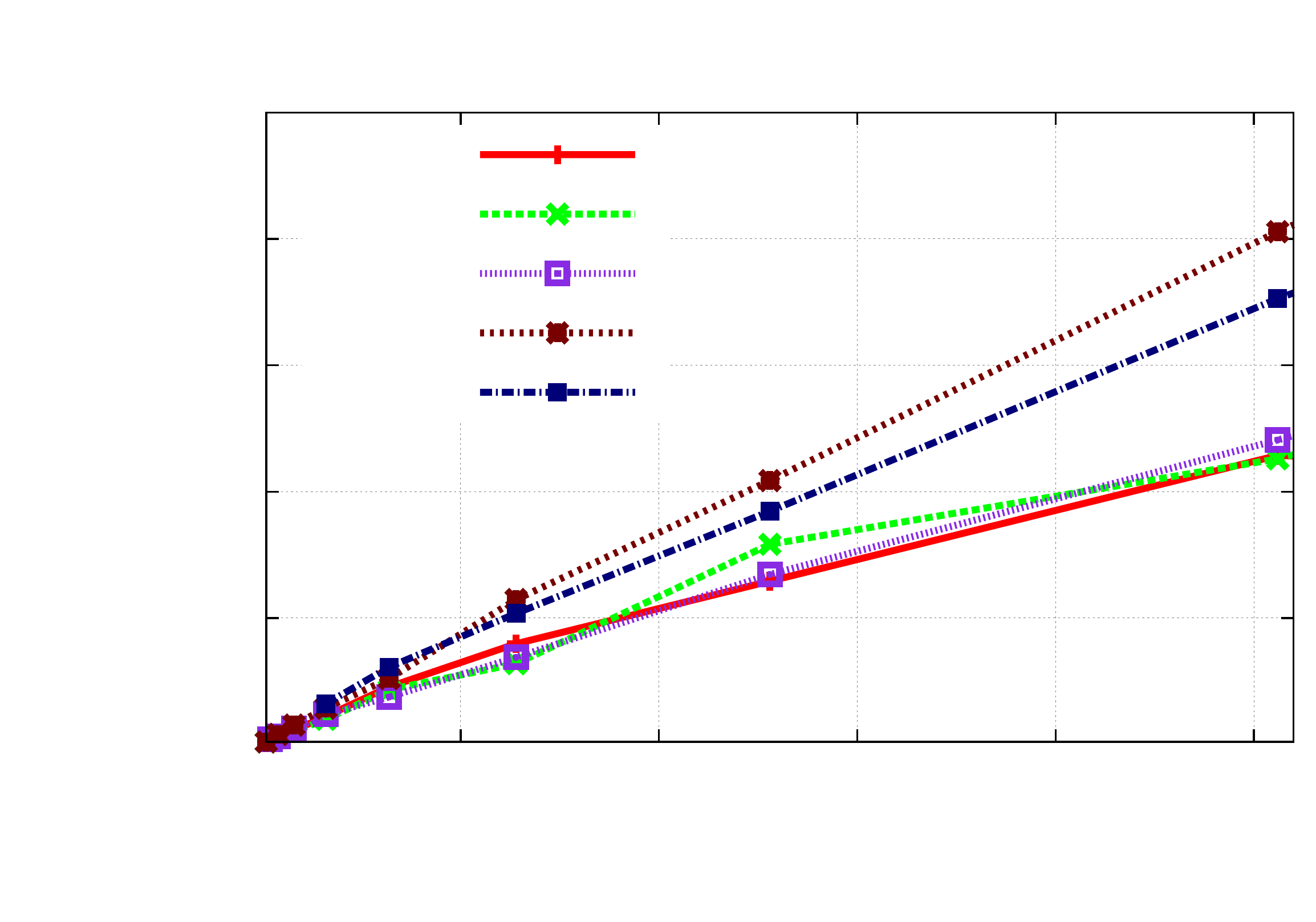}}%
    \put(0.12504231,0.21592692){\makebox(0,0)[lb]{\smash{50}}}%
    \put(0.09994075,0.31191586){\makebox(0,0)[lb]{\smash{100}}}%
    \put(0.09994075,0.4079048){\makebox(0,0)[lb]{\smash{150}}}%
    \put(0.09994075,0.5040442){\makebox(0,0)[lb]{\smash{200}}}%
    \put(0.09994075,0.60003314){\makebox(0,0)[lb]{\smash{250}}}%
    \put(0.331241,0.07660755){\makebox(0,0)[lb]{\smash{50}}}%
    \put(0.46929355,0.07660755){\makebox(0,0)[lb]{\smash{100}}}%
    \put(0.62004734,0.07660755){\makebox(0,0)[lb]{\smash{150}}}%
    \put(0.77080113,0.07660755){\makebox(0,0)[lb]{\smash{200}}}%
    \put(0.92155492,0.07660755){\makebox(0,0)[lb]{\smash{250}}}%
    \put(0.03230917,0.28175307){\rotatebox{90}{\makebox(0,0)[lb]{\smash{speed-up}}}}%
    \put(0.49603204,0.00890375){\makebox(0,0)[lb]{\smash{MPI proc.}}}%
    \put(0.18964527,0.66773693){\makebox(0,0)[lb]{\smash{A)}}}%
    \put(0.33,0.56798667){\makebox(0,0)[rb]{\smash{7.0}}}%
    \put(0.33,0.52285081){\makebox(0,0)[rb]{\smash{10.0}}}%
    \put(0.33,0.47771494){\makebox(0,0)[rb]{\smash{15.8}}}%
    \put(0.33,0.43257908){\makebox(0,0)[rb]{\smash{22.1}}}%
    \put(0.33,0.38744321){\makebox(0,0)[rb]{\smash{25.9}}}%
  \end{picture}%
\endgroup%

%% file: graph_bg_pfft_sp.tex
\begingroup%
  \makeatletter%
  \providecommand\color[2][]{%
    \errmessage{(Inkscape) Color is used for the text in Inkscape, but the package 'color.sty' is not loaded}%
    \renewcommand\color[2][]{}%
  }%
  \providecommand\transparent[1]{%
    \errmessage{(Inkscape) Transparency is used (non-zero) for the text in Inkscape, but the package 'transparent.sty' is not loaded}%
    \renewcommand\transparent[1]{}%
  }%
  \providecommand\rotatebox[2]{#2}%
  \ifx\svgwidth\undefined%
    \setlength{\unitlength}{680.79790717bp}%
    \ifx\svgscale\undefined%
      \relax%
    \else%
      \setlength{\unitlength}{\unitlength * \real{\svgscale}}%
    \fi%
  \else%
    \setlength{\unitlength}{\svgwidth}%
  \fi%
  \global\let\svgwidth\undefined%
  \global\let\svgscale\undefined%
  \makeatother%
  \begin{picture}(1,0.68398975)%
    \put(0,0){\includegraphics[width=\unitlength]{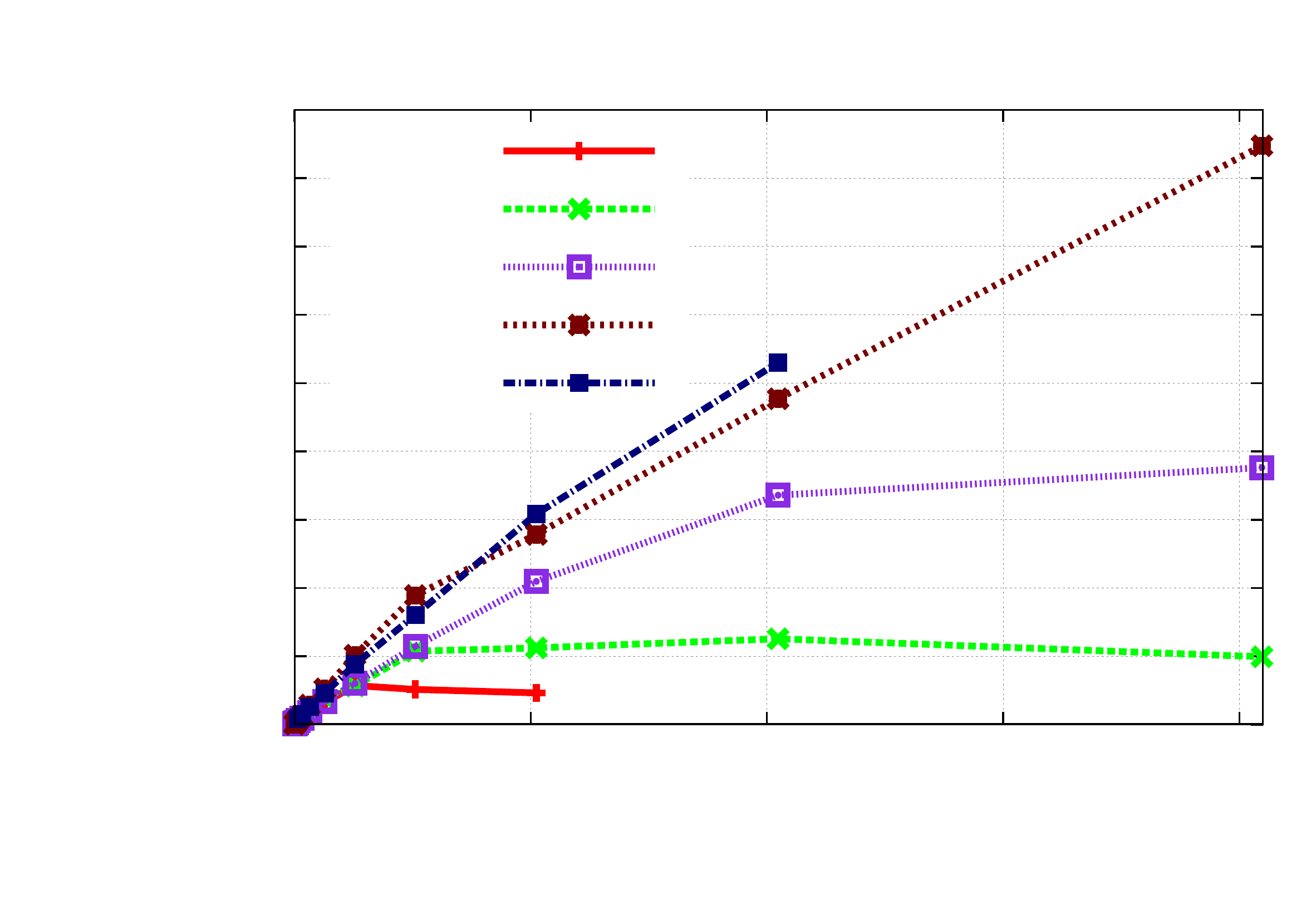}}%
    \put(0.17302581,0.11856377){\makebox(0,0)[lb]{\smash{0}}}%
    \put(0.12401126,0.17056158){\makebox(0,0)[lb]{\smash{200}}}%
    \put(0.12401126,0.22241251){\makebox(0,0)[lb]{\smash{400}}}%
    \put(0.12401126,0.27426343){\makebox(0,0)[lb]{\smash{600}}}%
    \put(0.12401126,0.32626124){\makebox(0,0)[lb]{\smash{800}}}%
    \put(0.09950473,0.37811216){\makebox(0,0)[lb]{\smash{1000}}}%
    \put(0.09950473,0.43010997){\makebox(0,0)[lb]{\smash{1200}}}%
    \put(0.09950473,0.4819609){\makebox(0,0)[lb]{\smash{1400}}}%
    \put(0.09950473,0.5339587){\makebox(0,0)[lb]{\smash{1600}}}%
    \put(0.09950473,0.58580963){\makebox(0,0)[lb]{\smash{1800}}}%
    \put(0.21769398,0.07479161){\makebox(0,0)[lb]{\smash{0}}}%
    \put(0.36042797,0.07479161){\makebox(0,0)[lb]{\smash{1000}}}%
    \put(0.53992324,0.07479161){\makebox(0,0)[lb]{\smash{2000}}}%
    \put(0.7194185,0.07479161){\makebox(0,0)[lb]{\smash{3000}}}%
    \put(0.89891376,0.07479161){\makebox(0,0)[lb]{\smash{4000}}}%
    \put(0.0315433,0.27507424){\rotatebox{90}{\makebox(0,0)[lb]{\smash{speed-up}}}}%
    \put(0.49749361,0.0086927){\makebox(0,0)[lb]{\smash{MPI proc.}}}%
    \put(0.1983696,0.65190854){\makebox(0,0)[lb]{\smash{B)}}}%
    \put(0.35,0.55452281){\makebox(0,0)[rb]{\smash{7.0}}}%
    \put(0.35,0.51045687){\makebox(0,0)[rb]{\smash{10.0}}}%
    \put(0.35,0.46639093){\makebox(0,0)[rb]{\smash{15.8}}}%
    \put(0.35,0.42232499){\makebox(0,0)[rb]{\smash{22.1}}}%
    \put(0.35,0.37825905){\makebox(0,0)[rb]{\smash{25.9}}}%
  \end{picture}%
\endgroup%

%% file: scheme.tex
\begingroup%
  \makeatletter%
  \providecommand\color[2][]{%
    \errmessage{(Inkscape) Color is used for the text in Inkscape, but the package 'color.sty' is not loaded}%
    \renewcommand\color[2][]{}%
  }%
  \providecommand\transparent[1]{%
    \errmessage{(Inkscape) Transparency is used (non-zero) for the text in Inkscape, but the package 'transparent.sty' is not loaded}%
    \renewcommand\transparent[1]{}%
  }%
  \providecommand\rotatebox[2]{#2}%
  \ifx\svgwidth\undefined%
    \setlength{\unitlength}{956.36004333bp}%
    \ifx\svgscale\undefined%
      \relax%
    \else%
      \setlength{\unitlength}{\unitlength * \real{\svgscale}}%
    \fi%
  \else%
    \setlength{\unitlength}{\svgwidth}%
  \fi%
  \global\let\svgwidth\undefined%
  \global\let\svgscale\undefined%
  \makeatother%
  \begin{picture}(1,1.48116641)%
    \put(0,0){\includegraphics[width=\unitlength]{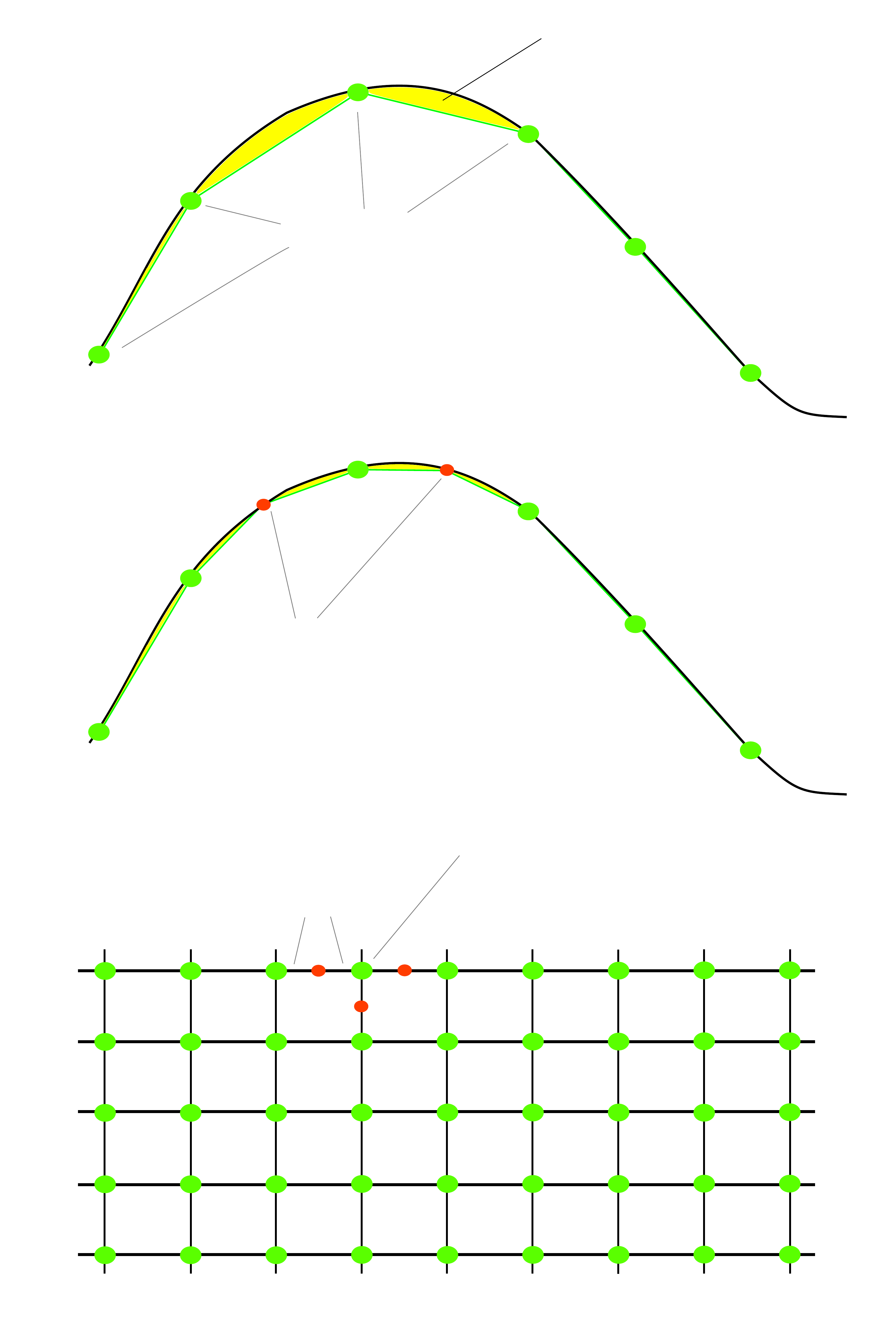}}%
    \put(-0.00187887,0.40676926){\color[rgb]{0,0,0}\makebox(0,0)[lb]{\smash{C)}}}%
    \put(0.47836763,0.53916635){\color[rgb]{0,0,0}\makebox(0,0)[lb]{\smash{point P}}}%
    \put(0.30958009,0.47202435){\color[rgb]{0,0,0}\makebox(0,0)[lb]{\smash{$L/2$}}}%
    \put(0.30535583,0.00695998){\color[rgb]{0,0,0}\makebox(0,0)[lb]{\smash{charges are equispaced}}}%
    \put(-0.00026141,1.35979077){\color[rgb]{0,0,0}\makebox(0,0)[lb]{\smash{A)}}}%
    \put(-0.00328394,0.95507376){\color[rgb]{0,0,0}\makebox(0,0)[lb]{\smash{B)}}}%
    \put(0.37485702,1.02594587){\color[rgb]{0,0,0}\makebox(0,0)[lb]{\smash{P}}}%
    \put(0.38045219,1.45677379){\color[rgb]{0,0,0}\makebox(0,0)[lb]{\smash{P}}}%
    \put(0.58187818,1.44744847){\color[rgb]{0,0,0}\makebox(0,0)[lb]{\smash{Error in the integral for V}}}%
    \put(0.34874626,1.20126112){\color[rgb]{0,0,0}\makebox(0,0)[lb]{\smash{P neighbours}}}%
    \put(0.2331128,0.74432248){\color[rgb]{0,0,0}\makebox(0,0)[lb]{\smash{P semi-neighbours}}}%
  \end{picture}%
\endgroup%

%% file: cells_img.tex
\begingroup%
  \makeatletter%
  \providecommand\color[2][]{%
    \errmessage{(Inkscape) Color is used for the text in Inkscape, but the package 'color.sty' is not loaded}%
    \renewcommand\color[2][]{}%
  }%
  \providecommand\transparent[1]{%
    \errmessage{(Inkscape) Transparency is used (non-zero) for the text in Inkscape, but the package 'transparent.sty' is not loaded}%
    \renewcommand\transparent[1]{}%
  }%
  \providecommand\rotatebox[2]{#2}%
  \ifx\svgwidth\undefined%
    \setlength{\unitlength}{1108.884235bp}%
    \ifx\svgscale\undefined%
      \relax%
    \else%
      \setlength{\unitlength}{\unitlength * \real{\svgscale}}%
    \fi%
  \else%
    \setlength{\unitlength}{\svgwidth}%
  \fi%
  \global\let\svgwidth\undefined%
  \global\let\svgscale\undefined%
  \makeatother%
  \begin{picture}(1,0.48680516)%
    \put(0,0){\includegraphics[width=\unitlength]{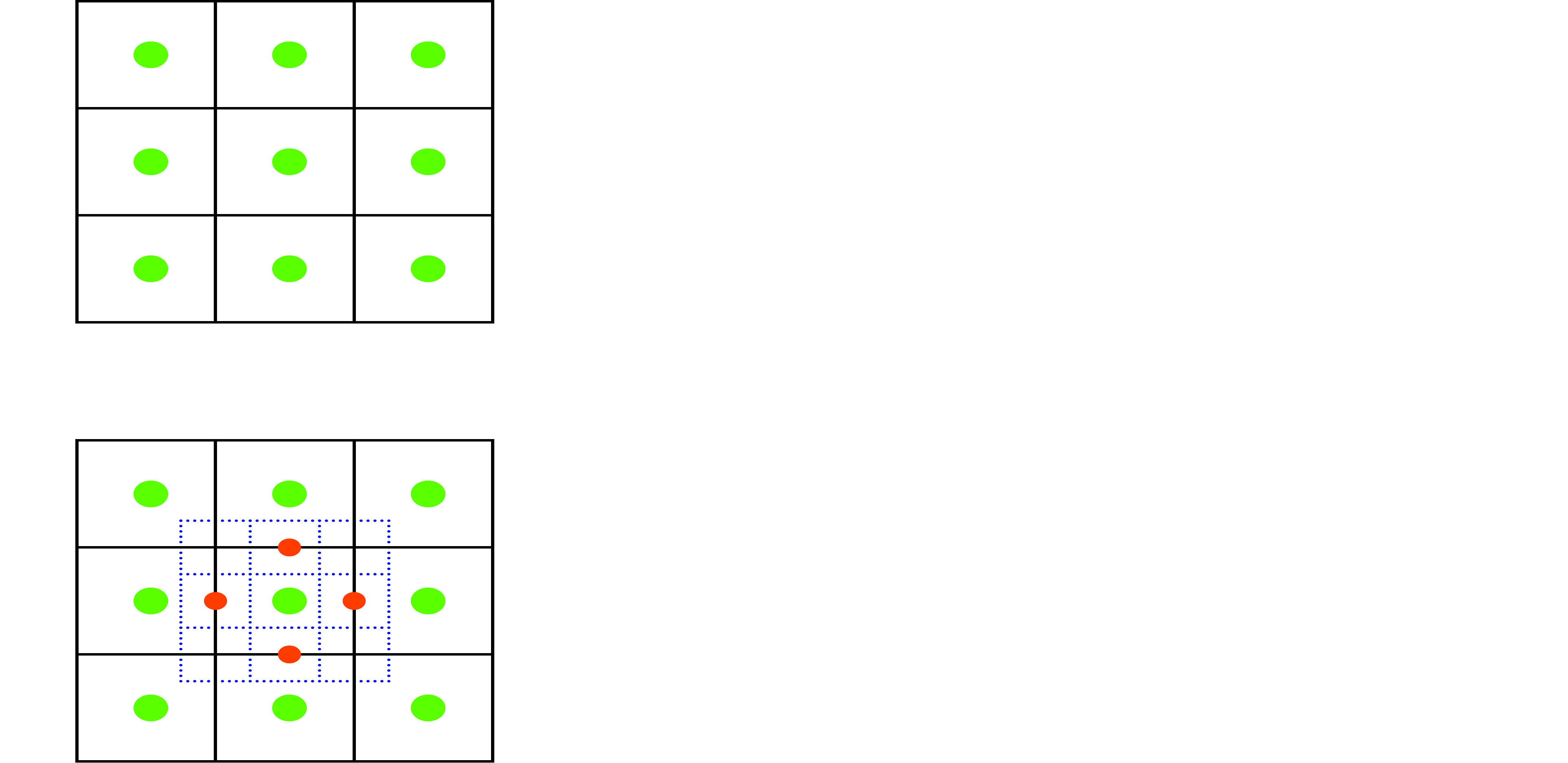}}%
    \put(-0.00022545,0.41513625){\color[rgb]{0,0,0}\makebox(0,0)[lb]{\smash{A)}}}%
    \put(-0.00283224,0.1698486){\color[rgb]{0,0,0}\makebox(0,0)[lb]{\smash{B)}}}%
    \put(0.34869779,0.37558981){\color[rgb]{0,0,0}\makebox(0,0)[lb]{\smash{Original box: all cell's size is $L^2$ (2D)}}}%
    \put(0.34748599,0.17769813){\color[rgb]{0,0,0}\makebox(0,0)[lb]{\smash{New: Semi-neighbours cell sizes are $L^2/4$ 
}}}%
    \put(0.43066082,0.12388273){\color[rgb]{0,0,0}\makebox(0,0)[lb]{\smash{Side cells sizes are $7/8 L^2$}}}%
    \put(0.43094263,0.06687657){\color[rgb]{0,0,0}\makebox(0,0)[lb]{\smash{Central cell size is $L^2/2$}}}%
    \put(0.4300831,0.01068934){\color[rgb]{0,0,0}\makebox(0,0)[lb]{\smash{(all in 2D)}}}%
  \end{picture}%
\endgroup%